\documentclass[conference]{IEEEtran}
\IEEEoverridecommandlockouts
\usepackage[T1]{fontenc}
\usepackage{cite}
\usepackage{amsmath,amssymb,amsfonts}
\usepackage{algorithmic}
\usepackage{graphicx}
\usepackage{textcomp}
\usepackage{xcolor}
\def\BibTeX{{\rm B\kern-.05em{\sc i\kern-.025em b}\kern-.08em
    T\kern-.1667em\lower.7ex\hbox{E}\kern-.125emX}}

\usepackage{adjustbox}
\usepackage[numbers,sort&compress]{natbib}
\usepackage[
    colorlinks=true,  
    linkcolor=blue,   
    citecolor=green,  
    urlcolor=cyan  
]{hyperref}
\usepackage{booktabs}
\usepackage{amsmath}
\usepackage[svgnames]{xcolor}
\usepackage{graphicx}  

\usepackage{threeparttable} 
\newcommand{\modelnote}{\tnote{*}}

\newcommand{\colornotetext}{\item[*] \footnotesize Did not evaluate all images due to content filtering. Metrics are reported on the evaluated subset. See the Results section for a breakdown. Green and red text indicate the highest and lowest F1 scores per column, respectively.}

\usepackage{booktabs}
\usepackage{multirow}

\usepackage{colortbl} %
\usepackage{xcolor}   
\usepackage{pgf}     
\usepackage{array}    

\definecolor{poorcorr}{HTML}{FFCDD2}   
\definecolor{weakcorr}{HTML}{FFF9C4}   
\definecolor{mediumcorr}{HTML}{E8F5E9} 
\definecolor{strongcorr}{HTML}{C8E6C9}  

\newcommand{\heatmapcell}[2][]{%
  \ifdim #2pt > 0.8pt \cellcolor{strongcorr} \else
  \ifdim #2pt > 0.6pt \cellcolor{mediumcorr} \else
  \ifdim #2pt > 0.4pt \cellcolor{weakcorr} \else
  \cellcolor{poorcorr} \fi\fi\fi
  \ifx#1b\bfseries\fi 
  #2
}

\definecolor{lowmae}{HTML}{C8E6C9}      
\definecolor{mediummae}{HTML}{E8F5E9}   
\definecolor{highmae}{HTML}{FFF9C4}     
\definecolor{veryhighmae}{HTML}{FFCDD2} 

\newcommand{\maeheatmapcell}[2][]{%
  \ifdim #2pt < 0.75pt \cellcolor{lowmae} \else
  \ifdim #2pt < 1.25pt \cellcolor{mediummae} \else
  \ifdim #2pt < 1.75pt \cellcolor{highmae} \else
  \cellcolor{veryhighmae} \fi\fi\fi
  \ifx#1b\bfseries\fi
  #2 
}

\usepackage{xurl}
\usepackage[caption=false,font=footnotesize]{subfig}
\usepackage{stfloats}
\usepackage{tcolorbox}
\newtcolorbox{promptbox}[1]{
  colback=gray!10!white,
  colframe=gray!50!black,
  title=#1,
  fonttitle=\bfseries
}

\makeatletter
\newcommand{\linebreakand}{%
  \end{@IEEEauthorhalign}
  \hfill\mbox{}\par
  \mbox{}\hfill\begin{@IEEEauthorhalign}
}
\makeatother

\begin{document}

\title{Visual Affect Analysis: Predicting Emotions of Image Viewers with Vision-Language Models\\
\thanks{This work was supported by the National Science Center in Poland (2020/39/B/HS6/00685, 2025/59/D/HS6/03333) and the IDUB program at Adam Mickiewicz University, Poznan (154/04/POB5/0001). The funders had no role in the study. We thank Przemysław Kazienko for feedback and PCSS for HPC access (grant pl0494-02).}
}

\author{
\IEEEauthorblockN{Filip Nowicki}
\IEEEauthorblockA{\textit{Adam Mickiewicz University} \\
Poznań, Poland \\
filip.nowicki@amu.edu.pl}
\and
\IEEEauthorblockN{Hubert Marciniak}
\IEEEauthorblockA{\textit{Adam Mickiewicz University} \\
Poznań, Poland \\
hubmar4@st.amu.edu.pl}
\and
\IEEEauthorblockN{Jakub Łączkowski}
\IEEEauthorblockA{\textit{Adam Mickiewicz University} \\
Poznań, Poland \\
jaklac3@st.amu.edu.pl}
\and
\IEEEauthorblockN{Bartosz Wilczek}
\IEEEauthorblockA{\textit{Adam Mickiewicz University} \\
Poznań, Poland \\
barwil2@st.amu.edu.pl}
\and
\IEEEauthorblockN{Krzysztof Jassem}
\IEEEauthorblockA{\textit{Adam Mickiewicz University} \\
Poznań, Poland \\
krzysztof.jassem@amu.edu.pl}
\and
\IEEEauthorblockN{Tomasz Górecki}
\IEEEauthorblockA{\textit{Adam Mickiewicz University} \\
Poznań, Poland \\
tomasz.gorecki@amu.edu.pl}
\and
\IEEEauthorblockN{Vimala Balakrishnan}
\IEEEauthorblockA{\textit{Universiti Malaya} \\
Kuala Lumpur, Malaysia \\
vimala.balakrishnan@um.edu.my}
\and
\IEEEauthorblockN{Desmond Ong}
\IEEEauthorblockA{\textit{University of Texas} \\
Austin, United States \\
desmond.ong@utexas.edu}

\linebreakand
\IEEEauthorblockN{Maciej Behnke}
\IEEEauthorblockA{\textit{Adam Mickiewicz University} \\
Poznań, Poland \\
maciej.behnke@amu.edu.pl}
}

\maketitle

\begin{abstract}
Vision-language models (VLMs) show promise for large-scale affect inference from images, but their alignment with human affective ratings remains unclear. We benchmarked 16 proprietary and open-source VLMs on three validated affective image datasets: the International Affective Picture System, the Nencki Affective Picture System, and the Library of AI-Generated Affective Images. In a zero-shot setting, models performed two tasks: (i) top-emotion classification and (ii) prediction of human ratings on 1--7/9 Likert scales for discrete emotions and affective dimensions. We also evaluated rater-conditioned prompting on the LAI-GAI dataset using de-identified participant metadata. Best models showed strong discrete emotion classification (60\%--80\% on six-emotion tasks; 60\%--75\% on a 12-category task), with anger and surprise predicted least accurately. For continuous ratings, models aligned moderately to strongly with human judgments ($r > 0.75$) but showed consistent biases, especially weaker performance on arousal and overestimation of response intensity. Rater-conditioned prompting had only small, inconsistent effects. Overall, VLMs captured broad affective patterns but sometimes lacked the nuance, although newer model generations performed better.
\end{abstract}

\begin{IEEEkeywords}
vision–language models, affective computing, emotion recognition
\end{IEEEkeywords}

\section{Introduction}
As large language models (LLMs) become embedded in everyday life, users increasingly rely on them for advice, support, and emotionally charged conversations. Although such interactions are a small share of total traffic, they occur at scale and can affect users’ well-being \cite{anthropic2025affective,phang2025openai_affective_use,roth2025chatgpt_prompts}. Socio-emotional use is already common: 24\% of U.S. adults reported using general-purpose LLMs for mental health purposes \cite{stade2025realworld_use}. This has prompted calls for \emph{socioaffective alignment}, emphasizing systems that support autonomy and well-being rather than merely optimize engagement \cite{kirk2025socioaffective}. In such settings, trust matters \cite{demszky2023using,picard2000affective}, especially because AI-generated responses may be perceived as less empathetic \cite{ong2025ai}.

A core question is whether the models can infer emotions in ways that match human judgments. This requires more than recognizing affect from faces, speech, or sentiment-laden text: affective cognition also depends on beliefs, expectations, and context \cite{ellsworth2003appraisal,ong2015affective,ong2019computational}. Although text-based foundation models can show human-like affective inference \cite{gandhi2024humanlike}, much less is known about whether vision--language models (VLMs) can predict how people respond to visual stimuli \cite{bhattacharyya2025evaluating,ogg2025large}. Images are central to both affective science and human--AI interaction: users share emotionally salient visual content, and images remain a common tool for emotion elicitation in research \cite{joseph2020manipulation}. Yet affective responses to images are often subtle, culturally shaped, and context-dependent \cite{barrett2022context}. 

We evaluate models’ ability to approximate human affective ratings by benchmarking open- and closed-source VLMs on three affective image datasets with existing human ratings: the International Affective Picture System (IAPS)~\cite{bradley2007international}, the Nencki Affective Picture System (NAPS)~\cite{marchewka2014nencki}, and the Library of AI-Generated Affective Images (LAI-GAI)~\cite{behnke2025using}. We also test whether conditioning prompts on de-identified rater background improves alignment with human ratings, extending silicon sampling to the multimodal setting \cite{argyle2023outofone,sun2024randomsiliconsampling}. We address three main research questions:
\renewcommand{\labelenumi}{\arabic{enumi}.}%
\begin{enumerate}
    \item How accurately do VLMs determine the top-rated discrete emotions assigned by human raters?
    \item How well do VLMs predict human Likert ratings across 18 discrete and dimensional emotion scales?
    \item Does conditioning on de-identified rater background (age, sex, country, and session-level initial emotional state) improve alignment between VLM predictions and human affective ratings?

\end{enumerate}

\section{Contributions}
Our contributions are: (1) a comprehensive evaluation of VLMs against psychometrically normed, multidimensional human ratings rather than categorical labels; (2) distributional prediction via $n{=}50$ sampling matched to rater-pool sizes, comparing both rating means and variability, and evidence of cross-dataset robustness across synthetic (LAI-GAI) and legacy (IAPS, NAPS) stimuli, including a reasoning-model reversal between them; (3) a content-filtering audit of proprietary models together with a test of rater-conditioned prompting; and (4) an open protocol with code, prompts, and evaluation scripts to support reproducibility.

\section{Related Work}
Affective scientists use AI to model emotion dynamics, classify affective behavior, build affect-aware robots, and analyze physiological signals \cite{coles2025big,cowen2021sixteen,saad,saganowski2022emotion,perz2025personalization}. Better affect understanding could support safer and more sensitive human--AI interaction in domains such as communication, healthcare, and creative work. Earlier methods relied on hand-engineered features or task-specific multimodal architectures for narrow tasks \cite{gu2018deep,lee2019context,liu2016emotion}. LLMs shifted attention toward general models for affect-related tasks. Recent benchmarks therefore examine not only reasoning but also socially grounded abilities such as affect recognition and empathy \cite{chen2024emotionqueen,feng2023affect,sorin2024large}. However, this work has focused mainly on text-only models, and may not extend to emotions elicited by images.

Early image-based affect analysis focused on visual features, object categories, and facial expressions \cite{zhao2014affective,zhao2014exploring,yao2020adaptive,rao2020learning,kundu2017advancements}. More recent work has begun to evaluate VLMs for emotion understanding \cite{zhang2025affective}. Although VLMs perform strongly on tasks such as visual question answering, multimodal reasoning, and captioning \cite{antol2015vqa,goyal2017making,alayrac2022flamingo,chen2022pali,li2023blip,zhang2021vinvl}, evidence on affect remains mixed. Bhattacharyya et al.~\cite{bhattacharyya2025evaluating} found that VLMs struggle with emotion prediction and are highly sensitive to prompt design, though some errors may reflect ambiguous ground-truth labels. Ogg et al.~\cite{ogg2025large} reported high alignment with human ratings for discrete emotions, but weaker performance on dimensional scales.

Performance in predicting image-elicited emotion depends not only on model capability but also on the evaluation dataset. Datasets differ in constructs, label quality, cultural diversity, and domain shift, so psychology- and computer-science-based benchmarks may probe different capabilities. Psychology-based affective image libraries are designed for controlled experiments and provide standardized protocols, use richer affective scales often grounded in newer theories, and provide normative ratings from relatively large samples \cite{behnke2025using}. Widely used datasets such as IAPS and NAPS contain about 1{,}000 images each, with ratings based on roughly 30--60 participants per image \cite{lang1997international,mikels2005emotional,marchewka2014nencki,riegel2016characterization}. LAI-GAI follows the same psychometric approach while using AI-generated, culturally adapted stimuli normed on an international sample \cite{behnke2025using}. In contrast, many computer-science affective image datasets, such as EmoSet, FI, Abstract, ArtPhoto, and Emotion6, prioritize scale and visual diversity, often relying on web crawling, crowdsourcing, or automatic labeling \cite{yang2023emoset,you2016building,machajdik2010affective,eomotion6}. They typically offer broader coverage but lighter validation, with more categorical or weakly supervised labels and smaller rater pools. This can increase label noise, so strong performance on such datasets may not translate to psychometrically grounded human affect ratings.

Given these differences, we focused on psychology-based datasets because they provide standardized, psychometrically grounded ratings that better fit our evaluation tasks. We therefore used IAPS and NAPS, and included LAI-GAI because it also provided participant-level metadata and allowed exact replication of the original rating protocol. We did not use large-scale computer-science datasets as primary benchmarks, as their labels are typically categorical, based on smaller rater pools or automatic pipelines, and not directly comparable to normative multidimensional ratings. Still, such datasets remain valuable for training and for testing scalability and robustness.

\section{Methods}
\subsection{Datasets}
We evaluated models on three psychology-based affective image datasets with normative ratings collected under standardized protocols. For each dataset, we used all available images and affective scales, inheriting the original taxonomies and rating formats.

\subsubsection{International Affective Picture System (IAPS)}
From IAPS~\cite{lang1997international}, we evaluated 692 images showing a variety of everyday and emotionally stimulating scenes. The ground truth for each image includes ratings of valence, arousal, and discrete emotions (anger, disgust, fear, happiness, sadness, and surprise) on 9-point scales (60 raters per image)~\cite{mikels2005emotional}.

\subsubsection{Nencki Affective Picture System (NAPS)}
From NAPS~\cite{marchewka2014nencki}, we evaluated 504 images across various semantic categories, including people, faces, animals, objects, and landscapes. The ground truth for each image includes ratings of valence, arousal, and discrete emotions (anger, disgust, fear, happiness, sadness, and surprise) on 7-point scales (39--44 raters per image)~\cite{riegel2016characterization}.

\subsubsection{Library of AI-Generated Affective Images (LAI-GAI)}
From LAI-GAI~\cite{behnke2025using}, we evaluated 480 AI-generated images designed to elicit 12 emotions. The ground truth for each image includes ratings of valence (positive and negative), arousal (arousing and calming), and motivational tendencies (approaching and avoiding), along with 12 ratings of discrete emotions (amusement, anger, attachment love, awe, craving, disgust, excitement, fear, joy, neutral, nurturant love, and sadness) on 7-point scales (50--70 raters per image).

\subsection{Tasks}
We evaluated VLMs on three tasks aligned with the ground truth available in IAPS, NAPS, and LAI-GAI. All tasks were performed in a zero-shot setting using standardized prompts and response constraints across models and datasets to support direct comparison under a reproducible protocol (see Supplementary Materials, Section I). As a robustness check, we also tested multiple prompt variants after the main study (see Supplementary Materials, Section I). Open-source models were run locally on NVIDIA H100 and RTX 4090 GPUs using vLLM \cite{kwon2023vllm}, whereas proprietary models were accessed through their official APIs. 

\subsubsection{Task 1: Emotion Classification}
Models predicted the discrete emotion with the highest average human rating in the original human studies. Each model produced one prediction per image, using a temperature of 0.0 and a fixed seed of 42.

\subsubsection{Task 2: Affective Dimension Prediction}
Models predicted affective ratings, namely how much of a given emotion (e.g., amusement) the image would elicit in an average person. For each image, models produced a single integer rating on scales matching the original datasets: 1--7 for NAPS and LAI-GAI, and 1--9 for IAPS. Generation was limited to one token. Ratings were generated using $n$-sampling with $n=50$ and temperature $=0.5$, with a fixed seed of 42. We chose $n=50$ to approximate the average number of human raters per image and temperature $=0.5$ to balance output diversity (i.e., sufficient stochasticity to calculate a mean and standard deviation, used as a proxy for, not a model of, human inter-rater variability) with stability, as preliminary tests showed poorer performance and instruction non-compliance in smaller models ($<9$B) at higher temperatures.

\subsubsection{Task 3: Rater-conditioned Prompting}
Models predicted affective ratings conditioned on de-identified rater background information. We compared baseline predictions without participant information (Table~\ref{tab:side_by_side}) with three prompting variants: ``Full Background'' (age, gender, country, emotional state), ``Demographic Background'' (age, gender, country), and ``Emotional Background'' (emotional state only). For each human rating, we generated a corresponding model prediction for affective dimensions using the same parameters as in Task~2. This task was conducted only on LAI-GAI, the only dataset with the required rater-level metadata.

\subsection{Models}
We evaluated state-of-the-art VLMs selected in July 2025, including both proprietary and open-source systems spanning multiple architectures and scales. All models used pre-trained LLMs as language backbones. A pilot study first tested nine proprietary models on LAI-GAI; based on these results, the latest Gemini and GPT variants were selected for the main evaluation. The proprietary models were evaluated in October 2025, reflecting the API behavior and checkpoint versions available at that time. To focus on core vision--language perception and instruction following, the main benchmark used non-thinking variants only, although thinking models were added in later evaluations (March 2026). Thinking (or reasoning) models~\cite{openai2025reasoning,google2025gemini3} are a class of VLMs that employ explicit chain-of-thought reasoning at inference time, generating intermediate reasoning steps before producing a final answer.

We used the following proprietary and open-source models:

\subsubsection{Proprietary:}
\begin{itemize} 
    \item \textbf{GPT-4.1}~\cite{openai_gpt4_1}: The most innovative non-thinking model from OpenAI available at the time of testing, excelling at instruction following and image understanding.
    \item \textbf{GPT-5.4}~\cite{openai2026gpt54}: A recent model from OpenAI that advances multimodal reasoning and instruction following capabilities beyond its predecessors.
    \item \textbf{Gemini-2.5-Flash}~\cite{comanici2025gemini}: A lightweight Gemini 2.5 model from Google DeepMind designed for fast multimodal inference while retaining multimodal capabilities.
    \item \textbf{Gemini-3-Pro}~\cite{google2025gemini3}: A flagship Gemini 3 model from Google DeepMind for multimodal understanding, reasoning, and generation.
    \item \textbf{Gemini-3-Flash}~\cite{google2025gemini3flash}: A lightweight Gemini 3 model from Google DeepMind optimized for fast multimodal inference while retaining multimodal capabilities.
\end{itemize}
\subsubsection{Open-source:}
\begin{itemize} 
    \item \textbf{Qwen-VL-2.5 72B \& 7B}~\cite{Qwen2.5-VL}: a flagship VLMs from Qwen, proficient in analyzing complex visual inputs, including text, charts, and graphics within images.
    \item \textbf{Qwen-VL-3-Thinking 30B \& 8B}~\cite{Qwen3-VL}: a reasoning-enhanced VLM from Qwen with chain-of-thought capabilities for complex visual understanding tasks.
    \item \textbf{Intern-VL-3.5 38B \& 8B}~\cite{wang2025internvl35}: an open-source multimodal models designed by OpenGVLab for enhanced versatility, reasoning, and efficiency.
    \item \textbf{Gemma-3 27B}~\cite{kamath2025gemma3}: a lightweight, open-weight model from Google built from the technology used to create Gemini models.
    \item \textbf{Kimi-VL-A3B-Instruct}~\cite{wu2025kimivl}: an open-source Mixture-of-Experts (MoE) VLM from Moonshot AI.
    \item \textbf{GLM-4.1-V-Base}~\cite{wu2025kimivl}: an open-source VLM from Zhipu AI and Tsinghua University designed for general-purpose multimodal understanding.
    \item \textbf{LLaVA-Mistral}~\cite{llavamistral}: a LLaVA variant built on the Mistral-7B language model for multimodal tasks.
    \item \textbf{LLaVA-Vicuna}~\cite{llavavicuna}: a pioneering open-source multimodal model combining a pre-trained CLIP vision encoder with the Vicuna language model.
\end{itemize}

All models were first evaluated on emotion classification across all datasets. The six top-performing models (two proprietary and four open-source) were selected for affective dimension prediction to reduce computational cost and energy use. We note that classification and rating calibration are distinct capabilities, so this Task~1--based selection may overlook models that rate well despite weaker classification. For the rater-conditioned analysis, we selected the best open- and closed-source models on the LAI-GAI dataset: Gemma-3 27B and Gemini-2.5-Flash. We also evaluated several thinking-enhanced models on emotion classification to test whether explicit reasoning improves performance, including Gemini-3-Pro, Gemini-3-Flash, GPT-5.4, and Qwen-VL-3-Thinking (8B, 30B). Of these, only Gemini-3-Flash was additionally evaluated on affective dimension prediction because it offered the best balance of performance and efficiency.

\subsection{Evaluation Metrics}
For emotion classification, we report per-emotion F1 scores, with ground truth defined as the top-rated emotion in the original validation studies (see Supplementary Materials for other metrics, Section III). Because emotion categories were often imbalanced, we report both overall and class-level performance. For continuous affective ratings, we report alignment between model and human ratings using Pearson correlation ($r$) and mean absolute error (MAE). For rater-conditioned prompting, we report effects using Cohen's $d$ with 95\% confidence intervals \cite{lakens2013calculating}.

\section{Results}
Before reporting task performance, we note an evaluation constraint affecting proprietary models: API-level content filtering. Closed-source models occasionally refused to process images due to internal safety policies. Gemini-2.5-Flash blocked 88 images across datasets (58 IAPS, 15 NAPS, 15 LAI-GAI), whereas GPT-4.1 blocked 34 images, all from IAPS. In contrast, newer proprietary models (GPT-5.4, Gemini-3-Pro, and Gemini-3-Flash) did not block any images, suggesting less restrictive safety policies in these later versions (including those from IAPS that contain nude content). Reported metrics for the older proprietary models were therefore computed on the evaluated subsets; detailed non-filtered subset-specific results are provided in the Supplementary Materials (see Section IV). On identical non-blocked subsets the substantive conclusions hold, with rank shifts occurring only between statistically tied models (e.g., Gemini-3-Flash still leads on the 634 unblocked IAPS images: $71.8$ vs.\ $66.9$, $p<.001$); nonetheless, filtered models effectively see a partially different, likely less intense, stimulus distribution. To illustrate visual-to-affective mapping, we provide t-SNE visualizations of the affective stimulus space comparing human and model emotion labels in the Supplementary Materials (see Section~II).

\subsubsection{Task 1: Emotion Classification}
Overall weighted F1 was highest on NAPS (six emotions), where the best-performing standard models reached $79$--$80\%$ (e.g., Qwen-VL-2.5-72B: $80.42\%$; GPT-4.1: $79.38\%$). Weighted F1 on LAI-GAI (twelve emotions) was lower, with leading models reaching approximately $70$--$75\%$ (e.g., Gemma-3-27B: $74.79\%$; Gemini-2.5-Flash: $70.48\%$). Performance was lowest on IAPS (six emotions), where the best models reached approximately $65$--$67\%$ (e.g., Gemini-2.5-Flash: $66.94\%$; GPT-4.1: $65.62\%$). Across datasets, per-class F1 was consistently weakest for anger and surprise. Among the thinking models, Gemini-3-Flash-Preview achieved the highest weighted F1 on both IAPS ($68.51\%$) and NAPS ($84.18\%$), surpassing all standard models. GPT-5.4 also improved over its predecessor GPT-4.1 on both datasets. However, the open-source Qwen-VL-3-Thinking variants (30B and 8B) did not benefit from thinking in the same way, both falling below the non-thinking Qwen-VL-2.5-72B (Tables~\ref{tab:combined_accuracy_summary}--\ref{tab:laigai_accuracy_summary}). Bootstrap 95\% confidence intervals and paired permutation tests (see Supplementary Materials, Section~VII) indicate that Qwen-VL-2.5-72B's NAPS lead over GPT-4.1 is not significant ($+1.0$, $p=.49$), whereas the generational gains are: Gemini-3-Flash over the best standard model ($+3.8$, $p=.009$), GPT-5.4 over GPT-4.1 ($+3.7$, $p=.003$), and the LAI-GAI reasoning reversal ($-10.3$, $p<.001$).
\begin{table}[!tbh]
    \centering
    \begin{adjustbox}{width=\columnwidth,center}
        \begin{threeparttable}
            \caption{Per-Emotion F1 Scores (\%) on IAPS and NAPS Datasets}
            \label{tab:combined_accuracy_summary}
            \begin{tabular}{lcccccc|c}
                \toprule
                \textbf{Model} & \textbf{Anger} & \textbf{Disgust} & \textbf{Fear} & \textbf{Happiness} & \textbf{Sadness} & \textbf{Surprise} & \textbf{Weighted F1} \\
                \midrule
                
                \multicolumn{8}{c}{\textbf{IAPS Dataset}} \\
                \textit{Support ($N$)} & \textit{(22)} & \textit{(67)} & \textit{(67)} & \textit{(372)} & \textit{(103)} & \textit{(61)} & \textit{(692)} \\
                \midrule
                GPT-4.1\modelnote & 11.11 & 58.33 & 55.24 & 79.68 & 59.30 & 14.74 & 65.62 \\
                GPT-5.4 & 18.18 & \textcolor{green}{61.04} & 55.02 & 82.56 & 58.38 & \textcolor{green}{27.37} & 67.30 \\
                Gemini-2.5-Flash\modelnote & 18.18 & 59.13 & 60.44 & 77.85 & \textcolor{green}{67.02} & 16.82 & 66.94 \\
                Gemini-3-Flash-Preview & \textcolor{red}{0.00} & 57.32 & \textcolor{green}{61.86} & \textcolor{green}{87.38} & 60.78 & 10.81 & \textcolor{green}{68.51} \\
                Gemini-3-Pro-Preview & 12.90 & 55.70 & 56.04 & 85.59 & 53.81 & 10.81 & 66.20 \\
                Qwen-VL-2.5-72B & 14.63 & 56.10 & 59.04 & 82.98 & 57.58 & 6.06 & 65.33 \\
                Qwen-VL-3-Thinking-30B & 6.25 & 54.05 & 54.63 & 83.45 & 51.69 & \textcolor{red}{2.78} & 63.52 \\
                Intern-VL-3.5-38B & 5.56 & 43.55 & 43.28 & 79.88 & 50.81 & 10.42 & 60.01 \\
                Gemma-3-27B & \textcolor{red}{0.00} & 54.55 & 55.88 & 69.15 & 56.00 & 19.58 & 57.93 \\
                Kimi-VL-A3B-16B & 7.14 & 50.57 & 58.75 & 75.58 & 62.77 & 12.90 & 61.92 \\
                GLM-V-4.1-9B & \textcolor{green}{24.62} & 44.44 & 46.70 & 80.30 & 44.83 & 5.33 & 59.92 \\
                Intern-VL-3.5-8B & 14.29 & 35.85 & 51.61 & 78.97 & 57.25 & 8.51 & 60.65 \\
                Qwen-VL-3-Thinking-8B & 16.22 & 39.06 & 50.00 & 80.35 & 51.64 & 9.30 & 60.84 \\
                Qwen-VL-2.5-7B & 23.73 & 50.00 & 43.33 & 84.34 & 58.71 & 9.52 & 64.71 \\
                LLaVA-Mistral-7B & 6.25 & 32.10 & 44.35 & 73.73 & 53.27 & 11.32 & 56.16 \\
                LLaVA-Vicuna-7B & 7.41 & \textcolor{red}{26.97} & \textcolor{red}{34.04} & \textcolor{red}{46.96} & \textcolor{red}{31.52} & 13.61 & \textcolor{red}{37.28} \\
                \midrule
                \textbf{IAPS Average} & \textbf{11.65} & \textbf{48.67} & \textbf{51.89} & \textbf{78.05} & \textbf{54.71} & \textbf{11.87} & \textbf{61.43} \\
                \midrule

                \multicolumn{8}{c}{\textbf{NAPS Dataset}} \\
                \textit{Support ($N$)} & \textit{(7)} & \textit{(85)} & \textit{(23)} & \textit{(268)} & \textit{(102)} & \textit{(19)} & \textit{(504)} \\
                \midrule
                GPT-4.1\modelnote & \textcolor{red}{0.00} & 82.29 & 65.67 & 86.61 & 78.85 & 13.04 & 79.38 \\
                GPT-5.4 & \textcolor{red}{0.00} & 80.79 & 69.70 & 92.49 & 81.05 & 18.75 & 83.10 \\
                Gemini-2.5-Flash\modelnote & 25.00 & 71.50 & \textcolor{green}{79.17} & 82.35 & 79.23 & 4.88 & 76.15 \\
                Gemini-3-Flash-Preview & \textcolor{red}{0.00} & 83.25 & 73.68 & \textcolor{green}{93.59} & 82.30 & 9.52 & \textcolor{green}{84.18} \\
                Gemini-3-Pro-Preview & \textcolor{red}{0.00} & \textcolor{green}{84.54} & 68.75 & 92.91 & 77.51 & 9.52 & 82.85 \\
                Qwen-VL-2.5-72B & 18.18 & 77.08 & 71.43 & 87.60 & \textcolor{green}{82.95} & 14.29 & 80.42 \\
                Qwen-VL-3-Thinking-30B & 15.38 & 73.73 & 75.00 & 88.18 & 71.36 & \textcolor{red}{0.00} & 77.40 \\
                Intern-VL-3.5-38B & \textcolor{red}{0.00} & 80.98 & 48.84 & 89.11 & 77.21 & \textcolor{green}{21.62} & 79.71 \\
                Gemma-VL-3-27B & \textcolor{red}{0.00} & 82.00 & 73.08 & 78.00 & 73.55 & 12.50 & 74.00 \\
                Kimi-VL-A3B-16B & 22.22 & 72.82 & 70.83 & 82.83 & 76.14 & 13.70 & 75.79 \\
                GLM-V-4.1-9B & \textcolor{green}{28.57} & 66.09 & 40.58 & 84.89 & 65.17 & \textcolor{red}{0.00} & 71.72 \\
                Intern-VL-3.5-8B & 20.00 & 63.57 & 76.60 & 87.57 & 69.78 & 20.00 & 75.94 \\
                Qwen-VL-3-Thinking-8B & 26.67 & 77.46 & 52.31 & 87.73 & 73.30 & 18.18 & 77.99 \\
                Qwen-VL-2.5-7B & 14.29 & 73.40 & 34.48 & 88.63 & 78.92 & \textcolor{red}{0.00} & 77.25 \\
                LLaVA-Mistral-7B & 20.00 & 55.00 & 25.35 & 77.78 & 75.44 & 15.87 & 67.94 \\
                LLaVA-Vicuna-7B & \textcolor{red}{0.00} & \textcolor{red}{47.03} & \textcolor{red}{22.22} & \textcolor{red}{58.64} & \textcolor{red}{47.30} & 7.50 & \textcolor{red}{49.98} \\
                \midrule
                \textbf{NAPS Average} & \textbf{11.89} & \textbf{73.22} & \textbf{59.23} & \textbf{84.93} & \textbf{74.38} & \textbf{11.21} & \textbf{75.86} \\
                \bottomrule
            \end{tabular}
            \begin{tablenotes}
                \colornotetext
            \end{tablenotes}
        \end{threeparttable}
    \end{adjustbox}
\end{table}

\subsubsection{Task 2: Affective Dimension Prediction}

In the affective dimension prediction task, we assessed alignment between model and human ratings using Pearson correlation and mean absolute error (MAE). As shown in Table~\ref{tab:side_by_side}, model ratings were positively correlated with human judgments across all three datasets. The highest average correlation was observed for LAI-GAI ($r = 0.82$), followed by NAPS ($r = 0.76$) and IAPS ($r = 0.71$). Alignment was especially strong for discrete negative emotions like disgust, fear, and sadness ($r > 0.79$), as well as for dimensional valence ($r > 0.84$). However, all models consistently struggled with arousal and surprise, with correlations for arousal ranging from $r = 0.44$ (LAI-GAI) to $r = 0.52$ (IAPS), and surprise reaching only $r = 0.39$ on IAPS and $r = 0.54$ on NAPS.

\begin{table*}[!t]
    \centering
    \begin{adjustbox}{max width=\linewidth,center}
        \begin{threeparttable}
            \caption{Per-Emotion F1 Scores (\%) on LAI-GAI Dataset}
            \label{tab:laigai_accuracy_summary}
            \begin{tabular}{lcccccccccccc | c}
                \toprule
                \textbf{Model} & \textbf{Amusement} & \textbf{Anger} & \textbf{Attach. Love} & \textbf{Awe} & \textbf{Craving} & \textbf{Disgust} & \textbf{Excitement} & \textbf{Fear} & \textbf{Joy} & \textbf{Neutral} & \textbf{Nurt. Love} & \textbf{Sadness} & \textbf{Weighted F1} \\
                \textit{Support ($N$)} & \textit{(25)} & \textit{(14)} & \textit{(8)} & \textit{(30)} & \textit{(40)} & \textit{(35)} & \textit{(20)} & \textit{(42)} & \textit{(150)} & \textit{(42)} & \textit{(14)} & \textit{(60)} & \textit{(480)} \\
                \midrule
                GPT-4.1\modelnote & 74.58 & 23.53 & 25.00 & 66.67 & \textcolor{green}{100.00} & 76.40 & \textcolor{green}{80.00} & 80.81 & 50.24 & 62.86 & 52.17 & 71.70 & 65.15 \\
                GPT-5.4 & 73.33 & 44.44 & 26.32 & 68.49 & \textcolor{green}{100.00} & 75.86 & 68.97 & 87.23 & 39.38 & 68.49 & 34.21 & 75.93 & 63.03 \\
                Gemini-2.5-Flash\modelnote & 69.84 & 57.14 & 29.41 & 73.53 & \textcolor{green}{100.00} & 76.19 & 70.59 & \textcolor{green}{91.95} & 55.96 & \textcolor{green}{74.07} & 38.10 & 83.17 & 70.48 \\
                Gemini-3-Flash-Preview & 69.84 & 57.14 & 22.22 & 62.79 & \textcolor{green}{100.00} & 84.62 & 78.43 & 87.91 & \textcolor{red}{28.41} & 62.69 & 30.00 & 85.95 & 61.04 \\
                Gemini-3-Pro-Preview & 66.67 & \textcolor{green}{66.67} & 32.26 & 63.29 & 98.77 & 81.48 & 70.18 & 88.37 & 32.04 & 69.33 & 29.89 & 80.70 & 61.78 \\
                Qwen-VL-2.5-72B & 73.33 & 58.33 & 31.37 & 68.49 & 59.65 & 81.48 & 65.38 & 85.71 & 64.00 & 54.84 & \textcolor{green}{60.87} & 82.05 & 68.29 \\
                Qwen-VL-3-Thinking-30B & 72.73 & 13.33 & 33.33 & 71.23 & 87.32 & 69.57 & 66.67 & 86.32 & 72.20 & 58.67 & 37.04 & 70.48 & 69.45 \\
                Intern-VL-3.5-38B & 45.71 & 25.00 & 25.45 & 66.67 & 90.41 & 85.29 & 66.67 & 78.85 & 66.20 & 58.46 & \textcolor{red}{0.00} & 83.64 & 67.39 \\
                Gemma-3-27B & \textcolor{green}{75.00} & 11.76 & 45.45 & 67.53 & 98.77 & \textcolor{green}{86.84} & 66.67 & 87.91 & 74.16 & 64.79 & 52.63 & 81.20 & \textcolor{green}{74.79} \\
                Kimi-VL-A3B-16B & 57.14 & 25.00 & \textcolor{green}{50.00} & \textcolor{green}{76.36} & 98.77 & 62.96 & 77.27 & 76.74 & \textcolor{green}{77.51} & 70.27 & 10.00 & \textcolor{red}{63.83} & 70.71 \\
                GLM-V-4.1-9B & 41.03 & 56.25 & 26.32 & \textcolor{red}{59.52} & 91.89 & 70.45 & 53.33 & 72.92 & 53.11 & 54.29 & 15.38 & 66.02 & 59.38 \\
                Intern-VL-3.5-8B & 57.89 & 30.00 & 27.27 & 73.24 & 26.09 & 75.76 & 60.87 & 74.67 & 73.32 & 39.29 & \textcolor{red}{0.00} & \textcolor{green}{87.41} & 62.96 \\
                Qwen-VL-3-Thinking-8B & 65.12 & \textcolor{red}{11.11} & 21.33 & 58.06 & 90.41 & 77.65 & 54.05 & 80.85 & 61.13 & 38.46 & 9.09 & 75.73 & 62.42 \\
                Qwen-VL-2.5-7B & 65.62 & 40.00 & 18.60 & 74.58 & 84.51 & 72.94 & 61.82 & 62.50 & 53.44 & 37.04 & \textcolor{red}{0.00} & 79.37 & 59.82 \\
                LLaVA-Mistral-7B & 58.54 & \textcolor{red}{11.11} & 13.33 & 60.24 & \textcolor{red}{26.09} & 75.76 & 37.04 & 72.53 & 70.59 & \textcolor{red}{31.37} & \textcolor{red}{0.00} & 76.39 & 57.30 \\
                LLaVA-Vicuna-7B & \textcolor{red}{19.35} & \textcolor{red}{11.11} & \textcolor{red}{0.00} & 18.59 & \textcolor{red}{26.09} & \textcolor{red}{62.75} & \textcolor{red}{0.00} & \textcolor{red}{4.44} & 53.15 & \textcolor{red}{0.00} & \textcolor{red}{0.00} & 73.83 & \textcolor{red}{35.47} \\
                \midrule
                \textbf{Average} & \textbf{61.61} & \textbf{33.87} & \textbf{26.73} & \textbf{64.33} & \textbf{79.92} & \textbf{76.00} & \textbf{61.12} & \textbf{76.23} & \textbf{57.80} & \textbf{52.81} & \textbf{23.09} & \textbf{77.34} & \textbf{63.07} \\
                \bottomrule
            \end{tabular}
            \begin{tablenotes}
                \colornotetext
            \end{tablenotes}
        \end{threeparttable}
    \end{adjustbox}
    \vspace{0.1cm} 
    \scriptsize 
    \setlength{\tabcolsep}{3pt} 
    \renewcommand{\arraystretch}{1.05} 
    \newcommand{\up}{\textcolor{gray}{$\uparrow$}}
    \newcommand{\down}{\textcolor{gray}{$\downarrow$}}
    \newcommand{\rot}[1]{\multicolumn{1}{c}{\rotatebox{45}{\textbf{#1}}}}
    
    \begin{adjustbox}{max width=\linewidth,center} 
        \begin{threeparttable}
            \caption{Combined Performance: Side-by-Side Comparison of Pearson's Correlation ($r$) and Mean Absolute Error (MAE)}
            \label{tab:side_by_side}
            
            \begin{tabular}{ll | ccccccc cc | ccccccc c}
                \toprule
                \multicolumn{2}{c}{} & 
                \multicolumn{9}{c}{\textbf{Pearson's Correlation ($r$)} \up} & 
                \multicolumn{8}{c}{\textbf{Mean Absolute Error (MAE)} \down} \\
                \cmidrule(lr){3-11} \cmidrule(lr){12-19}
                
                \multicolumn{2}{c}{\textbf{Scale}} & 
                \rot{Gemini 3 Flash} & \rot{GPT-4.1\modelnote} & \rot{Gemini 2.5\modelnote} & \rot{Qwen 72B} & \rot{Gemma 27B} & \rot{Intern 8B} & \rot{Qwen 7B} & \rot{Average} & \rot{Human ceiling} & 
                \rot{Gemini 3 Flash} & \rot{GPT-4.1\modelnote} & \rot{Gemini 2.5\modelnote} & \rot{Qwen 72B} & \rot{Gemma 27B} & \rot{Intern 8B} & \rot{Qwen 7B} & \rot{Average} \\
                \midrule
                
                \multirow{19}{*}{\rotatebox{90}{\textbf{LAI--GAI}}}
                & Amusement
                & \heatmapcell{0.65} & \heatmapcell{0.76} & \heatmapcell{0.74} & \heatmapcell{0.80} & \heatmapcell{0.79} & \heatmapcell{0.73} & \heatmapcell{0.74} & \heatmapcell[b]{0.74} & 0.94
                & \maeheatmapcell{1.37} & \maeheatmapcell{1.24} & \maeheatmapcell{1.47} & \maeheatmapcell{0.90} & \maeheatmapcell{0.91} & \maeheatmapcell{1.27} & \maeheatmapcell{1.03} & \maeheatmapcell[b]{1.17} \\
                
                & Anger 
                & \heatmapcell{0.88} & \heatmapcell{0.88} & \heatmapcell{0.91} & \heatmapcell{0.75} & \heatmapcell{0.81} & \heatmapcell{0.64} & \heatmapcell{0.54} & \heatmapcell[b]{0.77} & 0.97
                & \maeheatmapcell{0.45} & \maeheatmapcell{0.59} & \maeheatmapcell{0.71} & \maeheatmapcell{0.60} & \maeheatmapcell{0.57} & \maeheatmapcell{0.67} & \maeheatmapcell{0.75} & \maeheatmapcell[b]{0.62} \\
                
                & Attach. Love 
                & \heatmapcell{0.86} & \heatmapcell{0.86} & \heatmapcell{0.89} & \heatmapcell{0.88} & \heatmapcell{0.88} & \heatmapcell{0.88} & \heatmapcell{0.87} & \heatmapcell[b]{0.87} & 0.93
                & \maeheatmapcell{0.91} & \maeheatmapcell{1.32} & \maeheatmapcell{1.10} & \maeheatmapcell{0.99} & \maeheatmapcell{0.88} & \maeheatmapcell{1.33} & \maeheatmapcell{1.25} & \maeheatmapcell[b]{1.11} \\
                
                & Awe 
                & \heatmapcell{0.75} & \heatmapcell{0.56} & \heatmapcell{0.69} & \heatmapcell{0.73} & \heatmapcell{0.71} & \heatmapcell{0.72} & \heatmapcell{0.74} & \heatmapcell[b]{0.70} & 0.89
                & \maeheatmapcell{0.99} & \maeheatmapcell{1.17} & \maeheatmapcell{1.18} & \maeheatmapcell{0.91} & \maeheatmapcell{0.71} & \maeheatmapcell{1.49} & \maeheatmapcell{0.98} & \maeheatmapcell[b]{1.06} \\
                
                & Craving 
                & \heatmapcell{0.89} & \heatmapcell{0.88} & \heatmapcell{0.88} & \heatmapcell{0.88} & \heatmapcell{0.90} & \heatmapcell{0.83} & \heatmapcell{0.91} & \heatmapcell[b]{0.88} & 0.94
                & \maeheatmapcell{0.81} & \maeheatmapcell{0.89} & \maeheatmapcell{0.87} & \maeheatmapcell{0.66} & \maeheatmapcell{0.44} & \maeheatmapcell{0.72} & \maeheatmapcell{0.48} & \maeheatmapcell[b]{0.70} \\
                
                & Disgust 
                & \heatmapcell{0.96} & \heatmapcell{0.95} & \heatmapcell{0.91} & \heatmapcell{0.92} & \heatmapcell{0.93} & \heatmapcell{0.90} & \heatmapcell{0.92} & \heatmapcell[b]{0.93} & 0.98
                & \maeheatmapcell{0.40} & \maeheatmapcell{0.70} & \maeheatmapcell{0.61} & \maeheatmapcell{0.55} & \maeheatmapcell{0.64} & \maeheatmapcell{0.76} & \maeheatmapcell{0.49} & \maeheatmapcell[b]{0.59} \\
                
                & Excitement 
                & \heatmapcell{0.63} & \heatmapcell{0.60} & \heatmapcell{0.72} & \heatmapcell{0.58} & \heatmapcell{0.59} & \heatmapcell{0.70} & \heatmapcell{0.71} & \heatmapcell[b]{0.65} & 0.94
                & \maeheatmapcell{0.99} & \maeheatmapcell{1.21} & \maeheatmapcell{1.10} & \maeheatmapcell{1.15} & \maeheatmapcell{0.96} & \maeheatmapcell{1.45} & \maeheatmapcell{0.99} & \maeheatmapcell[b]{1.12} \\
                
                & Fear 
                & \heatmapcell{0.93} & \heatmapcell{0.96} & \heatmapcell{0.96} & \heatmapcell{0.95} & \heatmapcell{0.94} & \heatmapcell{0.92} & \heatmapcell{0.89} & \heatmapcell[b]{0.94} & 0.97
                & \maeheatmapcell{0.47} & \maeheatmapcell{0.78} & \maeheatmapcell{0.70} & \maeheatmapcell{0.56} & \maeheatmapcell{0.62} & \maeheatmapcell{0.81} & \maeheatmapcell{0.51} & \maeheatmapcell[b]{0.64} \\
                
                & Joy 
                & \heatmapcell{0.95} & \heatmapcell{0.95} & \heatmapcell{0.95} & \heatmapcell{0.95} & \heatmapcell{0.93} & \heatmapcell{0.95} & \heatmapcell{0.90} & \heatmapcell[b]{0.94} & 0.97
                & \maeheatmapcell{0.66} & \maeheatmapcell{1.16} & \maeheatmapcell{1.14} & \maeheatmapcell{0.86} & \maeheatmapcell{0.75} & \maeheatmapcell{1.46} & \maeheatmapcell{0.90} & \maeheatmapcell[b]{0.99} \\
                
                & Neutral 
                & \heatmapcell{0.76} & \heatmapcell{0.61} & \heatmapcell{0.70} & \heatmapcell{0.62} & \heatmapcell{0.66} & \heatmapcell{0.43} & \heatmapcell{0.60} & \heatmapcell[b]{0.63} & 0.86
                & \maeheatmapcell{0.92} & \maeheatmapcell{1.08} & \maeheatmapcell{1.11} & \maeheatmapcell{1.11} & \maeheatmapcell{0.58} & \maeheatmapcell{1.09} & \maeheatmapcell{0.52} & \maeheatmapcell[b]{0.92} \\
                
                & Nurt. Love 
                & \heatmapcell{0.83} & \heatmapcell{0.87} & \heatmapcell{0.86} & \heatmapcell{0.87} & \heatmapcell{0.88} & \heatmapcell{0.84} & \heatmapcell{0.85} & \heatmapcell[b]{0.86} & 0.92
                & \maeheatmapcell{1.23} & \maeheatmapcell{1.21} & \maeheatmapcell{1.26} & \maeheatmapcell{1.04} & \maeheatmapcell{0.81} & \maeheatmapcell{1.35} & \maeheatmapcell{1.03} & \maeheatmapcell[b]{1.13} \\
                
                & Sadness 
                & \heatmapcell{0.96} & \heatmapcell{0.95} & \heatmapcell{0.95} & \heatmapcell{0.92} & \heatmapcell{0.90} & \heatmapcell{0.91} & \heatmapcell{0.87} & \heatmapcell[b]{0.92} & 0.98
                & \maeheatmapcell{0.47} & \maeheatmapcell{0.52} & \maeheatmapcell{0.61} & \maeheatmapcell{0.65} & \maeheatmapcell{0.76} & \maeheatmapcell{0.81} & \maeheatmapcell{0.56} & \maeheatmapcell[b]{0.63} \\
                
                & Positive 
                & \heatmapcell{0.97} & \heatmapcell{0.96} & \heatmapcell{0.94} & \heatmapcell{0.97} & \heatmapcell{0.95} & \heatmapcell{0.96} & \heatmapcell{0.96} & \heatmapcell[b]{0.96} & 0.98
                & \maeheatmapcell{0.58} & \maeheatmapcell{0.97} & \maeheatmapcell{0.88} & \maeheatmapcell{0.60} & \maeheatmapcell{0.52} & \maeheatmapcell{1.00} & \maeheatmapcell{0.66} & \maeheatmapcell[b]{0.74} \\
                
                & Negative 
                & \heatmapcell{0.97} & \heatmapcell{0.97} & \heatmapcell{0.96} & \heatmapcell{0.96} & \heatmapcell{0.95} & \heatmapcell{0.94} & \heatmapcell{0.95} & \heatmapcell[b]{0.96} & 0.99
                & \maeheatmapcell{0.56} & \maeheatmapcell{0.73} & \maeheatmapcell{0.69} & \maeheatmapcell{0.64} & \maeheatmapcell{0.64} & \maeheatmapcell{0.70} & \maeheatmapcell{0.54} & \maeheatmapcell[b]{0.64} \\
                
                & Calm 
                & \heatmapcell{0.86} & \heatmapcell{0.87} & \heatmapcell{0.85} & \heatmapcell{0.82} & \heatmapcell{0.89} & \heatmapcell{0.80} & \heatmapcell{0.78} & \heatmapcell[b]{0.84} & 0.96
                & \maeheatmapcell{0.83} & \maeheatmapcell{0.93} & \maeheatmapcell{0.90} & \maeheatmapcell{1.00} & \maeheatmapcell{0.86} & \maeheatmapcell{1.26} & \maeheatmapcell{0.98} & \maeheatmapcell[b]{0.97} \\
                
                & Aroused 
                & \heatmapcell{0.39} & \heatmapcell{0.45} & \heatmapcell{0.41} & \heatmapcell{0.42} & \heatmapcell{0.32} & \heatmapcell{0.56} & \heatmapcell{0.53} & \heatmapcell[b]{0.44} & 0.69
                & \maeheatmapcell{1.52} & \maeheatmapcell{1.75} & \maeheatmapcell{2.19} & \maeheatmapcell{2.12} & \maeheatmapcell{1.27} & \maeheatmapcell{1.86} & \maeheatmapcell{1.87} & \maeheatmapcell[b]{1.80} \\
                
                & Motiv. Appr. 
                & \heatmapcell{0.92} & \heatmapcell{0.54} & \heatmapcell{0.87} & \heatmapcell{0.91} & \heatmapcell{0.87} & \heatmapcell{0.86} & \heatmapcell{0.87} & \heatmapcell[b]{0.83} & 0.95
                & \maeheatmapcell{0.70} & \maeheatmapcell{1.51} & \maeheatmapcell{1.24} & \maeheatmapcell{1.02} & \maeheatmapcell{0.76} & \maeheatmapcell{1.23} & \maeheatmapcell{0.86} & \maeheatmapcell[b]{1.05} \\
                
                & Motiv. Avoid 
                & \heatmapcell{0.96} & \heatmapcell{0.96} & \heatmapcell{0.94} & \heatmapcell{0.95} & \heatmapcell{0.93} & \heatmapcell{0.93} & \heatmapcell{0.93} & \heatmapcell[b]{0.94} & 0.97
                & \maeheatmapcell{0.76} & \maeheatmapcell{1.17} & \maeheatmapcell{0.93} & \maeheatmapcell{0.80} & \maeheatmapcell{0.85} & \maeheatmapcell{0.85} & \maeheatmapcell{0.52} & \maeheatmapcell[b]{0.84} \\
                
                & \textbf{Average} 
                & \heatmapcell[b]{0.84} & \heatmapcell[b]{0.81} & \heatmapcell[b]{0.84} & \heatmapcell[b]{0.83} & \heatmapcell[b]{0.82} & \heatmapcell[b]{0.81} & \heatmapcell[b]{0.81} & \heatmapcell[b]{0.82} & \textbf{0.93}
                & \maeheatmapcell[b]{0.81} & \maeheatmapcell[b]{1.05} & \maeheatmapcell[b]{1.04} & \maeheatmapcell[b]{0.90} & \maeheatmapcell[b]{0.75} & \maeheatmapcell[b]{1.06} & \maeheatmapcell[b]{0.83} & \maeheatmapcell[b]{0.93} \\
                \midrule

                \multirow{9}{*}{\rotatebox{90}{\textbf{IAPS}}}
                & Anger 
                & \heatmapcell{0.84} & \heatmapcell{0.83} & \heatmapcell{0.76} & \heatmapcell{0.62} & \heatmapcell{0.69} & \heatmapcell{0.58} & \heatmapcell{0.56} & \heatmapcell[b]{0.70} & 0.96
                & \maeheatmapcell{0.64} & \maeheatmapcell{0.81} & \maeheatmapcell{1.16} & \maeheatmapcell{1.04} & \maeheatmapcell{1.29} & \maeheatmapcell{0.99} & \maeheatmapcell{1.03} & \maeheatmapcell[b]{0.99} \\
                
                & Disgust 
                & \heatmapcell{0.86} & \heatmapcell{0.88} & \heatmapcell{0.82} & \heatmapcell{0.77} & \heatmapcell{0.81} & \heatmapcell{0.74} & \heatmapcell{0.78} & \heatmapcell[b]{0.81} & 0.97
                & \maeheatmapcell{0.80} & \maeheatmapcell{0.99} & \maeheatmapcell{0.94} & \maeheatmapcell{1.05} & \maeheatmapcell{1.35} & \maeheatmapcell{1.23} & \maeheatmapcell{1.00} & \maeheatmapcell[b]{1.05} \\
                
                & Fear 
                & \heatmapcell{0.88} & \heatmapcell{0.87} & \heatmapcell{0.83} & \heatmapcell{0.82} & \heatmapcell{0.81} & \heatmapcell{0.76} & \heatmapcell{0.77} & \heatmapcell[b]{0.82} & 0.95
                & \maeheatmapcell{0.95} & \maeheatmapcell{1.46} & \maeheatmapcell{1.42} & \maeheatmapcell{1.36} & \maeheatmapcell{1.86} & \maeheatmapcell{1.56} & \maeheatmapcell{1.16} & \maeheatmapcell[b]{1.40} \\
                
                & Happiness 
                & \heatmapcell{0.87} & \heatmapcell{0.87} & \heatmapcell{0.85} & \heatmapcell{0.83} & \heatmapcell{0.83} & \heatmapcell{0.78} & \heatmapcell{0.75} & \heatmapcell[b]{0.83} & 0.96
                & \maeheatmapcell{0.92} & \maeheatmapcell{1.14} & \maeheatmapcell{1.31} & \maeheatmapcell{1.15} & \maeheatmapcell{0.90} & \maeheatmapcell{1.49} & \maeheatmapcell{1.25} & \maeheatmapcell[b]{1.17} \\
                
                & Sadness 
                & \heatmapcell{0.87} & \heatmapcell{0.87} & \heatmapcell{0.82} & \heatmapcell{0.76} & \heatmapcell{0.80} & \heatmapcell{0.76} & \heatmapcell{0.67} & \heatmapcell[b]{0.79} & 0.97
                & \maeheatmapcell{0.78} & \maeheatmapcell{1.03} & \maeheatmapcell{1.21} & \maeheatmapcell{1.12} & \maeheatmapcell{1.51} & \maeheatmapcell{1.17} & \maeheatmapcell{1.04} & \maeheatmapcell[b]{1.12} \\
                
                & Surprise 
                & \heatmapcell{0.59} & \heatmapcell{0.53} & \heatmapcell{0.45} & \heatmapcell{0.24} & \heatmapcell{0.33} & \heatmapcell{0.43} & \heatmapcell{0.19} & \heatmapcell[b]{0.39} & 0.92
                & \maeheatmapcell{1.28} & \maeheatmapcell{1.62} & \maeheatmapcell{1.54} & \maeheatmapcell{1.47} & \maeheatmapcell{1.37} & \maeheatmapcell{1.47} & \maeheatmapcell{1.59} & \maeheatmapcell[b]{1.48} \\
                
                & Emot. Val. 
                & \heatmapcell{0.87} & \heatmapcell{0.86} & \heatmapcell{0.84} & \heatmapcell{0.84} & \heatmapcell{0.84} & \heatmapcell{0.82} & \heatmapcell{0.83} & \heatmapcell[b]{0.84} & 0.97
                & \maeheatmapcell{0.91} & \maeheatmapcell{1.34} & \maeheatmapcell{1.17} & \maeheatmapcell{1.17} & \maeheatmapcell{0.79} & \maeheatmapcell{1.07} & \maeheatmapcell{1.17} & \maeheatmapcell[b]{1.09} \\
                
                & Arousal 
                & \heatmapcell{0.58} & \heatmapcell{0.55} & \heatmapcell{0.52} & \heatmapcell{0.53} & \heatmapcell{0.53} & \heatmapcell{0.48} & \heatmapcell{0.48} & \heatmapcell[b]{0.52} & 0.85
                & \maeheatmapcell{1.65} & \maeheatmapcell{2.09} & \maeheatmapcell{1.95} & \maeheatmapcell{2.08} & \maeheatmapcell{2.20} & \maeheatmapcell{1.83} & \maeheatmapcell{2.46} & \maeheatmapcell[b]{2.04} \\
                
                & \textbf{Average} 
                & \heatmapcell[b]{0.80} & \heatmapcell[b]{0.78} & \heatmapcell[b]{0.74} & \heatmapcell[b]{0.68} & \heatmapcell[b]{0.71} & \heatmapcell[b]{0.67} & \heatmapcell[b]{0.63} & \heatmapcell[b]{0.71} & \textbf{0.94}
                & \maeheatmapcell[b]{0.99} & \maeheatmapcell[b]{1.31} & \maeheatmapcell[b]{1.34} & \maeheatmapcell[b]{1.31} & \maeheatmapcell[b]{1.41} & \maeheatmapcell[b]{1.35} & \maeheatmapcell[b]{1.34} & \maeheatmapcell[b]{1.29} \\
                \midrule

                \multirow{9}{*}{\rotatebox{90}{\textbf{NAPS}}}
                & Anger 
                & \heatmapcell{0.84} & \heatmapcell{0.78} & \heatmapcell{0.72} & \heatmapcell{0.57} & \heatmapcell{0.61} & \heatmapcell{0.49} & \heatmapcell{0.43} & \heatmapcell[b]{0.63} & --
                & \maeheatmapcell{0.31} & \maeheatmapcell{0.49} & \maeheatmapcell{0.87} & \maeheatmapcell{0.55} & \maeheatmapcell{0.87} & \maeheatmapcell{0.59} & \maeheatmapcell{0.48} & \maeheatmapcell[b]{0.59} \\
                
                & Disgust 
                & \heatmapcell{0.92} & \heatmapcell{0.86} & \heatmapcell{0.85} & \heatmapcell{0.86} & \heatmapcell{0.84} & \heatmapcell{0.80} & \heatmapcell{0.84} & \heatmapcell[b]{0.85} & --
                & \maeheatmapcell{0.62} & \maeheatmapcell{1.12} & \maeheatmapcell{1.04} & \maeheatmapcell{1.02} & \maeheatmapcell{1.28} & \maeheatmapcell{1.30} & \maeheatmapcell{0.86} & \maeheatmapcell[b]{1.03} \\
                
                & Fear 
                & \heatmapcell{0.90} & \heatmapcell{0.89} & \heatmapcell{0.86} & \heatmapcell{0.85} & \heatmapcell{0.85} & \heatmapcell{0.79} & \heatmapcell{0.76} & \heatmapcell[b]{0.84} & --
                & \maeheatmapcell{0.45} & \maeheatmapcell{0.87} & \maeheatmapcell{0.87} & \maeheatmapcell{0.84} & \maeheatmapcell{1.25} & \maeheatmapcell{1.05} & \maeheatmapcell{0.54} & \maeheatmapcell[b]{0.84} \\
                
                & Happiness 
                & \heatmapcell{0.93} & \heatmapcell{0.92} & \heatmapcell{0.91} & \heatmapcell{0.90} & \heatmapcell{0.88} & \heatmapcell{0.86} & \heatmapcell{0.88} & \heatmapcell[b]{0.90} & --
                & \maeheatmapcell{0.71} & \maeheatmapcell{1.01} & \maeheatmapcell{1.03} & \maeheatmapcell{0.98} & \maeheatmapcell{0.75} & \maeheatmapcell{1.44} & \maeheatmapcell{0.86} & \maeheatmapcell[b]{0.97} \\
                
                & Sadness 
                & \heatmapcell{0.94} & \heatmapcell{0.92} & \heatmapcell{0.89} & \heatmapcell{0.88} & \heatmapcell{0.87} & \heatmapcell{0.82} & \heatmapcell{0.81} & \heatmapcell[b]{0.88} & --
                & \maeheatmapcell{0.59} & \maeheatmapcell{0.96} & \maeheatmapcell{1.09} & \maeheatmapcell{0.94} & \maeheatmapcell{1.32} & \maeheatmapcell{1.19} & \maeheatmapcell{0.77} & \maeheatmapcell[b]{0.98} \\
                
                & Surprise 
                & \heatmapcell{0.82} & \heatmapcell{0.65} & \heatmapcell{0.63} & \heatmapcell{0.48} & \heatmapcell{0.48} & \heatmapcell{0.55} & \heatmapcell{0.15} & \heatmapcell[b]{0.54} & --
                & \maeheatmapcell{0.53} & \maeheatmapcell{1.04} & \maeheatmapcell{0.73} & \maeheatmapcell{0.68} & \maeheatmapcell{0.89} & \maeheatmapcell{1.06} & \maeheatmapcell{0.63} & \maeheatmapcell[b]{0.79} \\
                
                & Emot. Val. 
                & \heatmapcell{0.93} & \heatmapcell{0.90} & \heatmapcell{0.90} & \heatmapcell{0.91} & \heatmapcell{0.87} & \heatmapcell{0.83} & \heatmapcell{0.89} & \heatmapcell[b]{0.89} & --
                & \maeheatmapcell{1.23} & \maeheatmapcell{1.40} & \maeheatmapcell{1.09} & \maeheatmapcell{0.96} & \maeheatmapcell{1.24} & \maeheatmapcell{1.05} & \maeheatmapcell{1.09} & \maeheatmapcell[b]{1.15} \\
                
                & Arousal 
                & \heatmapcell{0.74} & \heatmapcell{0.68} & \heatmapcell{0.70} & \heatmapcell{0.52} & \heatmapcell{0.69} & \heatmapcell{0.25} & \heatmapcell{0.08} & \heatmapcell[b]{0.52} & --
                & \maeheatmapcell{0.66} & \maeheatmapcell{1.02} & \maeheatmapcell{1.08} & \maeheatmapcell{1.28} & \maeheatmapcell{0.94} & \maeheatmapcell{1.27} & \maeheatmapcell{1.35} & \maeheatmapcell[b]{1.09} \\
                
                & \textbf{Average} 
                & \heatmapcell[b]{0.88} & \heatmapcell[b]{0.83} & \heatmapcell[b]{0.81} & \heatmapcell[b]{0.75} & \heatmapcell[b]{0.76} & \heatmapcell[b]{0.67} & \heatmapcell[b]{0.61} & \heatmapcell[b]{0.76} & \textbf{--}
                & \maeheatmapcell[b]{0.64} & \maeheatmapcell[b]{0.99} & \maeheatmapcell[b]{0.98} & \maeheatmapcell[b]{0.91} & \maeheatmapcell[b]{1.07} & \maeheatmapcell[b]{1.12} & \maeheatmapcell[b]{0.82} & \maeheatmapcell[b]{0.93} \\
                
                \bottomrule
            \end{tabular}
            \begin{tablenotes}
                \item[*] Proprietary models evaluated only on non-blocked images. \textit{Human ceiling}: split-half human inter-rater reliability per scale ($r$); LAI-GAI uses official values~\cite{behnke2025using}, IAPS is estimated by variance decomposition, and NAPS is omitted (our copy provides only means). See Supplementary Materials, Section~VI.
            \end{tablenotes}
        \end{threeparttable}
    \end{adjustbox}
\end{table*}

The MAE analysis in Table~\ref{tab:side_by_side} reveals systematic deviations in rating magnitude. Average MAE was lowest for the 7-point scale datasets (LAI-GAI: $0.93$; NAPS: $0.93$) and higher for the 9-point IAPS ($1.29$). Arousal consistently had the largest errors (up to $2.04$ on IAPS), while discrete negative emotions were most accurately predicted (MAE often below $0.90$). Performance varied across models: on LAI-GAI, Gemma3 27B achieved the lowest MAE ($0.75$), while Intern3.5-VL 8B had the highest ($1.06$).

Gemini-3-Flash, the only thinking model evaluated on Task~2, achieved the highest average correlation across all three datasets ($r = 0.84$, $0.80$, and $0.88$ on LAI-GAI, IAPS, and NAPS, respectively) and the lowest average MAE ($0.81$, $0.99$, and $0.64$). Its advantage was particularly pronounced for dimensions that standard models found challenging, such as surprise on NAPS ($r = 0.82$ vs.\ $0.54$ average) and arousal ($r = 0.74$ vs.\ $0.52$ average).

These correlations should be read against the human ceiling, since the normative ratings are themselves noisy. Using split-half inter-rater reliability (official values for LAI-GAI~\cite{behnke2025using}; variance decomposition for IAPS; see Supplementary Materials, Section~VI), human agreement is high for valence and discrete emotions (split-half $r=.87$--$.99$) but markedly lower for arousal (LAI-GAI $.69$, IAPS $.85$). Relative to this ceiling, the best model reaches on average $89\%$ (LAI-GAI) and $84\%$ (IAPS) of human agreement---over $98\%$ for disgust, sadness, and valence---but only $56$--$68\%$ for arousal (Table~\ref{tab:side_by_side}). Model--human alignment is thus closer to the achievable ceiling than the raw correlations suggest, yet the arousal gap persists even against that ceiling: two independent halves of the rater pool agree at only $r=.69$ on LAI-GAI arousal, versus a model $r=.39$.

\subsubsection{Task 3: Rater-conditioned Prompting}
We treat this rater-conditioned analysis as an exploratory pilot (two models, one dataset) that cannot support strong conclusions. Rater-conditioned prompting produced mixed effects for Gemma-3 27B and Gemini-2.5-Flash. Overall, adding rater-specific background information was sometimes beneficial, but the direction and magnitude of the effects varied across scales and between models. ``Full Background'' was the most promising condition, but its benefits were inconsistent. For example, Gemini-2.5-Flash showed improved alignment for Amusement and Joy ($d \approx 0.10$), but not for Excitement or Nurturant Love. In several cases (e.g., Excitement, Fear, and Neutral), the two models even showed opposite effects. By contrast, ``Demographic Background'' generally reduced alignment with human ratings. For instance, Gemma-3 27B showed decreased alignment for Disgust ($d=-0.21$) and Sadness ($d=-0.27$), with smaller declines also observed for Gemini-2.5-Flash. For arousal, one of the most difficult dimensions, both models improved under ``Full Background'' (Gemma-3 27B: $d=0.34$; Gemini-2.5-Flash: $d=0.12$), whereas ``Demographic Background'' produced little change or reduced alignment. Taken together, these mixed and sometimes opposing effects (see Supplementary Materials, Section V) indicate that the models did not consistently incorporate rater-specific information in this prompting setup.

\section{Discussion}

\begin{figure*}[t]
    \centering
    \begin{minipage}[b]{0.48\linewidth}
        \centering
        \subfloat[LAI-GAI\label{fig:laigai_means}]{%
            \includegraphics[width=\linewidth]{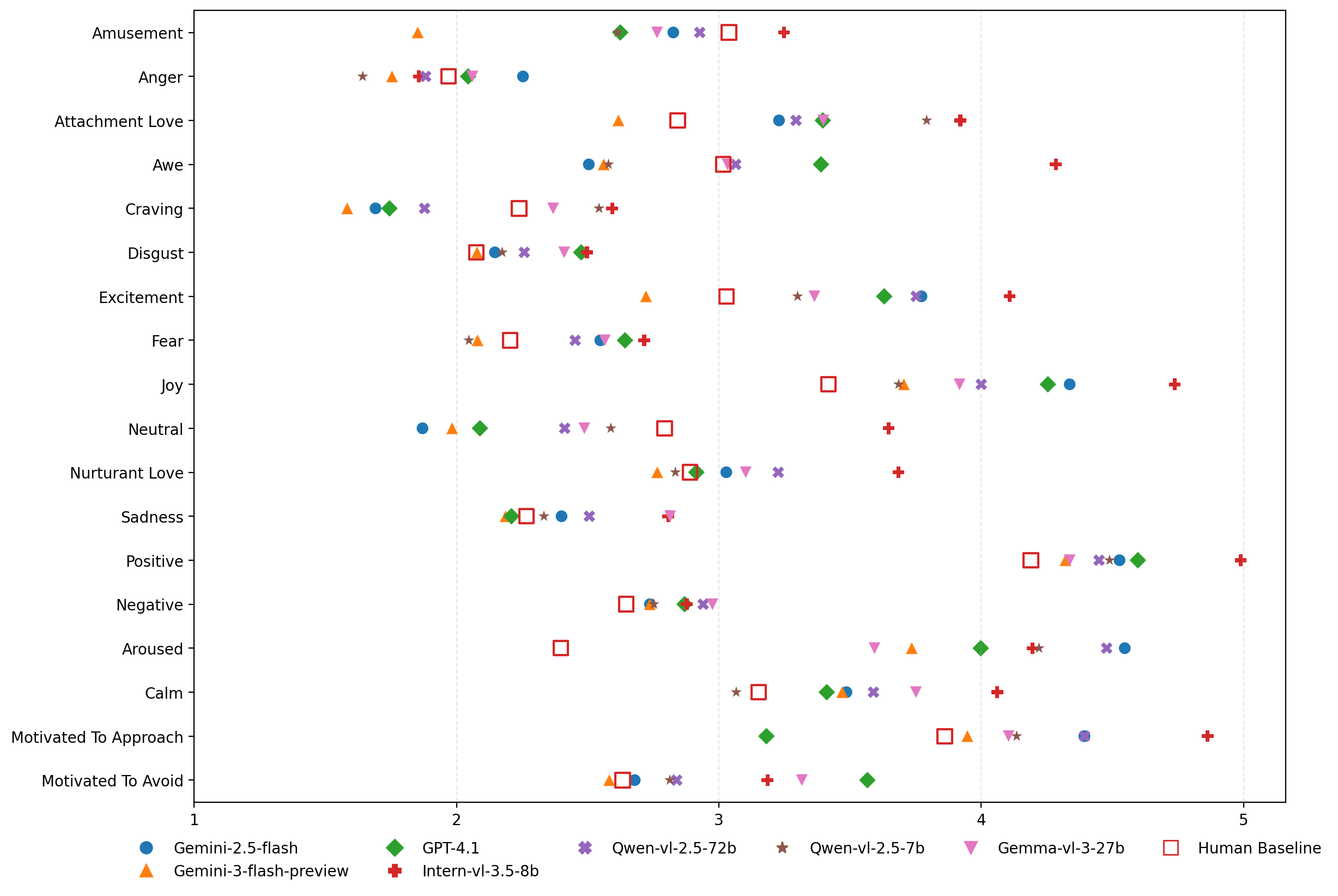}%
        }
    \end{minipage}
    \hfill
    \begin{minipage}[b]{0.48\linewidth}
        \centering
        \subfloat[NAPS\label{fig:naps_means}]{%
            \includegraphics[width=\linewidth]{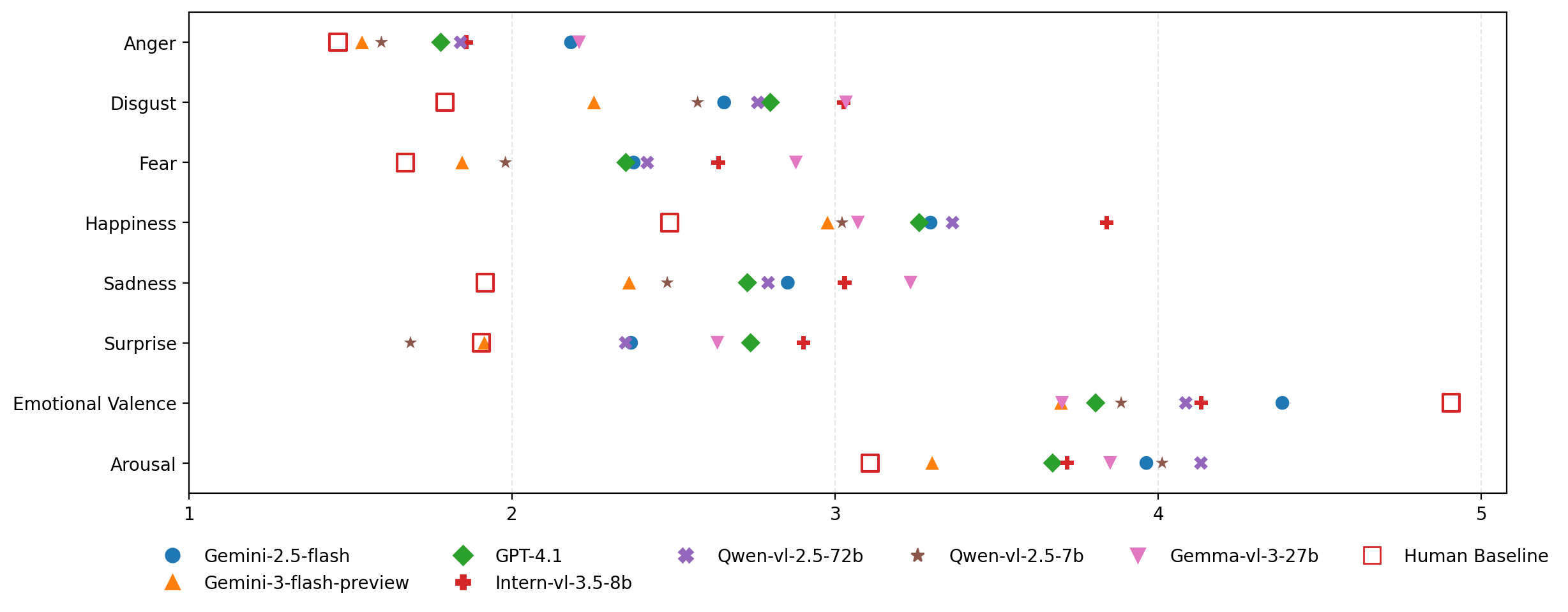}%
        }
        \vspace{-0.8em}
        \subfloat[IAPS\label{fig:iaps_means}]{%
            \includegraphics[width=\linewidth]{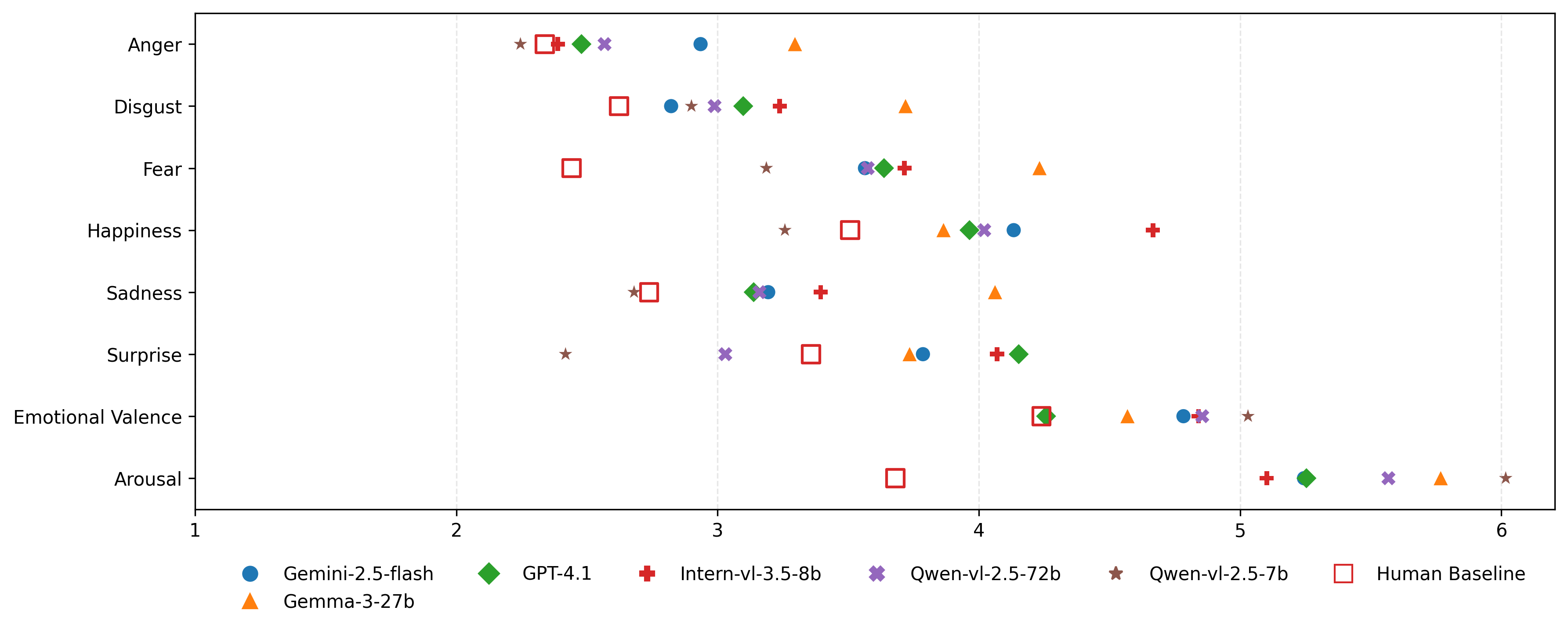}%
        }
    \end{minipage}
    \caption{
        Comparison of mean emotion ratings across datasets. Each panel displays a dot plot comparing the mean predicted emotion ratings from the models with the human baseline. The $y$-axis lists the emotions, and the $x$-axis represents the mean rating. Each marker shape and color corresponds to a specific model. Panels show results for (a) LAI-GAI, (b) NAPS, and (c) IAPS.
    }
    \label{fig:emotion_means_comparison}
\end{figure*}

We evaluated state-of-the-art VLMs on IAPS, NAPS, and LAI-GAI. Overall, models captured broad emotional patterns and aligned substantially with human ratings. However, they were less accurate for emotion intensity and tended to overestimate responses (Fig.~\ref{fig:emotion_means_comparison}). They struggled most with anger, surprise, and especially arousal across datasets. Finally, participant background information did not meaningfully improve performance, producing only small, inconsistent effects.

\subsection{Implications for Affective Science}
We found that current VLMs are better suited as assistive tools than as replacements for human psychometric assessment. They may help pre-screen stimuli, but their intensity estimates cannot replace normative ratings. This limitation is amplified in some closed-source models by API-level content filtering, which reduced the effective evaluation set and may bias downstream analyses (Gemini-2.5-Flash refused to process 88 images - 58 IAPS, 15 NAPS, 15 LAI-GAI- and GPT-4.1 refused 34 images from IAPS). Because such filtering is opaque and unstable over time, it raises reproducibility concerns for proprietary models. Newer models appeared less restrictive.

Our findings reinforce affective science principles. In line with theories of emotion construction \cite{barrett2022context}, affective meaning cannot be reduced to visual features alone. This yields a theory-derived prediction~\cite{ellsworth2003appraisal}: dimensions with reliable visual grounding (e.g., disgust, fear) should be easier to infer than those that depend on contextual appraisal (e.g., arousal, surprise), a pattern our results confirm. Anger and surprise are context-dependent: anger often reflects social appraisal, whereas surprise reflects violated expectations. Thus, static pictures rarely elicit clear anger, and models seldom identified anger as the top emotion, consistent with previous work \cite{behnke2022role,behnke2025using}. Likewise, surprise ratings in these datasets often appeared to depend on context rather than purely visual properties. For example, in IAPS, high surprise ratings for an image of two nude men likely reflected the unexpected social context and the lack of a more suitable label such as sexual arousal. Some apparent model errors may therefore reflect stimulus ambiguity or rating-scheme limits. Weak alignment for arousal likewise fits concerns that arousal is a broad, noisy construct that conflates multiple processes (i.e., physiological activation, attentional alerting, subjective energy, salience), \cite{sander2025arousal}.

\subsection{Implications for Computer Science}
Our findings suggest that general-purpose VLMs are not yet reliable drop-in tools for emotion recognition. Although our results are more optimistic, they remain consistent with works showing important gaps in affect understanding by VLMs \cite{bhattacharyya2025evaluating,ogg2025large}. Even the strongest models struggled with anger, arousal, and intensity judgments. These weaknesses seem to reflect difficulty with nuanced, context-dependent, and intensity-based affective cues. Such errors may reduce trust and effectiveness in affect-sensitive applications.

Larger and thinking-enhanced models generally outperformed smaller or non-thinking variants. Among proprietary models, Gemini-3-Flash-Preview achieved the highest weighted F1 on both IAPS and NAPS in Task~1 and the best Task~2 performance across all datasets, while GPT-5.4 improved over GPT-4.1 on both classification and regression. Across generations, the shift from Gemini-2.5-Flash to Gemini-3-Flash-Preview and from GPT-4.1 to GPT-5.4 indicates measurable gains in affective understanding, especially for dimensions that older models handled poorly, such as surprise and arousal. However, these comparisons are cross-sectional across model families, and the October~2025--March~2026 evaluation gap confounds reasoning ability with training-data recency; with closed models the two cannot be cleanly disentangled. Scaling and reasoning may partly explain these gains, but the factors are confounded.

Rater-conditioned prompting highlights a related challenge. In principle, VLMs should use culture or mood to disambiguate affective meaning. In practice, under our zero-shot setup, rater-specific background did not consistently improve predictions, and demographic information alone often reduced alignment. Thus, current models still struggle to use person- or situation-specific context, especially for arousal \cite{lei2024large}.

Benchmark choice shapes what VLM evaluations capture. Models performed better on LAI-GAI than on IAPS or NAPS, likely because LAI-GAI provides clearer stimuli, explicit emotion definitions, and more standardized instructions. By contrast, IAPS and NAPS contain more heterogeneous, context-dependent images that function as a stronger test of generalization. This pattern does not imply that legacy datasets are flawed; rather, it highlights a current limitation of VLMs in moving from relatively clear benchmarks to ecologically richer affective stimuli. Interestingly, the thinking models (Gemini-3-Pro, Gemini-3-Flash, and GPT-5.4) exhibited a reversed pattern: they improved substantially on IAPS and NAPS relative to their predecessors, yet performed worse on LAI-GAI. One possible explanation is that chain-of-thought reasoning helps with ambiguous, context-dependent content typical of legacy datasets. On LAI-GAI’s more prototypical stimuli, extended reasoning may instead introduce overthinking or second-guessing, although we did not test this mechanism directly. Understanding why reasoning benefits some affective benchmarks but not others remains an open question.

\subsection{Limitations and Future Directions}

Several limitations should be noted. First, although we used three datasets, they do not fully capture the diversity of affective stimuli. Future work should develop datasets that combine the scale and diversity of computer-science resources with the psychometric rigor of psychology-based libraries, including stronger cross-cultural validation. Second, we evaluated VLMs only in a zero-shot setting. Our results therefore reflect baseline performance and may underestimate what these models can achieve with affect-specific fine-tuning or structured adaptation. Third, we used multiple variants of prompts per task and did not test sensitivity to alternative prompting strategies. Future work should examine further prompt robustness more systematically, especially for rater-conditioned prediction and richer participant profiles. Fourth, the emotion classification task was affected by class imbalance. Some categories, especially anger, were represented by few images, increasing uncertainty and label noise. As a result, weaker performance on underrepresented emotions may reflect dataset limitations as much as model limitations. Fifth, continuous ratings used fixed sampling parameters (temperature $=0.5$, $n=50$, seed $=42$), chosen to approximate human rater counts while avoiding performance degradation and instruction non-compliance in smaller models ($<9$B parameters). We did not test the full parameter space, so optimal settings may differ across models. This stochastic sampling proxies inter-rater variability but does not model it directly; systematically comparing model dispersion with human rating standard deviations remains future work. Sixth, our model comparisons are cross-sectional across families and checkpoints evaluated months apart, so apparent generational progress may reflect reasoning, scaling, or training-data recency rather than a single cause.

\section{Conclusion} 
Taken together, our results show that modern VLMs have limited affective skills: they can imitate human emotional structures but struggle with nuance. VLMs therefore show potential as tools for affective science and affective computing, but realizing this potential will require careful testing, rigorous methods, and ongoing interdisciplinary collaboration.

\subsection{Declaration of AI use}
During the preparation of this work, the authors used Gemini and GPT models to improve readability and language. After using these tools/services, the authors reviewed and edited the content as necessary and take full responsibility for the publication.  

\subsection{Reproducibility Statement}
The code to regenerate all tables, figures, and results from raw predictions is available at \url{https://github.com/filnow/VAA-VLM}.

\section*{Ethical Impact Statement}

All datasets used in this study (IAPS, NAPS, LAI-GAI) were collected and distributed in accordance with approved ethical guidelines, with informed consent obtained from participants in the original studies. Our analyses rely solely on anonymized, publicly available ratings. Although vision–language models show potential for affective science, their outputs should not be viewed as clinical or diagnostic assessments. Given observed biases and limitations, responsible use requires careful validation and acknowledgment of the risks of misclassification, especially in sensitive applications related to human well-being.

\bibliographystyle{IEEEtran}
\bibliography{refs}

\twocolumn[{\begin{center}{\LARGE\bfseries Supplementary Materials}\end{center}\vspace{1em}}]
\setcounter{section}{0}
\renewcommand{\thesection}{\Roman{section}}
\renewcommand{\theHsection}{supp.\Roman{section}}
\setcounter{table}{0}
\setcounter{figure}{0}
\renewcommand{\thetable}{S\Roman{table}}
\renewcommand{\thefigure}{S\arabic{figure}}
\renewcommand{\theHtable}{supp.table.\arabic{table}}
\renewcommand{\theHfigure}{supp.figure.\arabic{figure}}

\section{Experimental Prompts}
This section provides the exact prompt templates used for the zero-shot evaluations. Variables enclosed in \texttt{\{brackets\}} were dynamically populated based on the specific dataset labels and definitions during inference.

\begin{figure*}[!b]
\begin{promptbox}{Task 1: Emotion Classification}
\textbf{System:} \\
You are an affective scientist tasked with classifying the primary emotion evoked by images in an average person. Your goal is to choose the single most prominent and fitting emotion from the list below that the provided picture would elicit.

Consider the following emotion labels and their definitions: \\
\textit{\{Definitions List Inserted Here\}}

Based on the image provided, select the ONE emotion label from the list above that best describes the primary affective response an average person would likely experience.

Output Format: \\
VERY IMPORTANT - Respond ONLY with the single lowercase emotion label itself. Do NOT include explanations, ratings, introductions, punctuation, or any other text. Just the single word label.

\rule{\linewidth}{0.4pt} 

\textbf{Assistant:} \\
Understood. I will analyze the image, consider the provided emotion definitions, and respond with the single most appropriate lowercase emotion label from the list, with no additional text.

\rule{\linewidth}{0.4pt} 

\textbf{User:} \\
Please classify the primary emotion this image evokes based on the provided list and definitions. Respond with only the single lowercase emotion label.
\end{promptbox}
\caption{The exact prompt template used for the zero-shot emotion classification task.}
\label{fig:prompt_template}
\end{figure*}

\begin{figure*}[t]
\begin{promptbox}{Task 2: Affective Dimension Prediction}
\textbf{System:} \\
You are an affective scientist tasked with rating images on specific affective dimensions. Your goal is to rate how much the specified emotion or affective dimension the provided picture would elicit in an average person.

Use a scale from 1 (very low) to \texttt{\{scale\_of\_emotion\}} (very high) for the rating.
Base your rating on the definition provided in the user prompt. Be precise in your rating, don't use fixed parameters or heuristics, focus on nuanced assessment.

Output Format: \\
VERY IMPORTANT - Respond ONLY with the single numerical rating (1--\texttt{\{scale\_of\_emotion\}}) for the dimension requested in the user prompt. Do NOT include the name of the dimension, any explanations, apologies, introductions, or any other text. Just one number.

\rule{\linewidth}{0.4pt} 

\textbf{Assistant:} \\
Understood. I will analyze the image and the requested affective dimension specified in the user prompt. I will provide exactly one numerical rating (1--\texttt{\{scale\_of\_emotion\}}) as the response, with no additional text.

\rule{\linewidth}{0.4pt} 

\textbf{User:} \\
Please rate this image for \texttt{\{emotion\_name\}} on a scale of 1--\texttt{\{scale\_of\_emotion\}} for an average person. \\
Definition: \texttt{\{emotion\_name\}} is \texttt{\{definition\}} \\
Provide only the single numerical rating (1--\texttt{\{scale\_of\_emotion\}}).
\end{promptbox}
\caption{The prompt template used for the affective dimension rating task. The scale variable was adjusted (e.g., 1-7 or 1-9) to match the ground truth of the specific dataset being evaluated.}
\label{fig:rating_prompt_template}
\end{figure*}

\begin{figure*}[t]
\begin{promptbox}{Task 3: Context-Aware Prediction}
\textbf{System:} \\
You are an affective scientist tasked with rating images on specific affective dimensions.
You will be given context about a human rater: \textbf{\textit{<Context Description>}} \\
Your task is to use this context to inform your rating.
You must rate how much the specified emotion or affective dimension the provided picture would elicit in an average person with a similar demographic profile and initial emotional state, using a scale from 1 (very low) to \texttt{\{scale\_of\_emotion\}} (very high).

Base your rating on the definition provided in the user prompt. Be precise in your rating, don't use fixed parameters or heuristics, focus on nuanced assessment.

Output Format: VERY IMPORTANT - Respond ONLY with the single numerical rating (1--\texttt{\{scale\_of\_emotion\}}). Do NOT include the name of the dimension, any explanations, apologies, introductions, or any other text. Just one number.

\rule{\linewidth}{0.4pt}

\textbf{Assistant:} \\
Understood. I will analyze the image, the rater context, and the requested affective dimension. I will provide exactly one numerical rating (1--\texttt{\{scale\_of\_emotion\}}) as the response, with no additional text.

\rule{\linewidth}{0.4pt}

\textbf{User:} \\
Rater Context: \\
\textbf{\textit{<Rater Context Block>}}

Please rate this image for \texttt{\{emotion\_name\}} on a scale of 1--\texttt{\{scale\_of\_emotion\}}. \\
Definition: \texttt{\{emotion\_name\}} is \texttt{\{definition\}} \\
Provide only the single numerical rating (1--\texttt{\{scale\_of\_emotion\}}).

\rule{\linewidth}{1.0pt} 

\textbf{Experimental Variations:} \\
\textbf{1. Full Context Condition} \\
\textit{Context Description:} ``their age, gender, country, and their self-reported emotional state for the target emotion before viewing the image.'' \\
\textit{Rater Context Block:} \\
Age: \texttt{\{age\}} \\
Gender: \texttt{\{gender\}} \\
Country: \texttt{\{country\}} \\
Self-reported feeling of \texttt{\{emotion\_name\}} before viewing the image (on a scale of 1--\texttt{\{scale\_of\_emotion\}}): \texttt{\{\_Before\_\{emotion\_name\}\}}

\vspace{0.2cm}

\textbf{2. Demographic Context Condition} \\
\textit{Context Description:} ``their age, gender and country.'' \\
\textit{Rater Context Block:} \\
Age: \texttt{\{age\}} \\
Gender: \texttt{\{gender\}} \\
Country: \texttt{\{country\}}

\vspace{0.2cm}

\textbf{3. Emotional Context Condition} \\
\textit{Context Description:} ``their self-reported emotional state for the target emotion before viewing the image.'' \\
\textit{Rater Context Block:} \\
Self-reported feeling of \texttt{\{emotion\_name\}} before viewing the image (on a scale of 1--\texttt{\{scale\_of\_emotion\}}): \texttt{\{\_Before\_\{emotion\_name\}\}}

\end{promptbox}
\caption{The prompt templates for the three Context-Aware experimental conditions. The core instructions remained constant, while the Context Description and Rater Context Block varied by condition.}
\label{fig:context_prompts}
\end{figure*}

\subsection{Emotion and Affective Dimension Definitions}

Tables~\ref{tab:def_laigai_cls}--\ref{tab:def_iaps_naps_reg} list the exact definitions provided to the models via the \texttt{\{definition\}} variable in the prompt templates (Figures~\ref{fig:prompt_template}--\ref{fig:rating_prompt_template}). Definitions were drawn from the original dataset articles and adapted where necessary to provide concise, self-contained descriptions suitable for zero-shot prompting.

\begin{table*}[t]
\centering
\caption{Emotion definitions used for the LAI-GAI dataset in the classification task (12 discrete emotion labels).}
\label{tab:def_laigai_cls}
\small
\begin{tabular}{@{}lp{15.5cm}@{}}
\toprule
\textbf{Emotion} & \textbf{Definition} \\
\midrule
Amusement       & An emotional experience associated with opportunities for humor and events that violate expectations. \\
Anger           & An emotional experience characterized by antagonism toward someone or something perceived as deliberately harmful or unfair. \\
Attachment love & An emotional experience tied to opportunities for affiliation, interdependence, and intimacy. \\
Awe             & An emotional experience triggered by stimuli that are vast and beyond the ordinary frame of reference, often involving wonder, admiration, and a sense of being small in the presence of something greater than oneself. \\
Craving         & An emotional experience associated with specific sensory pleasure that aids in distinguishing nutritional foods from nutritionally neutral objects and toxins. \\
Disgust         & An emotional experience involving strong aversion or repulsion toward something perceived as offensive, contaminating, or unpleasant. \\
Excitement      & A high-intensity emotional experience in response to novelty, challenge, or human excellence, often associated with some degree of risk. \\
Fear            & An emotional experience triggered by a perceived threat or danger. \\
Joy             & An emotional state brought about by good fortune and well-being. \\
Neutral         & An emotional state marked by the absence of strong emotions, either positive or negative, characterized by calmness and balance without significant arousal. \\
Nurturant love  & An emotional experience linked to the adaptive opportunities provided by offspring or other vulnerable kin. \\
Sadness         & An emotional experience characterized by feelings of disappointment, grief, or hopelessness. \\
\bottomrule
\end{tabular}
\end{table*}

\begin{table*}[t]
\centering
\caption{Affective dimension definitions for the LAI-GAI regression task, rated alongside the 12 discrete emotions in Table~\ref{tab:def_laigai_cls}.}
\label{tab:def_laigai_reg}
\small
\begin{tabular}{@{}lp{14.5cm}@{}}
\toprule
\textbf{Dimension} & \textbf{Definition} \\
\midrule
Positive              & Feelings of pleasure, satisfaction, or well-being, often associated with favorable situations. \\
Negative              & Feelings of discomfort, displeasure, or distress, often associated with unfavorable situations. \\
Calm                  & A sense of tranquility and composure, with low levels of emotional arousal or agitation. \\
Aroused               & Heightened physiological and psychological activity, such as increased heart rate, energy, or alertness. \\
Motivated to approach & An emotional drive to pursue a desired goal, object, or situation, often associated with curiosity, attraction, or reward. \\
Motivated to avoid    & An emotional drive to move away from an undesired or threatening goal, object, or situation, often associated with fear, discomfort, or aversion. \\
\bottomrule
\end{tabular}
\end{table*}

\begin{table*}[t]
\centering
\caption{Scales definitions for the IAPS and NAPS datasets. The discrete emotions were used in both classification and regression, whereas valence and arousal were used only in regression (adapted to match the original dataset instructions).}
\label{tab:def_iaps_naps_reg}
\small
\begin{tabular}{@{}llp{14cm}@{}}
\toprule
\textbf{Label} & \textbf{Dataset} & \textbf{Definition} \\
\midrule
\multicolumn{3}{@{}l}{\textit{Discrete emotions (shared across IAPS and NAPS)}} \\
\addlinespace[2pt]
Happiness & Both & A positive state characterized by feelings of joy, contentment, and wellbeing. It often arises in response to pleasant experiences and is typically accompanied by a sense of lightness, warmth, and optimism. \\
Surprise  & Both & The response to a sudden, unexpected event. \\
Sadness   & Both & Feelings of disappointment, grief, or hopelessness. \\
Anger     & Both & Antagonism toward someone or something perceived as deliberately harmful or unfair. \\
Disgust   & Both & Strong aversion or repulsion toward something perceived as offensive, contaminating, or unpleasant. \\
Fear      & Both & Triggered by a perceived threat or danger. \\
\midrule
\multicolumn{3}{@{}l}{\textit{Dimensional measures (regression task only)}} \\
\addlinespace[2pt]
Emotional valence & IAPS & Number 1 indicates you felt completely unhappy while viewing the picture; number 9 indicates you felt completely happy. \\
\addlinespace[2pt]
Emotional valence & NAPS & The extent of the positive or negative reaction evoked by a given picture, ranging from 1 to 9 (1 for very negative emotions and 9 for very positive emotions). \\
\addlinespace[2pt]
Arousal & IAPS & The number 1 indicates you felt completely relaxed or unaroused while viewing the picture, and number 9 indicates you felt completely stimulated or aroused. \\
\addlinespace[2pt]
Arousal & NAPS & The extent a particular picture made them feel unaroused or aroused, ranging from 1 to 9 (1 for unaroused/relaxed and 9 for very much aroused---e.g., jittery or excited). \\
\bottomrule
\end{tabular}
\end{table*}

\subsection{Prompt Sensitivity Analysis}

To assess whether model predictions are sensitive to prompt formulation, we tested 13~variations of the base affective dimension prediction prompt (Figure~\ref{fig:rating_prompt_template}) on two open-weight models: Gemma~3~27B and Qwen~2.5~VL~7B. Each variation modifies exactly one aspect of the original prompt while keeping all other components identical. The modifications fall into four categories described in Table~\ref{tab:prompt_variations}.

Tables~\ref{tab:prompt_mae_results} and~\ref{tab:prompt_corr_results} report, respectively, the mean MAE and mean Pearson correlation ($r$) across all affective dimensions for each prompt variant. Across both models and all three datasets, the standard deviation of mean MAE across the 14~prompt conditions (base + 13~variants) remains very low: $\sigma \leq 0.075$ for Gemma and $\sigma \leq 0.026$ for Qwen. The total MAE range between the best and worst prompt never exceeds 0.34 points. Pearson correlations are even more stable, with $\sigma \leq 0.007$ and a maximum range of only 0.027 across all conditions. These results indicate that the models' affective ratings are robust to the prompt formulations tested, and the choice of prompt wording does not meaningfully affect prediction quality.

\begin{table*}[t]
\centering
\caption{Summary of the 13~prompt variations tested for robustness of the affective dimension prediction task. Each variation modifies a single aspect of the base prompt (Figure~\ref{fig:rating_prompt_template}). All other prompt components remain unchanged.}
\label{tab:prompt_variations}
\small
\begin{tabular}{@{}clp{13.5cm}@{}}
\toprule
\textbf{\#} & \textbf{Category} & \textbf{Modification} \\
\midrule
\multicolumn{3}{@{}l}{\textit{Category 1: Definition Presentation (User Prompt)}} \\
\addlinespace[2pt]
1 & Definition removed & The definition line (``Definition: \texttt{\{emotion\_name\}} is \texttt{\{definition\}}'') is removed from the user prompt entirely. The model receives only the emotion name and scale. \\
\addlinespace[2pt]
2 & Synonyms instead of definition & The definition line is replaced with ``Synonyms/related terms are \texttt{\{definition\}}'', reframing the definitional cue as a set of related terms. \\
\midrule
\multicolumn{3}{@{}l}{\textit{Category 2: Assigned Persona (System Prompt)}} \\
\addlinespace[2pt]
3 & Study participant & The persona ``You are an affective scientist'' is replaced with ``You are a study participant''. \\
\addlinespace[2pt]
4 & Psychologist & Replaced with ``You are a psychologist''. \\
\addlinespace[2pt]
5 & Computer scientist & Replaced with ``You are a computer scientist''. \\
\addlinespace[2pt]
6 & Man & Replaced with ``You are a man''. \\
\addlinespace[2pt]
7 & Woman & Replaced with ``You are a woman''. \\
\midrule
\multicolumn{3}{@{}l}{\textit{Category 3: Rater Mindset Instruction (System Prompt)}} \\
\addlinespace[2pt]
8 & Sensitive to subtle cues & Adds: ``Rater mindset: Be sensitive to subtle cues. Consider fine-grained details in scene content, facial expressions, body language, context, and aesthetics that could influence the evoked emotion.'' \\
\addlinespace[2pt]
9 & Normative / consensus & Adds: ``Rater mindset (normative): Aim to match human consensus (the most typical rating an average group of viewers would give), not your personal reaction.'' \\
\addlinespace[2pt]
10 & Conservative & Adds: ``Rater mindset (conservative): Use the top end of the scale sparingly. Only give 6--7 when there are clear, strong cues that the image would strongly evoke the target emotion for most viewers.'' \\
\addlinespace[2pt]
11 & Anti-centrism & Adds: ``Rater mindset (anti-centrism): Do not default to the midpoint. Use the full scale when warranted.'' \\
\midrule
\multicolumn{3}{@{}l}{\textit{Category 4: Task Clarification (System Prompt)}} \\
\addlinespace[2pt]
12 & Evoked vs.\ displayed & Adds: ``Clarification: Rate the emotion the image would evoke in an average viewer, not the emotion displayed by people in the image.'' \\
\addlinespace[2pt]
13 & Content-only focus & Adds: ``Clarification: Consider only the content of the image and the definition of the emotion. The numeric scale is only for the output format.'' \\
\bottomrule
\end{tabular}
\end{table*}

\begin{table*}[t]
\centering
\caption{Mean MAE (across all affective dimensions) for each prompt variant on three datasets. Lower is better. ``Base'' denotes the original prompt from Figure~\ref{fig:rating_prompt_template}. Variant numbers correspond to Table~\ref{tab:prompt_variations}.}
\label{tab:prompt_mae_results}
\small
\begin{tabular}{@{}l ccc ccc@{}}
\toprule
 & \multicolumn{3}{c}{\textbf{Gemma 3 27B}} & \multicolumn{3}{c}{\textbf{Qwen 2.5 VL 7B}} \\
\cmidrule(lr){2-4} \cmidrule(lr){5-7}
\textbf{Prompt} & \textbf{IAPS} & \textbf{LAI-GAI} & \textbf{NAPS} & \textbf{IAPS} & \textbf{LAI-GAI} & \textbf{NAPS} \\
\midrule
Base              & 1.409 & 0.753 & 1.066 & 1.336 & 0.830 & 0.823 \\
\addlinespace[2pt]
\multicolumn{7}{@{}l}{\textit{Definition presentation}} \\
\quad 1. No definition      & 1.377 & 0.777 & 1.021 & 1.330 & 0.843 & 0.826 \\
\quad 2. Synonyms           & 1.402 & 0.756 & 1.046 & 1.336 & 0.865 & 0.834 \\
\addlinespace[2pt]
\multicolumn{7}{@{}l}{\textit{Assigned persona}} \\
\quad 3. Study participant  & 1.379 & 0.754 & 1.039 & 1.345 & 0.829 & 0.825 \\
\quad 4. Psychologist       & 1.413 & 0.754 & 1.072 & 1.347 & 0.833 & 0.829 \\
\quad 5. Computer scientist & 1.395 & 0.764 & 1.032 & 1.351 & 0.840 & 0.825 \\
\quad 6. Man                & 1.374 & 0.744 & 1.023 & 1.345 & 0.835 & 0.813 \\
\quad 7. Woman              & 1.377 & 0.745 & 1.036 & 1.337 & 0.829 & 0.817 \\
\addlinespace[2pt]
\multicolumn{7}{@{}l}{\textit{Rater mindset}} \\
\quad 8. Subtle cues        & 1.463 & 0.786 & 1.107 & 1.314 & 0.825 & 0.818 \\
\quad 9. Normative          & 1.342 & 0.722 & 1.013 & 1.299 & 0.817 & 0.812 \\
\quad 10. Conservative      & 1.165 & 0.671 & 0.965 & 1.397 & 0.920 & 0.871 \\
\quad 11. Anti-centrism     & 1.506 & 0.790 & 1.141 & 1.371 & 0.867 & 0.837 \\
\addlinespace[2pt]
\multicolumn{7}{@{}l}{\textit{Task clarification}} \\
\quad 12. Evoked emotion    & 1.386 & 0.724 & 1.033 & 1.333 & 0.828 & 0.817 \\
\quad 13. Content-only      & 1.370 & 0.747 & 1.013 & 1.345 & 0.854 & 0.820 \\
\midrule
Mean              & 1.383 & 0.749 & 1.043 & 1.342 & 0.844 & 0.826 \\
Std               & 0.072 & 0.029 & 0.041 & 0.022 & 0.026 & 0.014 \\
\bottomrule
\end{tabular}
\end{table*}

\begin{table*}[t]
\centering
\caption{Mean Pearson correlation ($r$) across all affective dimensions for each prompt variant on three datasets. Higher is better. ``Base'' denotes the original prompt from Figure~\ref{fig:rating_prompt_template}. Variant numbers correspond to Table~\ref{tab:prompt_variations}.}
\label{tab:prompt_corr_results}
\small
\begin{tabular}{@{}l ccc ccc@{}}
\toprule
 & \multicolumn{3}{c}{\textbf{Gemma 3 27B}} & \multicolumn{3}{c}{\textbf{Qwen 2.5 VL 7B}} \\
\cmidrule(lr){2-4} \cmidrule(lr){5-7}
\textbf{Prompt} & \textbf{IAPS} & \textbf{LAI-GAI} & \textbf{NAPS} & \textbf{IAPS} & \textbf{LAI-GAI} & \textbf{NAPS} \\
\midrule
Base              & .706 & .824 & .762 & .629 & .809 & .606 \\
\addlinespace[2pt]
\multicolumn{7}{@{}l}{\textit{Definition presentation}} \\
\quad 1. No definition      & .695 & .822 & .763 & .605 & .803 & .601 \\
\quad 2. Synonyms           & .699 & .824 & .759 & .625 & .799 & .608 \\
\addlinespace[2pt]
\multicolumn{7}{@{}l}{\textit{Assigned persona}} \\
\quad 3. Study participant  & .707 & .823 & .760 & .631 & .807 & .606 \\
\quad 4. Psychologist       & .702 & .822 & .761 & .630 & .804 & .607 \\
\quad 5. Computer scientist & .694 & .822 & .756 & .629 & .807 & .599 \\
\quad 6. Man                & .697 & .824 & .756 & .628 & .805 & .605 \\
\quad 7. Woman              & .695 & .827 & .754 & .627 & .806 & .601 \\
\addlinespace[2pt]
\multicolumn{7}{@{}l}{\textit{Rater mindset}} \\
\quad 8. Subtle cues        & .687 & .823 & .750 & .631 & .808 & .607 \\
\quad 9. Normative          & .703 & .823 & .761 & .628 & .807 & .606 \\
\quad 10. Conservative      & .696 & .817 & .746 & .619 & .801 & .613 \\
\quad 11. Anti-centrism     & .707 & .824 & .771 & .625 & .803 & .605 \\
\addlinespace[2pt]
\multicolumn{7}{@{}l}{\textit{Task clarification}} \\
\quad 12. Evoked emotion    & .702 & .829 & .766 & .632 & .806 & .614 \\
\quad 13. Content-only      & .696 & .823 & .757 & .624 & .805 & .592 \\
\midrule
Mean              & .699 & .823 & .759 & .626 & .805 & .605 \\
Std               & .006 & .003 & .006 & .007 & .003 & .005 \\
\bottomrule
\end{tabular}
\end{table*}

\section{Visualization of the Affective Stimulus Space}

\begin{figure*}[p]
    \centering
    \includegraphics[width=\linewidth, height=0.29\textheight, keepaspectratio]{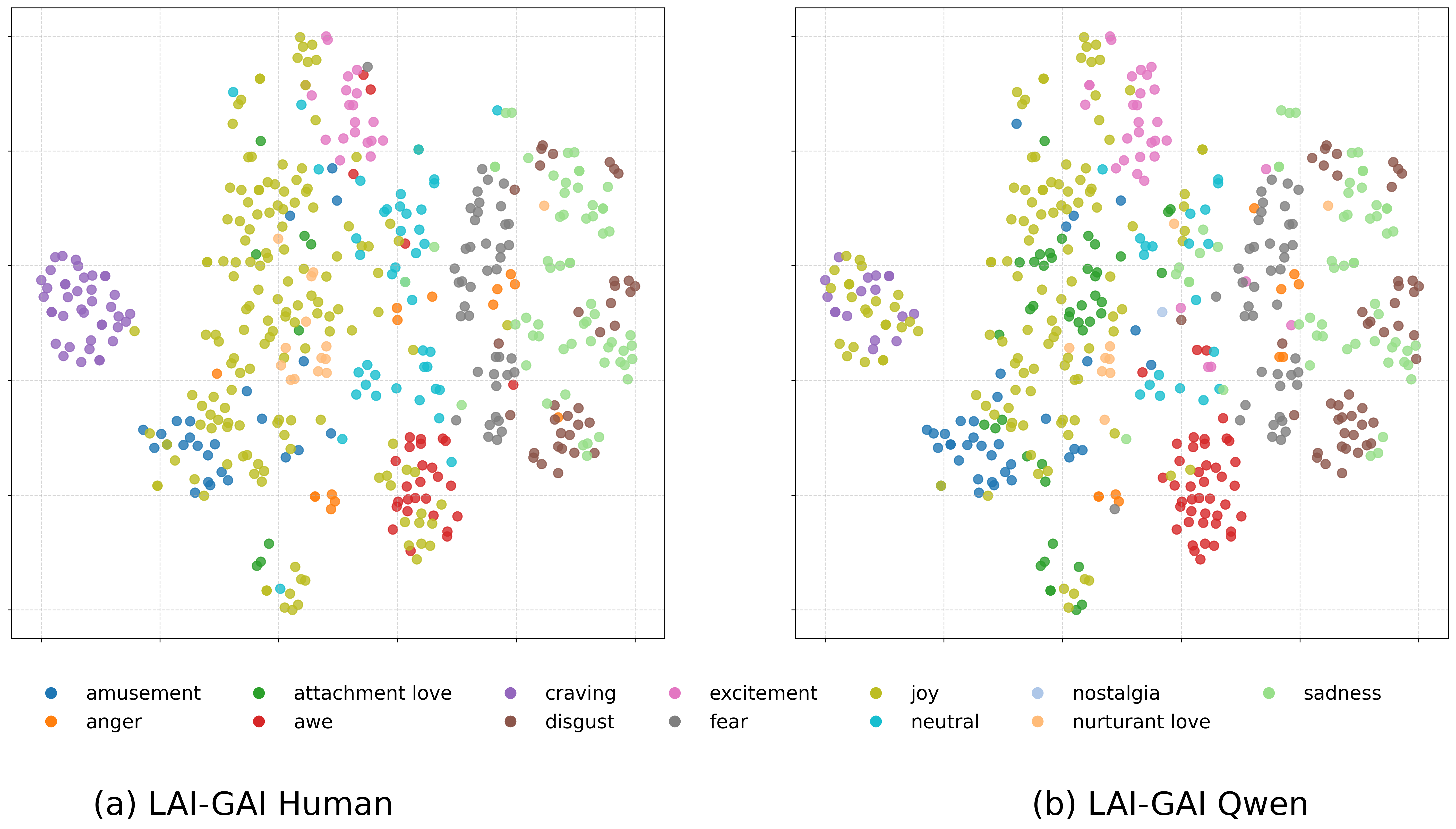}
    
    \vspace{0.5cm} 
    
    \includegraphics[width=\linewidth, height=0.29\textheight, keepaspectratio]{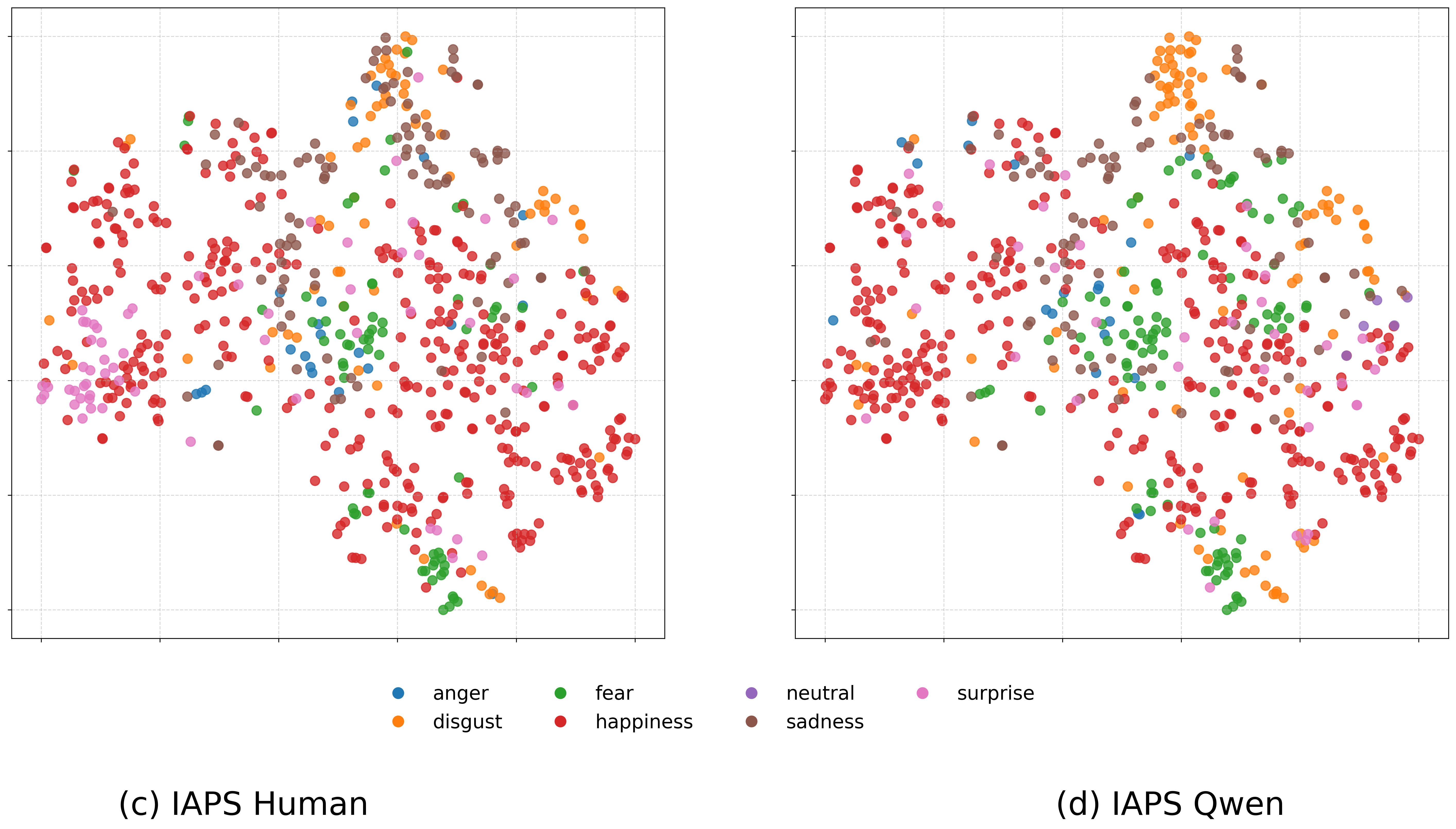}
    
    \vspace{0.5cm} 
    
    \includegraphics[width=\linewidth, height=0.29\textheight, keepaspectratio]{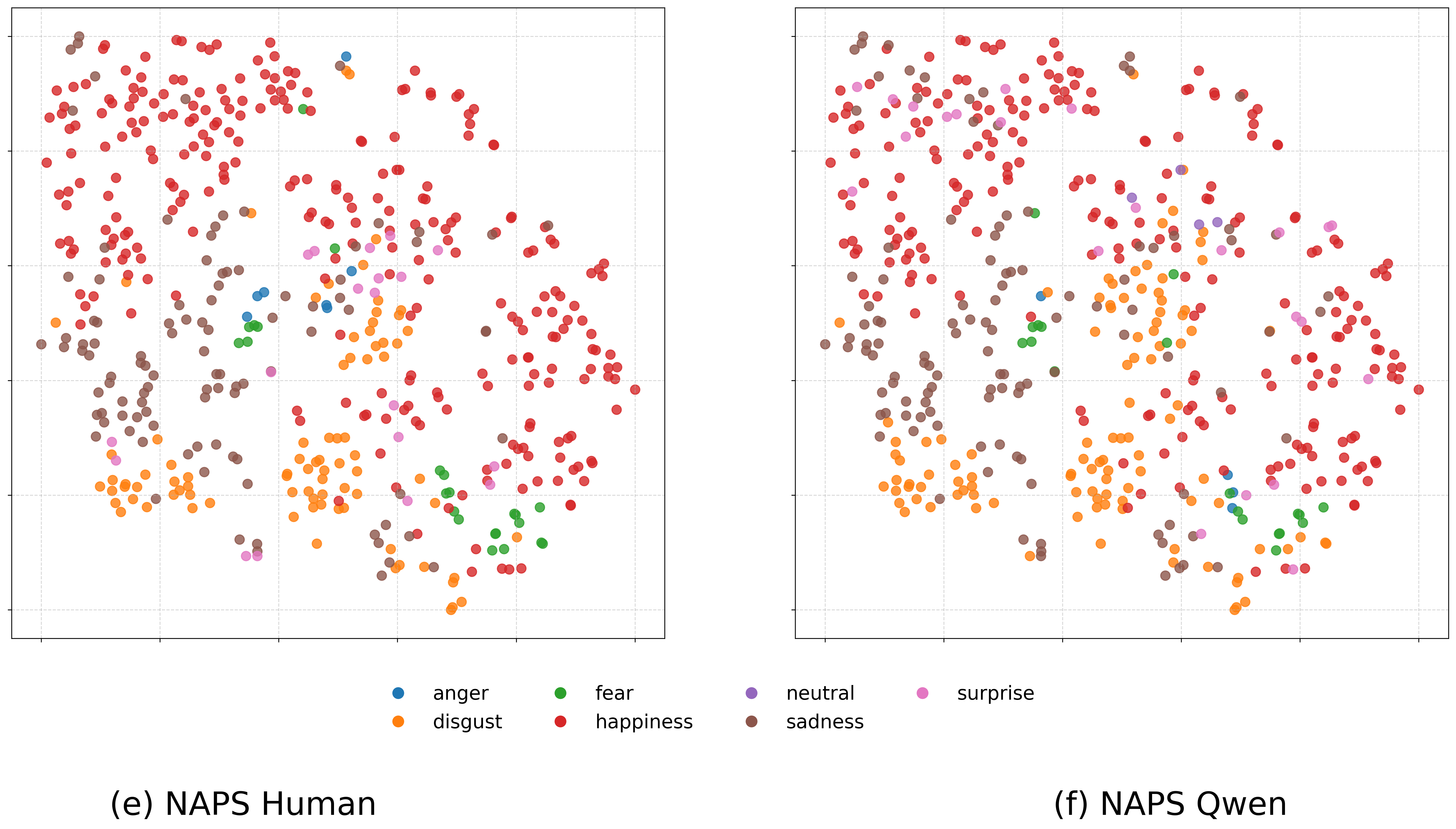}
    
    \caption{Visualization of the affective stimulus space via t-SNE embeddings. The figure compares ground-truth human emotion ratings with predictions from the Qwen-VL-2.5-72B model across three datasets. The panels are arranged as follows: (a) LAI-GAI, (c) IAPS, and (e) NAPS show the space colored by human-assigned emotion labels. Panels (b) LAI-GAI, (d) IAPS, and (f) NAPS show the same space colored by the model's predicted emotion labels. Each point represents an image, with its position determined by t-SNE applied to its L2-normalized CLIP vision embedding.}
    \label{fig:image_emb}
\end{figure*}

To illustrate the complexity of mapping visual content to affective meaning, we visualized the stimulus space as shown in Figure~\ref{fig:image_emb}. These visualizations compare the representations of human evaluation labels (left) and model evaluation labels (right) across the datasets. We present predictions from Qwen-VL-2.5-72B, which achieved strong performance across all datasets. To generate these plots, each image ($x_i$) was mapped into a CLIP embedding ($\pmb{v}_i = M_{\text{CLIP}}(x_i)$) and reduced via
\[
\pmb{v}_i \xrightarrow{\text{L2 Norm}} \hat{\pmb{v}}_i \xrightarrow{\text{t-SNE}} \pmb y_i^{(2D)} \xrightarrow{\text{MinMax}} \pmb{\tilde{y}}_i \in [0,1]^2,
\]
which yielded two-dimensional affective maps colored by the respective emotion labels. The maps showed that visually similar images elicited different emotional responses, and vice versa. These findings highlight the difficulty faced by VLMs: reliable affect recognition requires more than identifying surface-level visual features; it demands sensitivity to the stimuli's contextual and psychological aspects.

\section{Additional Classification Metrics}
\label{sec:classification_accuracy}

The main manuscript reports weighted F1 score as the primary classification metric, since it jointly summarises precision and recall and is robust to the strong class imbalance present in all three datasets (e.g., joy alone accounts for 31\% of LAI-GAI, and happiness for 54\% of IAPS and 53\% of NAPS). For completeness, this section reports pure classification accuracy alongside granular per-emotion precision and recall metrics. Reading these tables gives a more complete picture of where each model's competence lies.

\subsection{Overall Accuracy}

Table~\ref{tab:full_accuracy} reports overall accuracy per model per dataset on the full, unfiltered image pool, along with a cross-dataset average. Accuracy is calculated as the fraction of images for which the predicted label exactly matches the ground-truth label. Unlike F1, accuracy weights every sample equally and is therefore dominated by the majority class; a model that correctly identifies the most frequent emotion and fails on the rare ones can still achieve a high accuracy score. We note that for proprietary models (GPT-4.1, Gemini-2.5-Flash), accuracy is computed on the subset of images that were not blocked by the safety filters; the Content Safety Filters section (Section~\ref{sec:content_safety}) provides a matched-subset comparison that removes this confound.

\begin{table*}[t]
    \centering
    \begin{threeparttable}
        \caption{Overall classification accuracy (\%). \textbf{Bold} indicates the best model within each compute tier.}
        \label{tab:full_accuracy}
        \begin{tabular}{lrrr|r}
            \toprule
            Model & \multicolumn{4}{c}{Accuracy (\%)} \\
            \cmidrule(lr){2-5}
            & LAI-GAI & NAPS & IAPS & Average \\
            \midrule
            \textbf{Proprietary} \\
            GPT-4.1\modelnote            & 65.42 & 77.98 & 63.37 & 68.92 \\
            Gemini-2.5-Flash\modelnote   & 69.89 & 74.03 & 63.72 & 69.21 \\
            \midrule
            \textbf{Proprietary (Thinking)} \\
            GPT-5.4                      & 64.09 & 83.13 & 67.05 & 71.42 \\
            Gemini-3-Pro-Preview         & 63.33 & 83.73 & 67.49 & 71.52 \\
            Gemini-3-Flash-Preview       & \textbf{63.12} & \textbf{85.52} & \textbf{70.52} & \textbf{73.05} \\
            \midrule
            \textbf{On H100} \\
            Qwen-VL-2.5-72B   & 67.71 & \textbf{79.76} & \textbf{65.46} & \textbf{70.98} \\
            Intern-VL-3.5-38B & 67.50 & 78.37 & 58.67 & 68.18 \\
            Gemma-3-27B       & \textbf{75.42} & 72.62 & 56.36 & 68.13 \\
            Kimi-VL-A3B-16B   & 71.25 & 72.62 & 58.67 & 67.51 \\
            \midrule
            \textbf{On H100 (Thinking)} \\
            Qwen-VL-3-Thinking-30B & \textbf{69.79} & \textbf{77.38} & \textbf{64.60} & \textbf{70.59} \\
            \midrule
            \textbf{On RTX 4090} \\
            GLM-V-4.1-9B      & 58.75 & 70.63 & 59.25 & 62.88 \\
            Intern-VL-3.5-8B  & \textbf{67.08} & 75.79 & 60.84 & \textbf{67.90} \\
            Qwen-VL-2.5-7B    & 58.75 & \textbf{77.98} & \textbf{65.32} & 67.35 \\
            LLaVA-Mistral-7B  & 61.88 & 62.30 & 54.05 & 59.41 \\
            LLaVA-Vicuna-7B   & 35.42 & 46.83 & 33.09 & 38.45 \\
            \midrule
            \textbf{On RTX 4090 (Thinking)} \\
            Qwen-VL-3-Thinking-8B & \textbf{61.46} & \textbf{76.98} & \textbf{60.55} & \textbf{66.33} \\
            \bottomrule
        \end{tabular}
        \begin{tablenotes}
            \item[*] Proprietary models evaluated only on non-blocked images.
        \end{tablenotes}
    \end{threeparttable}
\end{table*}

Three main patterns emerge. First, cross-dataset \emph{average} accuracy is led by Qwen-VL-2.5-72B (70.98\%), narrowly ahead of the two proprietary models (Gemini-2.5-Flash 69.21\%, GPT-4.1 68.92\%); the strongest open 7-billion-parameter model (Intern-VL-3.5-8B) lands within three points of GPT-4.1 on average. Second, the ranking shifts noticeably across datasets: Gemma-3-27B achieves the highest accuracy on LAI-GAI (75.42\%) despite its lower 27B-class weighted-F1 on IAPS, reflecting that high accuracy on a dataset with a dominant majority class (joy in LAI-GAI) is easier to attain than high F1. Third, comparing Table~\ref{tab:full_accuracy} against the weighted F1 scores confirms that accuracy consistently \emph{exceeds} weighted F1 for all models on every dataset. The gap is widest for models that perform well on the majority class but fail on rare emotions (e.g., Kimi-VL-A3B-16B on IAPS: 58.67\% accuracy vs.\ 61.92\% weighted F1, and Qwen-VL-2.5-7B on IAPS: 65.32\% accuracy vs.\ 64.71\% weighted F1). This confirms that accuracy alone would overstate the models' ability to recognise minority emotions such as anger and surprise.

\subsection{Precision}

Precision measures the proportion of positive identifications that were actually correct ($\mathrm{TP}_e / (\mathrm{TP}_e + \mathrm{FP}_e)$). Tables~\ref{tab:precision_laigai} and~\ref{tab:precision_iaps_naps} report the per-emotion precision scores across the datasets, highlighting the models' ability to avoid false positives for specific emotions.

\begin{table*}[t]
    \centering
    \begin{adjustbox}{max width=\linewidth,center}
        \begin{threeparttable}
            \caption{Per-Emotion Precision Scores (\%) on LAI-GAI Dataset}
            \label{tab:precision_laigai}
            \begin{tabular}{lcccccccccccc | c}
                \toprule
                \textbf{Model} & \textbf{Amusement} & \textbf{Anger} & \textbf{Attach. Love} & \textbf{Awe} & \textbf{Craving} & \textbf{Disgust} & \textbf{Excitement} & \textbf{Fear} & \textbf{Joy} & \textbf{Neutral} & \textbf{Nurt. Love} & \textbf{Sadness} & \textbf{Weighted Precision} \\
                \textit{Support ($N$)} & \textit{(25)} & \textit{(14)} & \textit{(8)} & \textit{(30)} & \textit{(40)} & \textit{(35)} & \textit{(20)} & \textit{(42)} & \textit{(150)} & \textit{(42)} & \textit{(14)} & \textit{(60)} & \textit{(480)} \\
                \midrule
                GPT-4.1\modelnote & 64.71 & 66.67 & 14.29 & 54.17 & \textcolor{green}{100.00} & 62.96 & 72.00 & 70.18 & 91.23 & 78.57 & 37.50 & 82.61 & 77.81 \\
                GPT-5.4\modelnote & 62.86 & 80.00 & 16.67 & 58.14 & \textcolor{green}{100.00} & 63.46 & 52.63 & 78.85 & 88.37 & 80.65 & 20.97 & 85.42 & 77.53 \\
                Gemini-3-Pro\modelnote & 55.26 & 80.00 & 21.74 & 51.02 & 97.56 & 71.74 & 54.05 & 86.36 & 93.55 & 78.79 & 17.81 & 85.19 & 79.23 \\
                Gemini-3-Flash\modelnote & 57.89 & 85.71 & 13.51 & 48.21 & \textcolor{green}{100.00} & 76.74 & 64.52 & 81.63 & \textcolor{green}{96.15} & 84.00 & 18.18 & 85.25 & \textcolor{green}{81.10} \\
                Gemini-2.5-Flash\modelnote & 57.89 & 85.71 & 19.23 & 65.79 & \textcolor{green}{100.00} & 64.86 & 58.06 & 88.89 & 89.71 & 76.92 & 24.49 & 89.36 & 80.02 \\
                Qwen-VL-2.5-72B & 62.86 & 70.00 & 18.60 & 58.14 & \textcolor{green}{100.00} & 71.74 & 53.12 & 85.71 & 70.40 & 85.00 & \textcolor{green}{77.78} & 84.21 & 74.77 \\
                Qwen-VL-3-Think-30B & 66.67 & \textcolor{green}{100.00} & 22.73 & 60.47 & \textcolor{green}{100.00} & 56.14 & 60.00 & 77.36 & 78.74 & 66.67 & 38.46 & 82.22 & 74.08 \\
                Qwen-VL-3-Think-8B & 77.78 & \textcolor{red}{25.00} & 11.94 & 42.86 & \textcolor{green}{100.00} & 66.00 & 58.82 & 73.08 & 70.43 & \textcolor{green}{100.00} & 12.50 & \textcolor{green}{90.70} & 72.11 \\
                Intern-VL-3.5-38B & 80.00 & \textcolor{green}{100.00} & 14.89 & 55.56 & \textcolor{green}{100.00} & 87.88 & 51.35 & 66.13 & 69.34 & 82.61 & \textcolor{red}{0.00} & 92.00 & 73.87 \\
                Intern-VL-3.5-8B & \textcolor{green}{84.62} & 50.00 & 21.43 & 63.41 & \textcolor{green}{100.00} & 80.65 & 53.85 & 84.85 & 61.54 & 78.57 & \textcolor{red}{0.00} & 78.67 & 70.01 \\
                Gemma-3-27B & 78.26 & 33.33 & 35.71 & 55.32 & 97.56 & 80.49 & 68.42 & 81.63 & 84.62 & 79.31 & 41.67 & 73.97 & 76.94 \\
                Kimi-VL-A3B-16B & 42.31 & \textcolor{green}{100.00} & \textcolor{green}{50.00} & \textcolor{green}{84.00} & 97.56 & 46.58 & 70.83 & 75.00 & 80.58 & 81.25 & 16.67 & 88.24 & 76.05 \\
                GLM-V-4.1-9B & \textcolor{red}{30.19} & 50.00 & 16.67 & 46.30 & \textcolor{green}{100.00} & 58.49 & \textcolor{green}{80.00} & 64.81 & 70.33 & 67.86 & 16.67 & 79.07 & 66.09 \\
                LLaVA-Mistral-7B & 75.00 & \textcolor{red}{25.00} & 14.29 & 47.17 & \textcolor{green}{100.00} & 80.65 & 71.43 & 67.35 & \textcolor{red}{60.87} & \textcolor{green}{88.89} & \textcolor{red}{0.00} & 65.48 & 65.89 \\
                LLaVA-Vicuna-7B & 50.00 & \textcolor{red}{25.00} & \textcolor{red}{0.00} & \textcolor{red}{10.28} & \textcolor{green}{100.00} & \textcolor{green}{100.00} & \textcolor{red}{0.00} & \textcolor{red}{33.33} & 81.94 & \textcolor{red}{0.00} & \textcolor{red}{0.00} & \textcolor{red}{61.80} & \textcolor{red}{55.85} \\
                \midrule
                \textbf{Average} & \textbf{63.09} & \textbf{63.09} & \textbf{19.45} & \textbf{53.39} & \textbf{99.51} & \textbf{71.23} & \textbf{57.87} & \textbf{74.21} & \textbf{79.19} & \textbf{75.27} & \textbf{21.51} & \textbf{81.55} & \textbf{73.43} \\
                \bottomrule
            \end{tabular}
            \begin{tablenotes}
                \item[*] Proprietary models evaluated only on non-blocked images.
            \end{tablenotes}
        \end{threeparttable}
    \end{adjustbox}
\end{table*}

\begin{table*}[t]
    \centering
    \begin{adjustbox}{max width=\linewidth,center}
        \begin{threeparttable}
            \caption{Per-Emotion Precision Scores (\%) on IAPS and NAPS Datasets}
            \label{tab:precision_iaps_naps}
            \begin{tabular}{lcccccc|c}
                \toprule
                \textbf{Model} & \textbf{Anger} & \textbf{Disgust} & \textbf{Fear} & \textbf{Happiness} & \textbf{Sadness} & \textbf{Surprise} & \textbf{Weighted Precision} \\
                \midrule

                \multicolumn{8}{c}{\textbf{IAPS Dataset}} \\
                \textit{Support ($N$)} & \textit{(22)} & \textit{(67)} & \textit{(67)} & \textit{(372)} & \textit{(103)} & \textit{(61)} & \textit{(692)} \\
                \midrule
                GPT-4.1\modelnote & 14.29 & 53.85 & 40.56 & 93.96 & 61.46 & 12.73 & 71.81 \\
                GPT-5.4\modelnote & \textcolor{green}{27.27} & 54.02 & 38.89 & 89.87 & 65.85 & \textcolor{green}{38.24} & 71.35 \\
                Gemini-3-Pro\modelnote & 22.22 & 48.35 & 41.43 & 89.21 & 56.38 & 30.77 & 68.46 \\
                Gemini-3-Flash\modelnote & \textcolor{red}{0.00} & 48.45 & 47.24 & 90.26 & 61.39 & 30.77 & 69.63 \\
                Gemini-2.5-Flash\modelnote & 17.39 & 53.12 & 47.41 & 94.84 & 65.62 & 13.85 & \textcolor{green}{74.38} \\
                Qwen-VL-2.5-72B & 15.79 & 47.42 & 49.49 & 87.02 & 60.00 & \textcolor{red}{7.89} & 66.29 \\
                Qwen-VL-3-Think-30B & 10.00 & 42.37 & 39.60 & 89.78 & 61.33 & 9.09 & 66.45 \\
                Qwen-VL-3-Think-8B & 20.00 & 40.98 & 35.67 & 87.62 & 50.00 & 16.00 & 64.01 \\
                Intern-VL-3.5-38B & 7.14 & 47.37 & \textcolor{red}{28.86} & 89.63 & 57.32 & 14.29 & 65.58 \\
                Intern-VL-3.5-8B & 15.00 & 48.72 & 45.45 & 84.40 & 45.66 & 12.12 & \textcolor{red}{62.83} \\
                Gemma-3-27B & \textcolor{red}{0.00} & 45.92 & 41.61 & 93.58 & 47.62 & 17.07 & 67.37 \\
                Kimi-VL-A3B-16B & 16.67 & 41.12 & 50.54 & 89.67 & \textcolor{green}{69.41} & 9.60 & 68.79 \\
                GLM-V-4.1-9B & 18.60 & 33.59 & 35.38 & 89.44 & 54.93 & 14.29 & 64.78 \\
                LLaVA-Mistral-7B & 10.00 & \textcolor{green}{92.86} & 31.29 & \textcolor{red}{82.89} & 55.21 & 9.18 & 65.92 \\
                LLaVA-Vicuna-7B & 20.00 & \textcolor{red}{15.87} & \textcolor{green}{59.26} & \textcolor{green}{95.08} & \textcolor{red}{41.94} & 11.63 & 66.29 \\
                \midrule
                \textbf{IAPS Average} & \textbf{14.29} & \textbf{47.60} & \textbf{41.91} & \textbf{89.82} & \textbf{56.94} & \textbf{16.50} & \textbf{67.60} \\
                \midrule

                \multicolumn{8}{c}{\textbf{NAPS Dataset}} \\
                \textit{Support ($N$)} & \textit{(7)} & \textit{(85)} & \textit{(23)} & \textit{(268)} & \textit{(102)} & \textit{(19)} & \textit{(504)} \\
                \midrule
                GPT-4.1\modelnote & \textcolor{red}{0.00} & 73.83 & 50.00 & 98.57 & 77.36 & 11.11 & 83.22 \\
                GPT-5.4\modelnote & \textcolor{red}{0.00} & 69.49 & 53.49 & 98.32 & \textcolor{green}{87.50} & 23.08 & \textcolor{green}{85.02} \\
                Gemini-3-Pro\modelnote & \textcolor{red}{0.00} & 75.23 & 53.66 & 98.33 & 75.70 & \textcolor{green}{50.00} & 84.63 \\
                Gemini-3-Flash\modelnote & \textcolor{red}{0.00} & 73.21 & 61.76 & 97.57 & 80.37 & \textcolor{green}{50.00} & 85.20 \\
                Gemini-2.5-Flash\modelnote & 22.22 & 56.56 & 76.00 & 98.95 & 77.36 & 4.55 & 82.49 \\
                Qwen-VL-2.5-72B & 25.00 & 69.16 & 78.95 & 94.40 & 78.26 & 13.04 & 82.14 \\
                Qwen-VL-3-Think-30B & 16.67 & 60.61 & 72.00 & 95.24 & 73.20 & \textcolor{red}{0.00} & 79.19 \\
                Qwen-VL-3-Think-8B & 25.00 & 76.14 & 40.48 & 95.20 & 68.07 & 21.43 & 80.24 \\
                Intern-VL-3.5-38B & \textcolor{red}{0.00} & 84.62 & 33.33 & 96.93 & 73.45 & 22.22 & 83.04 \\
                Intern-VL-3.5-8B & 33.33 & 93.18 & 75.00 & 92.89 & \textcolor{red}{55.11} & 27.27 & 81.18 \\
                Gemma-VL-3-27B & \textcolor{red}{0.00} & 71.30 & 65.52 & \textcolor{green}{99.42} & 63.57 & 8.89 & 81.08 \\
                Kimi-VL-A3B-16B & \textcolor{green}{50.00} & 61.98 & 68.00 & 97.47 & 78.95 & 9.26 & 82.41 \\
                GLM-V-4.1-9B & 21.43 & 52.41 & 30.43 & 95.35 & 76.32 & \textcolor{red}{0.00} & \textcolor{red}{76.67} \\
                Qwen-VL-2.5-7B & 14.29 & 66.99 & \textcolor{green}{83.33} & \textcolor{red}{91.63} & 72.73 & \textcolor{red}{0.00} & 78.74 \\
                LLaVA-Mistral-7B & 33.33 & \textcolor{green}{94.29} & \textcolor{red}{18.75} & 96.15 & 68.25 & 9.35 & 82.52 \\
                LLaVA-Vicuna-7B & \textcolor{red}{0.00} & \textcolor{red}{30.97} & 75.00 & 98.25 & 76.09 & 4.92 & 76.47 \\
                \midrule
                \textbf{NAPS Average} & \textbf{15.08} & \textbf{69.37} & \textbf{58.42} & \textbf{96.54} & \textbf{73.89} & \textbf{15.95} & \textbf{81.52} \\
                \bottomrule
            \end{tabular}
            \begin{tablenotes}
                \item[*] Proprietary models evaluated only on non-blocked images.
            \end{tablenotes}
        \end{threeparttable}
    \end{adjustbox}
\end{table*}

\subsection{Recall (Per-Emotion Accuracy)}

We computed per-emotion recall as the fraction of images with a given ground-truth emotion that were correctly classified. By construction, this quantity coincides with per-class accuracy, since on the single-class subset where $y_{\text{true}} = e$, accuracy reduces to $\mathrm{TP}_e / (\mathrm{TP}_e + \mathrm{FN}_e)$. Tables~\ref{tab:recall_laigai} and~\ref{tab:recall_iaps_naps} provide this information, which can be read directly as per-emotion accuracy. 

\begin{table*}[t]
    \centering
    \begin{adjustbox}{max width=\linewidth,center}
        \begin{threeparttable}
            \caption{Per-Emotion Recall Scores (\%) on LAI-GAI Dataset}
            \label{tab:recall_laigai}
            \begin{tabular}{lcccccccccccc | c}
                \toprule
                \textbf{Model} & \textbf{Amusement} & \textbf{Anger} & \textbf{Attach. Love} & \textbf{Awe} & \textbf{Craving} & \textbf{Disgust} & \textbf{Excitement} & \textbf{Fear} & \textbf{Joy} & \textbf{Neutral} & \textbf{Nurt. Love} & \textbf{Sadness} & \textbf{Weighted Recall} \\
                \textit{Support ($N$)} & \textit{(25)} & \textit{(14)} & \textit{(8)} & \textit{(30)} & \textit{(40)} & \textit{(35)} & \textit{(20)} & \textit{(42)} & \textit{(150)} & \textit{(42)} & \textit{(14)} & \textit{(60)} & \textit{(480)} \\
                \midrule
                GPT-4.1\modelnote & \textcolor{green}{88.00} & 14.29 & \textcolor{green}{100.00} & 86.67 & \textcolor{green}{100.00} & \textcolor{green}{97.14} & 90.00 & 95.24 & 34.67 & 52.38 & 85.71 & 63.33 & 65.42 \\
                GPT-5.4\modelnote & \textcolor{green}{88.00} & 30.77 & 62.50 & 83.33 & \textcolor{green}{100.00} & 94.29 & \textcolor{green}{100.00} & \textcolor{green}{97.62} & \textcolor{red}{25.33} & 59.52 & \textcolor{green}{92.86} & 68.33 & 64.09 \\
                Gemini-3-Pro\modelnote & 84.00 & 57.14 & 62.50 & 83.33 & \textcolor{green}{100.00} & 94.29 & \textcolor{green}{100.00} & 90.48 & 19.33 & 61.90 & \textcolor{green}{92.86} & 76.67 & 63.33 \\
                Gemini-3-Flash\modelnote & \textcolor{green}{88.00} & 42.86 & 62.50 & \textcolor{green}{90.00} & \textcolor{green}{100.00} & 94.29 & \textcolor{green}{100.00} & 95.24 & 16.67 & 50.00 & 85.71 & 86.67 & 63.12 \\
                Gemini-2.5-Flash\modelnote & \textcolor{green}{88.00} & 42.86 & 62.50 & 83.33 & \textcolor{green}{100.00} & 92.31 & 90.00 & 95.24 & 40.67 & \textcolor{green}{71.43} & 85.71 & 77.78 & 69.89 \\
                Qwen-VL-2.5-72B & \textcolor{green}{88.00} & 50.00 & \textcolor{green}{100.00} & 83.33 & 42.50 & 94.29 & 85.00 & 85.71 & 58.67 & 40.48 & 50.00 & 80.00 & 67.71 \\
                Qwen-VL-3-Think-30B & 80.00 & \textcolor{red}{7.14} & 62.50 & 86.67 & 77.50 & 91.43 & 75.00 & \textcolor{green}{97.62} & 66.67 & 52.38 & 35.71 & 61.67 & 69.79 \\
                Qwen-VL-3-Think-8B & 56.00 & \textcolor{red}{7.14} & \textcolor{green}{100.00} & \textcolor{green}{90.00} & 82.50 & 94.29 & 50.00 & 90.48 & 54.00 & 23.81 & 7.14 & 65.00 & 61.46 \\
                Intern-VL-3.5-38B & 32.00 & 14.29 & 87.50 & 83.33 & 82.50 & 82.86 & 95.00 & \textcolor{green}{97.62} & 63.33 & 45.24 & \textcolor{red}{0.00} & 76.67 & 67.50 \\
                Intern-VL-3.5-8B & 44.00 & 21.43 & 37.50 & 86.67 & 15.00 & 71.43 & 70.00 & 66.67 & \textcolor{green}{90.67} & 26.19 & \textcolor{red}{0.00} & \textcolor{green}{98.33} & 67.08 \\
                Gemma-3-27B & 72.00 & \textcolor{red}{7.14} & 62.50 & 86.67 & \textcolor{green}{100.00} & 94.29 & 65.00 & 95.24 & 66.00 & 54.76 & 71.43 & 90.00 & \textcolor{green}{75.42} \\
                Kimi-VL-A3B-16B & \textcolor{green}{88.00} & 14.29 & 50.00 & \textcolor{red}{70.00} & \textcolor{green}{100.00} & \textcolor{green}{97.14} & 85.00 & 78.57 & 74.67 & 61.90 & 7.14 & \textcolor{red}{50.00} & 71.25 \\
                GLM-V-4.1-9B & 64.00 & \textcolor{green}{64.29} & 62.50 & 83.33 & 85.00 & 88.57 & 40.00 & 83.33 & 42.67 & 45.24 & 14.29 & 56.67 & 58.75 \\
                LLaVA-Mistral-7B & 48.00 & \textcolor{red}{7.14} & 12.50 & 83.33 & \textcolor{red}{15.00} & 71.43 & 25.00 & 78.57 & 84.00 & 19.05 & \textcolor{red}{0.00} & 91.67 & 61.88 \\
                LLaVA-Vicuna-7B & \textcolor{red}{12.00} & \textcolor{red}{7.14} & \textcolor{red}{0.00} & \textcolor{green}{96.67} & \textcolor{red}{15.00} & \textcolor{red}{45.71} & \textcolor{red}{0.00} & \textcolor{red}{2.38} & 39.33 & \textcolor{red}{0.00} & \textcolor{red}{0.00} & 91.67 & \textcolor{red}{35.42} \\
                \midrule
                \textbf{Average} & \textbf{68.00} & \textbf{25.85} & \textbf{61.67} & \textbf{85.04} & \textbf{74.33} & \textbf{86.94} & \textbf{71.33} & \textbf{83.33} & \textbf{51.78} & \textbf{44.29} & \textbf{41.90} & \textbf{74.30} & \textbf{64.01} \\
                \bottomrule
            \end{tabular}
            \begin{tablenotes}
                \item[*] Proprietary models evaluated only on non-blocked images.
            \end{tablenotes}
        \end{threeparttable}
    \end{adjustbox}
\end{table*}

\begin{table*}[t]
    \centering
    \begin{adjustbox}{max width=\linewidth,center}
        \begin{threeparttable}
            \caption{Per-Emotion Recall Scores (\%) on IAPS and NAPS Datasets}
            \label{tab:recall_iaps_naps}
            \begin{tabular}{lcccccc|c}
                \toprule
                \textbf{Model} & \textbf{Anger} & \textbf{Disgust} & \textbf{Fear} & \textbf{Happiness} & \textbf{Sadness} & \textbf{Surprise} & \textbf{Weighted Recall} \\
                \midrule

                \multicolumn{8}{c}{\textbf{IAPS Dataset}} \\
                \textit{Support ($N$)} & \textit{(22)} & \textit{(67)} & \textit{(67)} & \textit{(372)} & \textit{(103)} & \textit{(61)} & \textit{(692)} \\
                \midrule
                GPT-4.1\modelnote & 9.09 & 63.64 & 86.57 & 69.17 & 57.28 & 17.50 & 63.37 \\
                GPT-5.4\modelnote & 13.64 & 70.15 & \textcolor{green}{94.03} & 76.34 & 52.43 & 21.31 & 67.05 \\
                Gemini-3-Pro\modelnote & 9.09 & 65.67 & 86.57 & 82.26 & 51.46 & 6.56 & 67.49 \\
                Gemini-3-Flash\modelnote & \textcolor{red}{0.00} & 70.15 & 89.55 & \textcolor{green}{84.68} & 60.19 & 6.56 & \textcolor{green}{70.52} \\
                Gemini-2.5-Flash\modelnote & 19.05 & 66.67 & 83.33 & 66.02 & 68.48 & 21.43 & 63.72 \\
                Qwen-VL-2.5-72B & 13.64 & 68.66 & 73.13 & 79.30 & 55.34 & 4.92 & 65.46 \\
                Qwen-VL-3-Think-30B & 4.55 & 74.63 & 88.06 & 77.96 & 44.66 & \textcolor{red}{1.64} & 64.60 \\
                Qwen-VL-3-Think-8B & 13.64 & 37.31 & 83.58 & 74.19 & 53.40 & 6.56 & 60.55 \\
                Intern-VL-3.5-38B & 4.55 & 40.30 & 86.57 & 72.04 & 45.63 & 8.20 & 58.67 \\
                Intern-VL-3.5-8B & 13.64 & 28.36 & 59.70 & 74.19 & \textcolor{green}{76.70} & 6.56 & 60.84 \\
                Gemma-3-27B & \textcolor{red}{0.00} & 67.16 & 85.07 & 54.84 & 67.96 & \textcolor{green}{22.95} & 56.36 \\
                Kimi-VL-A3B-16B & 4.55 & 65.67 & 70.15 & 65.32 & 57.28 & 19.67 & 58.67 \\
                GLM-V-4.1-9B & \textcolor{green}{36.36} & 65.67 & 68.66 & 72.85 & 37.86 & 3.28 & 59.25 \\
                LLaVA-Mistral-7B & 4.55 & \textcolor{red}{19.40} & 76.12 & 66.40 & 51.46 & 14.75 & 54.05 \\
                LLaVA-Vicuna-7B & 4.55 & \textcolor{green}{89.55} & \textcolor{red}{23.88} & \textcolor{red}{31.18} & \textcolor{red}{25.24} & 16.39 & \textcolor{red}{33.09} \\
                \midrule
                \textbf{IAPS Average} & \textbf{10.06} & \textbf{59.53} & \textbf{77.00} & \textbf{69.77} & \textbf{53.69} & \textbf{11.88} & \textbf{60.22} \\
                \midrule

                \multicolumn{8}{c}{\textbf{NAPS Dataset}} \\
                \textit{Support ($N$)} & \textit{(7)} & \textit{(85)} & \textit{(23)} & \textit{(268)} & \textit{(102)} & \textit{(19)} & \textit{(504)} \\
                \midrule
                GPT-4.1\modelnote & \textcolor{red}{0.00} & 92.94 & 95.65 & 77.24 & 80.39 & 15.79 & 77.98 \\
                GPT-5.4\modelnote & \textcolor{red}{0.00} & 96.47 & \textcolor{green}{100.00} & 87.31 & 75.49 & 15.79 & 83.13 \\
                Gemini-3-Pro\modelnote & \textcolor{red}{0.00} & 96.47 & 95.65 & 88.06 & 79.41 & 5.26 & 83.73 \\
                Gemini-3-Flash\modelnote & \textcolor{red}{0.00} & 96.47 & 91.30 & 89.93 & 84.31 & 5.26 & \textcolor{green}{85.52} \\
                Gemini-2.5-Flash\modelnote & 28.57 & 97.18 & 82.61 & 70.52 & 81.19 & 5.26 & 74.03 \\
                Qwen-VL-2.5-72B & 14.29 & 87.06 & 65.22 & 81.72 & 88.24 & 15.79 & 79.76 \\
                Qwen-VL-3-Think-30B & 14.29 & 94.12 & 78.26 & 82.09 & 69.61 & \textcolor{red}{0.00} & 77.38 \\
                Qwen-VL-3-Think-8B & 28.57 & 78.82 & 73.91 & 81.34 & 79.41 & 15.79 & 76.98 \\
                Intern-VL-3.5-38B & \textcolor{red}{0.00} & 77.65 & 91.30 & 82.46 & 81.37 & 21.05 & 78.37 \\
                Intern-VL-3.5-8B & 14.29 & \textcolor{red}{48.24} & 78.26 & 82.84 & \textcolor{green}{95.10} & 15.79 & 75.79 \\
                Gemma-VL-3-27B & \textcolor{red}{0.00} & 96.47 & 82.61 & 64.18 & 87.25 & 21.05 & 72.62 \\
                Kimi-VL-A3B-16B & 14.29 & 88.24 & 73.91 & 72.01 & 73.53 & 26.32 & 72.62 \\
                GLM-V-4.1-9B & \textcolor{green}{42.86} & 89.41 & 60.87 & 76.49 & 56.86 & \textcolor{red}{0.00} & 70.63 \\
                Qwen-VL-2.5-7B & 14.29 & 81.18 & 21.74 & 85.82 & 86.27 & \textcolor{red}{0.00} & 77.98 \\
                LLaVA-Mistral-7B & 14.29 & 38.82 & 39.13 & 65.30 & 84.31 & \textcolor{green}{52.63} & 62.30 \\
                LLaVA-Vicuna-7B & \textcolor{red}{0.00} & \textcolor{green}{97.65} & \textcolor{red}{13.04} & \textcolor{red}{41.79} & \textcolor{red}{34.31} & 15.79 & \textcolor{red}{46.83} \\
                \midrule
                \textbf{NAPS Average} & \textbf{10.98} & \textbf{84.76} & \textbf{71.47} & \textbf{76.69} & \textbf{77.31} & \textbf{14.47} & \textbf{74.73} \\
                \bottomrule
            \end{tabular}
            \begin{tablenotes}
                \item[*] Proprietary models evaluated only on non-blocked images.
            \end{tablenotes}
        \end{threeparttable}
    \end{adjustbox}
\end{table*}

\raggedbottom
\section{Impact of Content Safety Filters}
\label{sec:content_safety}

As noted in the main manuscript, proprietary models (GPT-4.1 and Gemini-2.5-Flash) triggered safety filters for certain images, refusing to generate an output. A qualitative review of the filtered images revealed that the safety mechanisms primarily targeted explicit sexual content and extreme graphic violence. 

Specifically, GPT-4.1 blocked images exclusively containing explicit erotic scenes, nudity, and exposed intimate body parts (observed in the IAPS dataset). Gemini-2.5-Flash applied a more aggressive filter; in addition to the sexually explicit content blocked by GPT-4.1, Gemini also filtered images depicting severe graphic violence and gore across all datasets. This included gruesome scenes of human mutilation, accident victims, corpses, a bloodied child, severe facial and body lacerations, and exposed entrails (in IAPS, LAI-GAI, and NAPS). Furthermore, Gemini blocked animal gore, such as flayed animal carcasses, a dog with exposed entrails, and predators consuming prey (in NAPS and LAI-GAI).

To ensure that the comparisons between proprietary and open-source models were not skewed by the removal of these specific extreme images, we re-evaluated all models on the restricted subsets.

\subsection{Performance on the GPT-4.1 Evaluated Subset}
The tables below present performance metrics calculated \textbf{only} on the subset of images that GPT-4.1 successfully processed (excluding the 34 blocked images from the IAPS dataset).

\begin{table*}[!b]
    \centering
    \begin{minipage}[t]{0.3\linewidth}
        \centering
        \caption{Comparison of Model Accuracy (GPT-4.1 Subset)}
        \label{tab:gpt_subset_acc}
        \begin{tabular}{lr}
            \toprule
            Model & Acc. (\%) \\
            \cmidrule(lr){2-2}
            & IAPS \\
            \midrule
            \textbf{Proprietary} & \\
            Gemini-2.5 & 64.10 \\
            \midrule
            \textbf{On H100} & \\
            Qwen-72B & 66.87 \\
            Intern-38B & \textbf{67.50} \\
            Gemma3-27B & 56.99 \\
            Kimi-16B & 59.88 \\
            \midrule
            \textbf{On 4090} & \\
            GLM-4.1-9B & 60.33 \\
            Intern-8B & 62.01 \\
            Qwen-7B & \textbf{66.87} \\
            Mistral-7B & 55.02 \\
            Vicuna-7B & 34.50 \\
            \bottomrule
        \end{tabular}
    \end{minipage}%
    \hfill 
    \begin{minipage}[t]{0.68\linewidth}
        \centering
        \caption{Per-Emotion F1 Scores (\%) on IAPS (GPT-4.1 Subset)}
        \label{tab:gpt_subset_f1}
        \begin{tabular}{lcccccc|c}
            \toprule
            \textbf{Model} & \textbf{Anger} & \textbf{Disg.} & \textbf{Fear} & \textbf{Happ.} & \textbf{Sad.} & \textbf{Surp.} & \textbf{W. F1} \\
            \midrule
            \multicolumn{8}{c}{\textbf{IAPS Dataset}} \\
            \textit{Support ($N$)} & \textit{(22)} & \textit{(66)} & \textit{(67)} & \textit{(360)} & \textit{(103)} & \textit{(40)} & \textit{(658)} \\
            \midrule
            Gemini-2.5-Flash & 18.18 & \textcolor{green}{60.18} & \textcolor{green}{60.44} & 78.15 & \textcolor{green}{67.02} & \textcolor{green}{16.16} & 67.57 \\
            Qwen-VL-2.5-72B & 14.63 & 55.90 & 59.04 & 84.86 & 57.58 & 7.69 & \textcolor{green}{68.01} \\
            Intern-VL-3.5-38B & 5.56 & 48.15 & 43.28 & 80.56 & 51.37 & 13.51 & 62.36 \\
            Gemma-3-27B & \textcolor{red}{0.00} & 57.14 & 55.88 & 69.27 & 56.68 & 15.52 & 59.14 \\
            Kimi-VL-A3B-16B & 7.14 & 53.42 & 58.75 & 76.07 & 63.10 & 14.55 & 63.96 \\
            GLM-V-4.1-9B & \textcolor{green}{24.62} & 45.74 & 46.70 & 81.06 & 45.09 & \textcolor{red}{7.41} & 62.03 \\
            Intern-VL-3.5-8B & 15.00 & 35.29 & 51.61 & 80.00 & 57.66 & 10.96 & 62.76 \\
            Qwen-VL-2.5-7B & 23.73 & 50.93 & 43.33 & \textcolor{green}{86.08} & 58.71 & 12.90 & 67.38 \\
            LLaVA-Mistral-7B & 6.25 & 32.50 & 44.35 & 75.20 & 53.27 & 13.14 & 58.26 \\
            LLaVA-Vicuna-7B & 7.41 & \textcolor{red}{28.85} & \textcolor{red}{34.04} & \textcolor{red}{47.92} & \textcolor{red}{31.71} & 14.63 & \textcolor{red}{38.68} \\
            \midrule
            \textbf{Average} & \textbf{12.15} & \textbf{47.86} & \textbf{50.24} & \textbf{76.26} & \textbf{54.68} & \textbf{12.84} & \textbf{61.43} \\
            \bottomrule
        \end{tabular}
    \end{minipage}
\end{table*}

\begin{table*}[!t]
    \centering
    \scriptsize 
    \setlength{\tabcolsep}{3pt} 
    \renewcommand{\arraystretch}{1.5} 
    \newcommand{\up}{\textcolor{gray}{$\uparrow$}}
    \newcommand{\down}{\textcolor{gray}{$\downarrow$}}
    \newcommand{\rot}[1]{\multicolumn{1}{c}{\rotatebox{45}{\textbf{#1}}}}
    \begin{threeparttable}
        \caption{Side-by-Side Comparison of Pearson's Correlation ($r$) and MAE (GPT-4.1 Subset)}
        \label{tab:gpt_subset_corr_mae}
        
        \begin{tabular}{ll | ccccc c | ccccc c}
            \toprule
            \multicolumn{2}{c}{} & 
            \multicolumn{6}{c}{\textbf{Pearson's Correlation ($r$)} \up} & 
            \multicolumn{6}{c}{\textbf{Mean Absolute Error (MAE)} \down} \\
            \cmidrule(lr){3-8} \cmidrule(lr){9-14}
            
            \multicolumn{2}{c}{\textbf{Scale}} & 
            \rot{Gemini 2.5} & \rot{Qwen 72B} & \rot{Gemma 27B} & \rot{Intern 8B} & \rot{Qwen 7B} & \rot{Average} & 
            \rot{Gemini 2.5} & \rot{Qwen 72B} & \rot{Gemma 27B} & \rot{Intern 8B} & \rot{Qwen 7B} & \rot{Average} \\
            \midrule
            
            \multirow{9}{*}{\rotatebox{90}{\textbf{IAPS}}}
            & Anger 
            & \heatmapcell{0.76} & \heatmapcell{0.62} & \heatmapcell{0.69} & \heatmapcell{0.57} & \heatmapcell{0.55} & \heatmapcell[b]{0.64} 
            & \maeheatmapcell{1.17} & \maeheatmapcell{1.06} & \maeheatmapcell{1.33} & \maeheatmapcell{1.01} & \maeheatmapcell{1.05} & \maeheatmapcell[b]{1.12} \\
            
            & Disgust 
            & \heatmapcell{0.82} & \heatmapcell{0.78} & \heatmapcell{0.82} & \heatmapcell{0.75} & \heatmapcell{0.79} & \heatmapcell[b]{0.79} 
            & \maeheatmapcell{0.93} & \maeheatmapcell{1.06} & \maeheatmapcell{1.37} & \maeheatmapcell{1.20} & \maeheatmapcell{1.01} & \maeheatmapcell[b]{1.11} \\
            
            & Fear 
            & \heatmapcell{0.83} & \heatmapcell{0.82} & \heatmapcell{0.81} & \heatmapcell{0.75} & \heatmapcell{0.77} & \heatmapcell[b]{0.80} 
            & \maeheatmapcell{1.42} & \maeheatmapcell{1.41} & \maeheatmapcell{1.91} & \maeheatmapcell{1.62} & \maeheatmapcell{1.21} & \maeheatmapcell[b]{1.51} \\
            
            & Happiness 
            & \heatmapcell{0.85} & \heatmapcell{0.83} & \heatmapcell{0.84} & \heatmapcell{0.79} & \heatmapcell{0.77} & \heatmapcell[b]{0.82} 
            & \maeheatmapcell{1.30} & \maeheatmapcell{1.13} & \maeheatmapcell{0.89} & \maeheatmapcell{1.46} & \maeheatmapcell{1.20} & \maeheatmapcell[b]{1.20} \\
            
            & Sadness 
            & \heatmapcell{0.82} & \heatmapcell{0.76} & \heatmapcell{0.80} & \heatmapcell{0.77} & \heatmapcell{0.67} & \heatmapcell[b]{0.76} 
            & \maeheatmapcell{1.22} & \maeheatmapcell{1.14} & \maeheatmapcell{1.51} & \maeheatmapcell{1.17} & \maeheatmapcell{1.06} & \maeheatmapcell[b]{1.22} \\
            
            & Surprise 
            & \heatmapcell{0.48} & \heatmapcell{0.31} & \heatmapcell{0.40} & \heatmapcell{0.49} & \heatmapcell{0.24} & \heatmapcell[b]{0.38} 
            & \maeheatmapcell{1.52} & \maeheatmapcell{1.38} & \maeheatmapcell{1.31} & \maeheatmapcell{1.44} & \maeheatmapcell{1.49} & \maeheatmapcell[b]{1.43} \\
            
            & Emot.\ Val.\ 
            & \heatmapcell{0.86} & \heatmapcell{0.85} & \heatmapcell{0.83} & \heatmapcell{0.84} & \heatmapcell{0.84} & \heatmapcell[b]{0.84} 
            & \maeheatmapcell{1.16} & \maeheatmapcell{1.11} & \maeheatmapcell{0.77} & \maeheatmapcell{1.06} & \maeheatmapcell{1.15} & \maeheatmapcell[b]{1.05} \\
            
            & Arousal 
            & \heatmapcell{0.52} & \heatmapcell{0.52} & \heatmapcell{0.51} & \heatmapcell{0.45} & \heatmapcell{0.45} & \heatmapcell[b]{0.49} 
            & \maeheatmapcell{1.94} & \maeheatmapcell{2.06} & \maeheatmapcell{2.18} & \maeheatmapcell{1.80} & \maeheatmapcell{2.45} & \maeheatmapcell[b]{2.09} \\
            
            & \textbf{Average} 
            & \heatmapcell[b]{0.74} & \heatmapcell[b]{0.69} & \heatmapcell[b]{0.71} & \heatmapcell[b]{0.67} & \heatmapcell[b]{0.64} & \heatmapcell[b]{0.69} 
            & \maeheatmapcell[b]{1.33} & \maeheatmapcell[b]{1.29} & \maeheatmapcell[b]{1.41} & \maeheatmapcell[b]{1.35} & \maeheatmapcell[b]{1.33} & \maeheatmapcell[b]{1.34} \\
            
            \bottomrule
        \end{tabular}
    \end{threeparttable}
\end{table*}

\subsection{Performance on the Gemini-2.5-Flash Evaluated Subset}
The tables below present performance metrics calculated \textbf{only} on the subset of images that Gemini-2.5-Flash successfully processed (excluding 58 images from IAPS, 15 from NAPS, and 15 from LAI-GAI).

\begin{table*}[t]
    \centering
    \begin{threeparttable}
        \caption{Comparison of Model Accuracy (Gemini Subset)}
        \label{tab:gemini_subset_acc_table}
        \begin{tabular}{lrrr|r}
            \toprule
            Model & \multicolumn{4}{c}{Accuracy (\%)} \\
            \cmidrule(lr){2-5}
            & LAI-GAI & NAPS & IAPS & Average \\
            \midrule
            \textbf{Proprietary} \\
            GPT-4.1 & 65.16 & \textbf{77.71} & \textbf{64.32} & 69.06 \\
            \midrule
            \textbf{On H100} \\
            Qwen-VL-2.5-72B & 67.31 & \textbf{79.35} & \textbf{67.19} & \textbf{71.28} \\
            Intern-VL-3.5-38B & 67.96 & 79.14 & 60.88 & 69.33 \\
            Gemma3-27B & \textbf{75.05} & 71.98 & 56.94 & 67.99 \\
            Kimi-VL-A3B-16B & 71.40 & 72.39 & 59.46 & 67.75 \\
            \midrule
            \textbf{On RTX 4090} \\
            GLM-V-4.1-9B & 59.14 & 70.76 & 60.73 & 63.54 \\
            Intern-VL-3.5-8B & \textbf{66.24} & 76.69 & 62.46 & \textbf{68.46} \\
            Qwen-VL-2.5-7B & 58.28 & \textbf{77.71} & \textbf{67.51} & 67.83 \\
            LLaVA-Mistral-7B & 61.72 & 63.60 & 55.99 & 60.44 \\
            LLaVA-Vicuna-7B & 33.98 & 45.40 & 33.91 & 37.76 \\
            \bottomrule
        \end{tabular}
    \end{threeparttable}

    \vspace{0.8cm} 

    \begin{threeparttable}
        \caption{Per-Emotion F1 Scores (\%) on IAPS and NAPS Datasets (Gemini Subset)}
        \label{tab:gemini_subset_f1_iaps_naps}
        \begin{tabular}{lcccccc|c}
            \toprule
            \textbf{Model} & \textbf{Anger} & \textbf{Disgust} & \textbf{Fear} & \textbf{Happiness} & \textbf{Sadness} & \textbf{Surprise} & \textbf{Weighted F1} \\
            \midrule
            
            \multicolumn{8}{c}{\textbf{IAPS Dataset}} \\
            \textit{Support ($N$)} & \textit{(21)} & \textit{(51)} & \textit{(66)} & \textit{(362)} & \textit{(92)} & \textit{(42)} & \textit{(634)} \\
            \midrule
            GPT-4.1 & 11.43 & \textcolor{green}{55.24} & 56.59 & 79.87 & \textcolor{green}{62.37} & 15.38 & 66.81 \\
            Qwen-VL-2.5-72B & 15.00 & 50.85 & \textcolor{green}{59.39} & 84.44 & 61.29 & \textcolor{red}{5.06} & \textcolor{green}{68.22} \\
            Intern-VL-3.5-38B & 5.71 & 42.35 & 45.78 & 80.37 & 54.34 & 8.00 & 62.67 \\
            Gemma-3-27B & \textcolor{red}{0.00} & 53.10 & 56.44 & 69.14 & 57.87 & \textcolor{green}{16.67} & 59.12 \\
            Kimi-VL-A3B-16B & 7.41 & 46.40 & 59.12 & 76.02 & 63.95 & 14.37 & 63.77 \\
            GLM-V-4.1-9B & \textcolor{green}{25.40} & 40.00 & 46.88 & 81.06 & 47.85 & 7.14 & 62.64 \\
            Intern-VL-3.5-8B & 14.63 & 28.95 & 52.29 & 80.12 & 58.30 & 10.67 & 63.17 \\
            Qwen-VL-2.5-7B & 24.14 & 46.67 & 44.07 & \textcolor{green}{85.79} & 61.70 & 12.50 & 67.91 \\
            LLaVA-Mistral-7B & 6.45 & 29.51 & 48.08 & 74.88 & 54.35 & 11.59 & 59.00 \\
            LLaVA-Vicuna-7B & 7.69 & \textcolor{red}{24.60} & \textcolor{red}{34.41} & \textcolor{red}{47.93} & \textcolor{red}{34.21} & 15.75 & \textcolor{red}{39.19} \\
            \midrule
            \textbf{IAPS Average} & \textbf{12.37} & \textbf{43.35} & \textbf{51.23} & \textbf{75.13} & \textbf{56.66} & \textbf{12.18} & \textbf{61.77} \\
            \midrule
            \midrule

            \multicolumn{8}{c}{\textbf{NAPS Dataset}} \\
            \textit{Support ($N$)} & \textit{(7)} & \textit{(71)} & \textit{(23)} & \textit{(268)} & \textit{(101)} & \textit{(19)} & \textit{(489)} \\
            \midrule
            GPT-4.1 & \textcolor{red}{0.00} & 80.49 & 66.67 & 86.61 & 79.23 & 13.04 & 79.16 \\
            Qwen-VL-2.5-72B & 18.18 & 73.62 & 71.43 & 87.60 & \textcolor{green}{83.33} & 14.29 & 80.09 \\
            Intern-VL-3.5-38B & \textcolor{red}{0.00} & \textcolor{green}{82.86} & 52.50 & \textcolor{green}{89.11} & 77.57 & \textcolor{green}{21.62} & \textcolor{green}{80.20} \\
            Gemma-VL-3-27B & \textcolor{red}{0.00} & 79.53 & 73.08 & 78.00 & 73.86 & 12.50 & 73.48 \\
            Kimi-VL-A3B-16B & 22.22 & 70.39 & 70.83 & 82.83 & 77.32 & 13.70 & 75.77 \\
            GLM-V-4.1-9B & \textcolor{green}{28.57} & 64.39 & 43.08 & 84.89 & 65.54 & \textcolor{red}{0.00} & 71.84 \\
            Intern-VL-3.5-8B & 20.00 & 64.22 & \textcolor{green}{76.60} & 87.57 & 71.64 & 20.00 & 76.78 \\
            Qwen-VL-2.5-7B & 14.29 & 70.00 & 34.48 & 88.80 & 79.28 & \textcolor{red}{0.00} & 77.03 \\
            LLaVA-Mistral-7B & 20.00 & 58.25 & 29.51 & 77.78 & 76.11 & 16.00 & 69.10 \\
            LLaVA-Vicuna-7B & \textcolor{red}{0.00} & \textcolor{red}{42.59} & \textcolor{red}{22.22} & \textcolor{red}{58.64} & \textcolor{red}{47.62} & 7.50 & \textcolor{red}{49.49} \\
            \midrule
            \textbf{NAPS Average} & \textbf{13.48} & \textbf{68.90} & \textbf{56.32} & \textbf{82.20} & \textbf{73.70} & \textbf{11.23} & \textbf{73.55} \\
            \bottomrule
        \end{tabular}
    \end{threeparttable}
\end{table*}

\begin{table*}[t]
    \centering
    \begin{adjustbox}{max width=\linewidth,center}
        \begin{threeparttable}
            \caption{Per-Emotion F1 Scores (\%) on LAI-GAI Dataset (Gemini Subset)}
            \label{tab:gemini_subset_f1_laigai}
            \begin{tabular}{lcccccccccccc | c}
                \toprule
                \textbf{Model} & \textbf{Amusement} & \textbf{Anger} & \textbf{Attach. Love} & \textbf{Awe} & \textbf{Craving} & \textbf{Disgust} & \textbf{Excitement} & \textbf{Fear} & \textbf{Joy} & \textbf{Neutral} & \textbf{Nurt. Love} & \textbf{Sadness} & \textbf{Weighted F1} \\
                \textit{Support ($N$)} & \textit{(25)} & \textit{(14)} & \textit{(8)} & \textit{(30)} & \textit{(40)} & \textit{(26)} & \textit{(20)} & \textit{(42)} & \textit{(150)} & \textit{(42)} & \textit{(14)} & \textit{(54)} & \textit{(465)} \\
                \midrule
                GPT-4.1 & 74.58 & 23.53 & 25.00 & 66.67 & \textcolor{green}{100.00} & 73.53 & \textcolor{green}{80.00} & 81.63 & \textcolor{red}{50.24} & 62.86 & 52.17 & 73.47 & 64.96 \\
                Qwen-VL-2.5-72B & 73.33 & \textcolor{green}{58.33} & 31.37 & 68.49 & 59.65 & 80.00 & 65.38 & 85.71 & 64.00 & 54.84 & \textcolor{green}{60.87} & 83.33 & 67.92 \\
                Intern-VL-3.5-38B & 45.71 & 25.00 & 25.45 & 66.67 & 90.41 & \textcolor{green}{86.27} & 66.67 & 83.67 & 66.20 & 58.46 & \textcolor{red}{0.00} & \textcolor{green}{87.38} & 67.76 \\
                Gemma-3-27B & \textcolor{green}{75.00} & 11.76 & 45.45 & 67.53 & 98.77 & 85.71 & 66.67 & 87.91 & 74.16 & 64.79 & 52.63 & 81.30 & \textcolor{green}{74.42} \\
                Kimi-VL-A3B-16B & 57.14 & 25.00 & \textcolor{green}{50.00} & \textcolor{green}{76.36} & 98.77 & 58.82 & 77.27 & 76.74 & \textcolor{green}{77.51} & 70.27 & 10.00 & \textcolor{red}{66.67} & 71.04 \\
                GLM-V-4.1-9B & 41.03 & 56.25 & 26.32 & 59.52 & 91.89 & 71.43 & 53.33 & 76.92 & 53.11 & 54.29 & 15.38 & 68.75 & 59.82 \\
                Intern-VL-3.5-8B & 57.89 & 30.00 & 27.27 & 73.24 & \textcolor{red}{26.09} & 69.39 & 60.87 & 74.67 & 73.32 & 39.29 & \textcolor{red}{0.00} & 86.89 & 61.98 \\
                Qwen-VL-2.5-7B & 65.62 & 40.00 & 18.60 & 74.58 & 84.51 & 70.77 & 61.82 & 63.49 & 53.44 & 37.04 & \textcolor{red}{0.00} & 80.34 & 59.40 \\
                LLaVA-Mistral-7B & 58.54 & \textcolor{red}{11.11} & 13.33 & 60.24 & \textcolor{red}{26.09} & 73.47 & 37.04 & 75.86 & 70.59 & 31.37 & \textcolor{red}{0.00} & 77.04 & 56.94 \\
                LLaVA-Vicuna-7B & \textcolor{red}{19.35} & \textcolor{red}{11.11} & \textcolor{red}{0.00} & \textcolor{red}{18.59} & \textcolor{red}{26.09} & \textcolor{red}{55.56} & \textcolor{red}{0.00} & \textcolor{red}{4.44} & 53.15 & \textcolor{red}{0.00} & \textcolor{red}{0.00} & 73.13 & \textcolor{red}{33.97} \\
                \midrule
                \textbf{Average} & \textbf{56.82} & \textbf{29.11} & \textbf{26.28} & \textbf{63.19} & \textbf{70.22} & \textbf{72.50} & \textbf{56.91} & \textbf{71.10} & \textbf{63.57} & \textbf{47.32} & \textbf{19.11} & \textbf{77.83} & \textbf{61.82} \\
                \bottomrule
            \end{tabular}
        \end{threeparttable}
    \end{adjustbox}
    \vspace{1.0cm} 
    
    \scriptsize 
    \setlength{\tabcolsep}{3pt}
    \renewcommand{\arraystretch}{1.2} 
    \newcommand{\up}{\textcolor{gray}{$\uparrow$}}
    \newcommand{\down}{\textcolor{gray}{$\downarrow$}}
    \newcommand{\rot}[1]{\multicolumn{1}{c}{\rotatebox{45}{\textbf{#1}}}}
    
    \begin{threeparttable}
        \caption{Side-by-Side Comparison of Pearson's Correlation ($r$) and MAE (Gemini Subset)}
        \label{tab:gemini_subset_corr_mae}
        
        \begin{tabular}{ll | ccccc c | ccccc c}
            \toprule
            \multicolumn{2}{c}{} & 
            \multicolumn{6}{c}{\textbf{Pearson's Correlation ($r$)} \up} & 
            \multicolumn{6}{c}{\textbf{Mean Absolute Error (MAE)} \down} \\
            \cmidrule(lr){3-8} \cmidrule(lr){9-14}
            
            \multicolumn{2}{c}{\textbf{Scale}} & 
            \rot{GPT-4.1} & \rot{Qwen 72B} & \rot{Gemma 27B} & \rot{Intern 8B} & \rot{Qwen 7B} & \rot{Average} & 
            \rot{GPT-4.1} & \rot{Qwen 72B} & \rot{Gemma 27B} & \rot{Intern 8B} & \rot{Qwen 7B} & \rot{Average} \\
            \midrule
            
            \multirow{19}{*}{\rotatebox{90}{\textbf{LAI--GAI}}}
            & Amusement 
            & \heatmapcell{0.75} & \heatmapcell{0.79} & \heatmapcell{0.78} & \heatmapcell{0.73} & \heatmapcell{0.73} & \heatmapcell[b]{0.76} 
            & \maeheatmapcell{1.27} & \maeheatmapcell{0.92} & \maeheatmapcell{0.93} & \maeheatmapcell{1.30} & \maeheatmapcell{1.06} & \maeheatmapcell[b]{1.10} \\
            
            & Anger 
            & \heatmapcell{0.89} & \heatmapcell{0.76} & \heatmapcell{0.81} & \heatmapcell{0.64} & \heatmapcell{0.56} & \heatmapcell[b]{0.73} 
            & \maeheatmapcell{0.58} & \maeheatmapcell{0.58} & \maeheatmapcell{0.55} & \maeheatmapcell{0.65} & \maeheatmapcell{0.71} & \maeheatmapcell[b]{0.61} \\
            
            & Attach.\ Love 
            & \heatmapcell{0.86} & \heatmapcell{0.88} & \heatmapcell{0.87} & \heatmapcell{0.88} & \heatmapcell{0.87} & \heatmapcell[b]{0.87} 
            & \maeheatmapcell{1.35} & \maeheatmapcell{1.01} & \maeheatmapcell{0.90} & \maeheatmapcell{1.36} & \maeheatmapcell{1.28} & \maeheatmapcell[b]{1.18} \\
            
            & Awe 
            & \heatmapcell{0.57} & \heatmapcell{0.72} & \heatmapcell{0.70} & \heatmapcell{0.71} & \heatmapcell{0.73} & \heatmapcell[b]{0.69} 
            & \maeheatmapcell{1.17} & \maeheatmapcell{0.92} & \maeheatmapcell{0.71} & \maeheatmapcell{1.52} & \maeheatmapcell{0.98} & \maeheatmapcell[b]{1.06} \\
            
            & Craving 
            & \heatmapcell{0.89} & \heatmapcell{0.88} & \heatmapcell{0.90} & \heatmapcell{0.82} & \heatmapcell{0.90} & \heatmapcell[b]{0.88} 
            & \maeheatmapcell{0.91} & \maeheatmapcell{0.68} & \maeheatmapcell{0.45} & \maeheatmapcell{0.74} & \maeheatmapcell{0.50} & \maeheatmapcell[b]{0.66} \\
            
            & Disgust 
            & \heatmapcell{0.95} & \heatmapcell{0.91} & \heatmapcell{0.92} & \heatmapcell{0.89} & \heatmapcell{0.90} & \heatmapcell[b]{0.91} 
            & \maeheatmapcell{0.67} & \maeheatmapcell{0.51} & \maeheatmapcell{0.60} & \maeheatmapcell{0.73} & \maeheatmapcell{0.45} & \maeheatmapcell[b]{0.59} \\
            
            & Excitement 
            & \heatmapcell{0.59} & \heatmapcell{0.54} & \heatmapcell{0.57} & \heatmapcell{0.67} & \heatmapcell{0.70} & \heatmapcell[b]{0.61} 
            & \maeheatmapcell{1.23} & \maeheatmapcell{1.17} & \maeheatmapcell{0.98} & \maeheatmapcell{1.49} & \maeheatmapcell{1.02} & \maeheatmapcell[b]{1.18} \\
            
            & Fear 
            & \heatmapcell{0.96} & \heatmapcell{0.94} & \heatmapcell{0.94} & \heatmapcell{0.92} & \heatmapcell{0.88} & \heatmapcell[b]{0.93} 
            & \maeheatmapcell{0.74} & \maeheatmapcell{0.52} & \maeheatmapcell{0.59} & \maeheatmapcell{0.77} & \maeheatmapcell{0.48} & \maeheatmapcell[b]{0.62} \\
            
            & Joy 
            & \heatmapcell{0.94} & \heatmapcell{0.94} & \heatmapcell{0.92} & \heatmapcell{0.94} & \heatmapcell{0.90} & \heatmapcell[b]{0.93} 
            & \maeheatmapcell{1.19} & \maeheatmapcell{0.88} & \maeheatmapcell{0.77} & \maeheatmapcell{1.51} & \maeheatmapcell{0.93} & \maeheatmapcell[b]{1.06} \\
            
            & Neutral 
            & \heatmapcell{0.60} & \heatmapcell{0.61} & \heatmapcell{0.63} & \heatmapcell{0.40} & \heatmapcell{0.56} & \heatmapcell[b]{0.56} 
            & \maeheatmapcell{1.09} & \maeheatmapcell{1.12} & \maeheatmapcell{0.58} & \maeheatmapcell{1.10} & \maeheatmapcell{0.52} & \maeheatmapcell[b]{0.88} \\
            
            & Nurt.\ Love 
            & \heatmapcell{0.86} & \heatmapcell{0.87} & \heatmapcell{0.87} & \heatmapcell{0.84} & \heatmapcell{0.85} & \heatmapcell[b]{0.86} 
            & \maeheatmapcell{1.23} & \maeheatmapcell{1.05} & \maeheatmapcell{0.81} & \maeheatmapcell{1.37} & \maeheatmapcell{1.04} & \maeheatmapcell[b]{1.10} \\
            
            & Sadness 
            & \heatmapcell{0.95} & \heatmapcell{0.91} & \heatmapcell{0.89} & \heatmapcell{0.90} & \heatmapcell{0.87} & \heatmapcell[b]{0.90} 
            & \maeheatmapcell{0.51} & \maeheatmapcell{0.63} & \maeheatmapcell{0.75} & \maeheatmapcell{0.77} & \maeheatmapcell{0.54} & \maeheatmapcell[b]{0.64} \\
            
            & Positive 
            & \heatmapcell{0.96} & \heatmapcell{0.96} & \heatmapcell{0.95} & \heatmapcell{0.95} & \heatmapcell{0.96} & \heatmapcell[b]{0.96} 
            & \maeheatmapcell{1.00} & \maeheatmapcell{0.61} & \maeheatmapcell{0.53} & \maeheatmapcell{1.02} & \maeheatmapcell{0.68} & \maeheatmapcell[b]{0.77} \\
            
            & Negative 
            & \heatmapcell{0.97} & \heatmapcell{0.96} & \heatmapcell{0.94} & \heatmapcell{0.94} & \heatmapcell{0.94} & \heatmapcell[b]{0.95} 
            & \maeheatmapcell{0.72} & \maeheatmapcell{0.62} & \maeheatmapcell{0.63} & \maeheatmapcell{0.68} & \maeheatmapcell{0.53} & \maeheatmapcell[b]{0.64} \\
            
            & Calm 
            & \heatmapcell{0.86} & \heatmapcell{0.80} & \heatmapcell{0.88} & \heatmapcell{0.79} & \heatmapcell{0.77} & \heatmapcell[b]{0.82} 
            & \maeheatmapcell{0.95} & \maeheatmapcell{1.02} & \maeheatmapcell{0.88} & \maeheatmapcell{1.28} & \maeheatmapcell{1.00} & \maeheatmapcell[b]{1.03} \\
            
            & Aroused 
            & \heatmapcell{0.47} & \heatmapcell{0.44} & \heatmapcell{0.35} & \heatmapcell{0.58} & \heatmapcell{0.54} & \heatmapcell[b]{0.48} 
            & \maeheatmapcell{1.70} & \maeheatmapcell{2.06} & \maeheatmapcell{1.19} & \maeheatmapcell{1.90} & \maeheatmapcell{1.87} & \maeheatmapcell[b]{1.74} \\
            
            & Motiv.\ Appr.\ 
            & \heatmapcell{0.52} & \heatmapcell{0.90} & \heatmapcell{0.86} & \heatmapcell{0.85} & \heatmapcell{0.86} & \heatmapcell[b]{0.80} 
            & \maeheatmapcell{1.54} & \maeheatmapcell{1.03} & \maeheatmapcell{0.77} & \maeheatmapcell{1.25} & \maeheatmapcell{0.87} & \maeheatmapcell[b]{1.09} \\
            
            & Motiv.\ Avoid 
            & \heatmapcell{0.96} & \heatmapcell{0.94} & \heatmapcell{0.93} & \heatmapcell{0.92} & \heatmapcell{0.92} & \heatmapcell[b]{0.93} 
            & \maeheatmapcell{1.15} & \maeheatmapcell{0.78} & \maeheatmapcell{0.83} & \maeheatmapcell{0.83} & \maeheatmapcell{0.50} & \maeheatmapcell[b]{0.82} \\
            
            & \textbf{Average} 
            & \heatmapcell[b]{0.82} & \heatmapcell[b]{0.83} & \heatmapcell[b]{0.83} & \heatmapcell[b]{0.80} & \heatmapcell[b]{0.81} & \heatmapcell[b]{0.82} 
            & \maeheatmapcell[b]{1.06} & \maeheatmapcell[b]{0.90} & \maeheatmapcell[b]{0.75} & \maeheatmapcell[b]{1.13} & \maeheatmapcell[b]{0.83} & \maeheatmapcell[b]{0.93} \\
            \midrule
            
            \multirow{9}{*}{\rotatebox{90}{\textbf{IAPS}}}
            & Anger 
            & \heatmapcell{0.82} & \heatmapcell{0.65} & \heatmapcell{0.68} & \heatmapcell{0.58} & \heatmapcell{0.58} & \heatmapcell[b]{0.66} 
            & \maeheatmapcell{0.79} & \maeheatmapcell{1.00} & \maeheatmapcell{1.29} & \maeheatmapcell{0.97} & \maeheatmapcell{0.97} & \maeheatmapcell[b]{1.00} \\
            
            & Disgust 
            & \heatmapcell{0.86} & \heatmapcell{0.73} & \heatmapcell{0.78} & \heatmapcell{0.69} & \heatmapcell{0.74} & \heatmapcell[b]{0.76} 
            & \maeheatmapcell{0.93} & \maeheatmapcell{1.01} & \maeheatmapcell{1.32} & \maeheatmapcell{1.15} & \maeheatmapcell{0.96} & \maeheatmapcell[b]{1.07} \\
            
            & Fear 
            & \heatmapcell{0.87} & \heatmapcell{0.81} & \heatmapcell{0.80} & \heatmapcell{0.74} & \heatmapcell{0.75} & \heatmapcell[b]{0.79} 
            & \maeheatmapcell{1.37} & \maeheatmapcell{1.31} & \maeheatmapcell{1.81} & \maeheatmapcell{1.51} & \maeheatmapcell{1.11} & \maeheatmapcell[b]{1.42} \\
            
            & Happiness 
            & \heatmapcell{0.86} & \heatmapcell{0.82} & \heatmapcell{0.82} & \heatmapcell{0.77} & \heatmapcell{0.76} & \heatmapcell[b]{0.81} 
            & \maeheatmapcell{1.17} & \maeheatmapcell{1.17} & \maeheatmapcell{0.92} & \maeheatmapcell{1.52} & \maeheatmapcell{1.25} & \maeheatmapcell[b]{1.21} \\
            
            & Sadness 
            & \heatmapcell{0.86} & \heatmapcell{0.72} & \heatmapcell{0.77} & \heatmapcell{0.72} & \heatmapcell{0.65} & \heatmapcell[b]{0.74} 
            & \maeheatmapcell{0.98} & \maeheatmapcell{1.12} & \maeheatmapcell{1.50} & \maeheatmapcell{1.13} & \maeheatmapcell{1.03} & \maeheatmapcell[b]{1.15} \\
            
            & Surprise 
            & \heatmapcell{0.50} & \heatmapcell{0.31} & \heatmapcell{0.42} & \heatmapcell{0.42} & \heatmapcell{0.31} & \heatmapcell[b]{0.39} 
            & \maeheatmapcell{1.61} & \maeheatmapcell{1.35} & \maeheatmapcell{1.27} & \maeheatmapcell{1.44} & \maeheatmapcell{1.41} & \maeheatmapcell[b]{1.42} \\
            
            & Emot.\ Val.\ 
            & \heatmapcell{0.85} & \heatmapcell{0.84} & \heatmapcell{0.84} & \heatmapcell{0.81} & \heatmapcell{0.82} & \heatmapcell[b]{0.83} 
            & \maeheatmapcell{1.37} & \maeheatmapcell{1.15} & \maeheatmapcell{0.79} & \maeheatmapcell{1.09} & \maeheatmapcell{1.19} & \maeheatmapcell[b]{1.12} \\
            
            & Arousal 
            & \heatmapcell{0.54} & \heatmapcell{0.51} & \heatmapcell{0.50} & \heatmapcell{0.49} & \heatmapcell{0.47} & \heatmapcell[b]{0.50} 
            & \maeheatmapcell{2.02} & \maeheatmapcell{1.99} & \maeheatmapcell{2.10} & \maeheatmapcell{1.77} & \maeheatmapcell{2.42} & \maeheatmapcell[b]{2.06} \\
            
            & \textbf{Average} 
            & \heatmapcell[b]{0.77} & \heatmapcell[b]{0.67} & \heatmapcell[b]{0.70} & \heatmapcell[b]{0.65} & \heatmapcell[b]{0.64} & \heatmapcell[b]{0.69} 
            & \maeheatmapcell[b]{1.28} & \maeheatmapcell[b]{1.26} & \maeheatmapcell[b]{1.38} & \maeheatmapcell[b]{1.32} & \maeheatmapcell[b]{1.29} & \maeheatmapcell[b]{1.31} \\
            \midrule
            
            \multirow{9}{*}{\rotatebox{90}{\textbf{NAPS}}}
            & Anger 
            & \heatmapcell{0.77} & \heatmapcell{0.58} & \heatmapcell{0.61} & \heatmapcell{0.49} & \heatmapcell{0.44} & \heatmapcell[b]{0.58} 
            & \maeheatmapcell{0.48} & \maeheatmapcell{0.55} & \maeheatmapcell{0.86} & \maeheatmapcell{0.60} & \maeheatmapcell{0.47} & \maeheatmapcell[b]{0.59} \\
            
            & Disgust 
            & \heatmapcell{0.85} & \heatmapcell{0.85} & \heatmapcell{0.83} & \heatmapcell{0.78} & \heatmapcell{0.81} & \heatmapcell[b]{0.82} 
            & \maeheatmapcell{1.08} & \maeheatmapcell{0.99} & \maeheatmapcell{1.24} & \maeheatmapcell{1.27} & \maeheatmapcell{0.84} & \maeheatmapcell[b]{1.08} \\
            
            & Fear 
            & \heatmapcell{0.88} & \heatmapcell{0.83} & \heatmapcell{0.84} & \heatmapcell{0.76} & \heatmapcell{0.73} & \heatmapcell[b]{0.81} 
            & \maeheatmapcell{0.83} & \maeheatmapcell{0.80} & \maeheatmapcell{1.20} & \maeheatmapcell{0.99} & \maeheatmapcell{0.53} & \maeheatmapcell[b]{0.87} \\
            
            & Happiness 
            & \heatmapcell{0.91} & \heatmapcell{0.89} & \heatmapcell{0.88} & \heatmapcell{0.86} & \heatmapcell{0.88} & \heatmapcell[b]{0.88} 
            & \maeheatmapcell{1.04} & \maeheatmapcell{1.01} & \maeheatmapcell{0.77} & \maeheatmapcell{1.47} & \maeheatmapcell{0.88} & \maeheatmapcell[b]{1.03} \\
            
            & Sadness 
            & \heatmapcell{0.92} & \heatmapcell{0.87} & \heatmapcell{0.87} & \heatmapcell{0.82} & \heatmapcell{0.83} & \heatmapcell[b]{0.86} 
            & \maeheatmapcell{0.93} & \maeheatmapcell{0.94} & \maeheatmapcell{1.29} & \maeheatmapcell{1.15} & \maeheatmapcell{0.76} & \maeheatmapcell[b]{1.01} \\
            
            & Surprise 
            & \heatmapcell{0.61} & \heatmapcell{0.49} & \heatmapcell{0.53} & \heatmapcell{0.53} & \heatmapcell{0.22} & \heatmapcell[b]{0.48} 
            & \maeheatmapcell{1.02} & \maeheatmapcell{0.68} & \maeheatmapcell{0.87} & \maeheatmapcell{1.06} & \maeheatmapcell{0.59} & \maeheatmapcell[b]{0.84} \\
            
            & Emot.\ Val.\ 
            & \heatmapcell{0.90} & \heatmapcell{0.90} & \heatmapcell{0.86} & \heatmapcell{0.82} & \heatmapcell{0.89} & \heatmapcell[b]{0.87} 
            & \maeheatmapcell{1.39} & \maeheatmapcell{0.95} & \maeheatmapcell{1.23} & \maeheatmapcell{1.05} & \maeheatmapcell{1.08} & \maeheatmapcell[b]{1.14} \\
            
            & Arousal 
            & \heatmapcell{0.64} & \heatmapcell{0.47} & \heatmapcell{0.66} & \heatmapcell{0.24} & \heatmapcell{0.07} & \heatmapcell[b]{0.42} 
            & \maeheatmapcell{1.01} & \maeheatmapcell{1.28} & \maeheatmapcell{0.93} & \maeheatmapcell{1.26} & \maeheatmapcell{1.34} & \maeheatmapcell[b]{1.16} \\
            
            & \textbf{Average} 
            & \heatmapcell[b]{0.81} & \heatmapcell[b]{0.74} & \heatmapcell[b]{0.76} & \heatmapcell[b]{0.66} & \heatmapcell[b]{0.61} & \heatmapcell[b]{0.72} 
            & \maeheatmapcell[b]{0.97} & \maeheatmapcell[b]{0.90} & \maeheatmapcell[b]{1.05} & \maeheatmapcell[b]{1.11} & \maeheatmapcell[b]{0.81} & \maeheatmapcell[b]{0.97} \\
            
            \bottomrule
        \end{tabular}
    \end{threeparttable}
\end{table*}
\section{Rater-conditioned Prompting}

\begin{figure*}[b]
    \begin{adjustbox}{width=\textwidth,center}
        \subfloat[Gemini with ``Full Background'']{
        \includegraphics[width=0.5\linewidth]{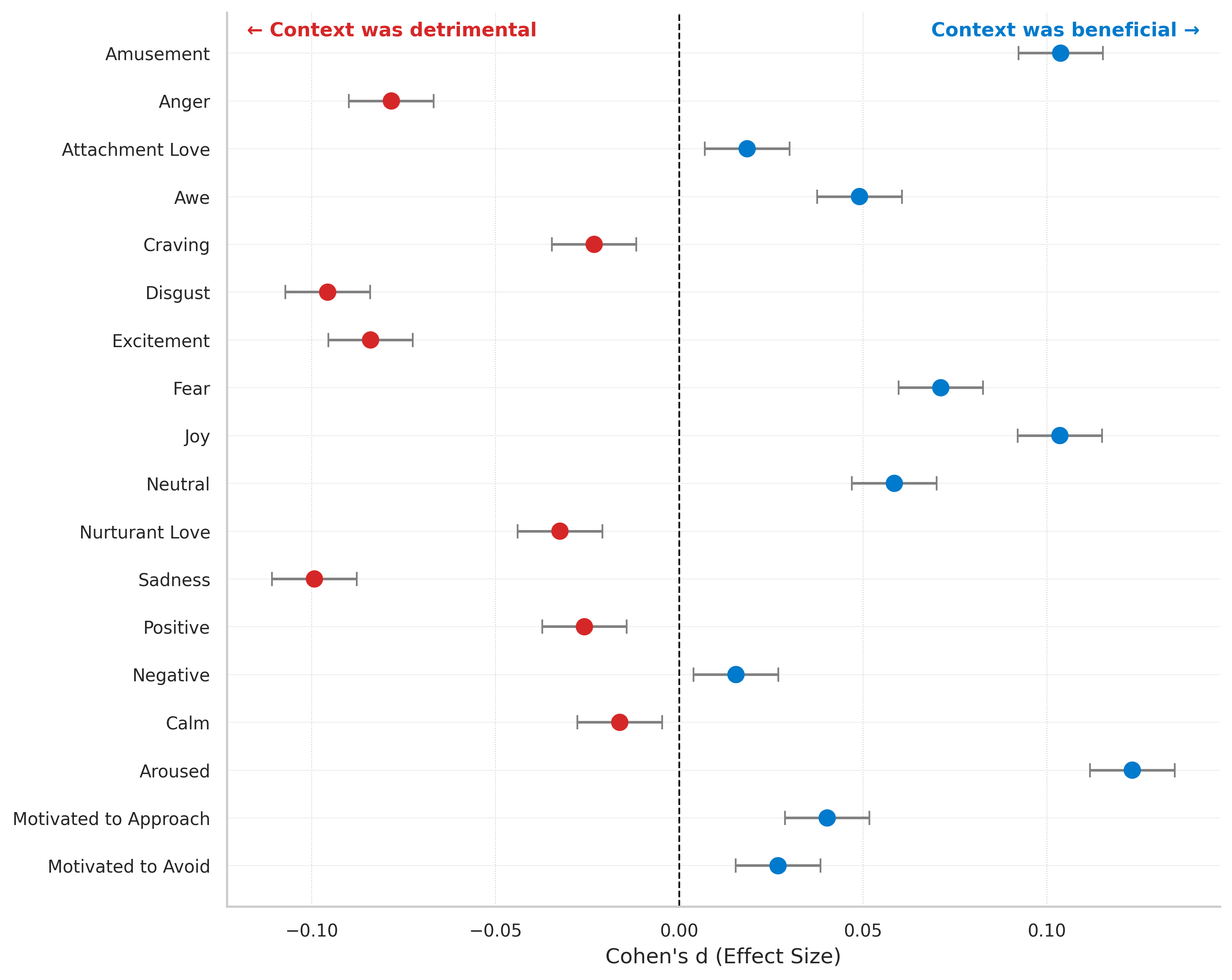}
        }
        \hfill
        \subfloat[Gemma with ``Full Background'']{
        \includegraphics[width=0.5\linewidth]{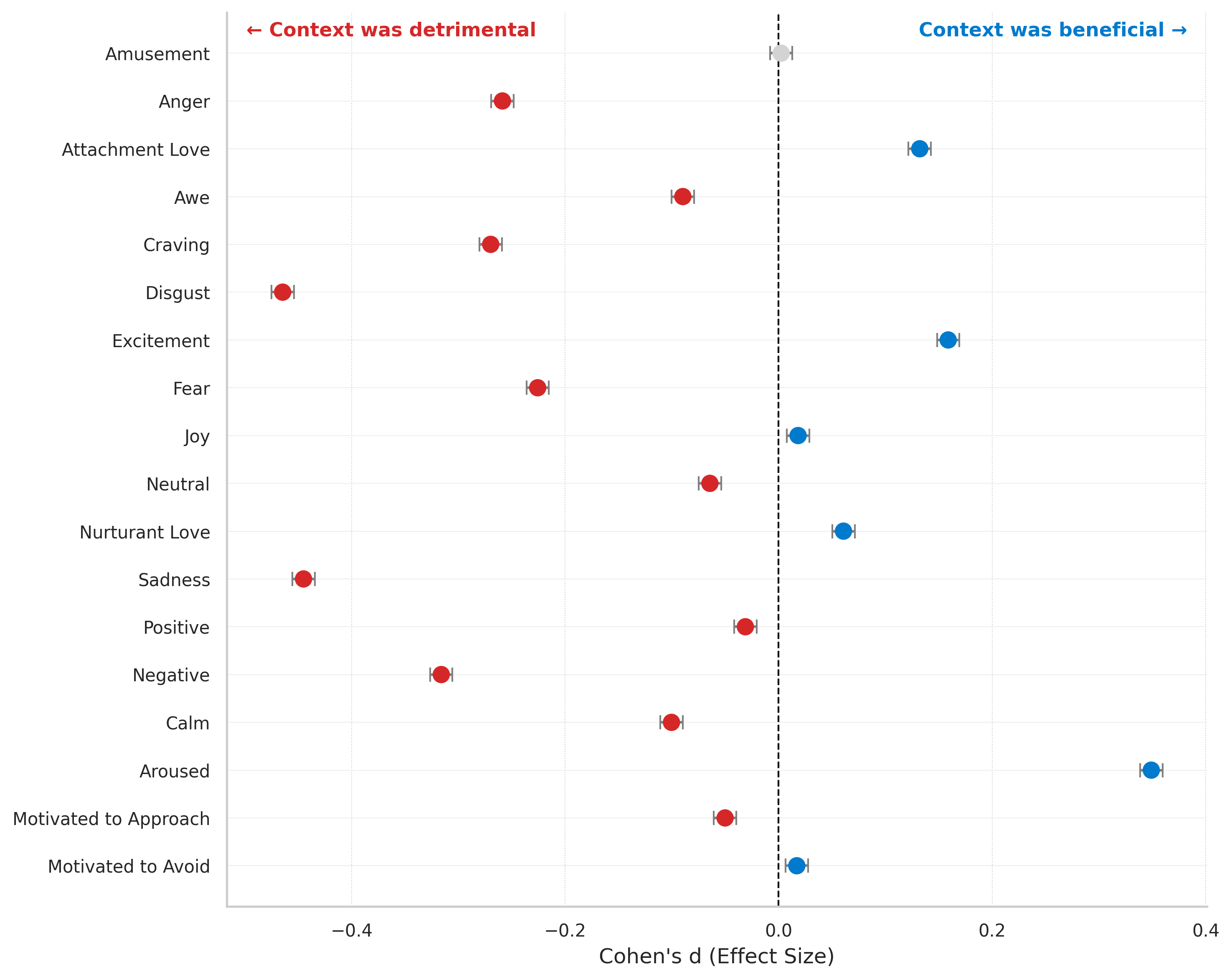}
        }  
    \end{adjustbox}
        \begin{adjustbox}{width=\textwidth,center}
        \subfloat[Gemini with ``Demographic Background'']{
            \includegraphics[width=0.5\linewidth]{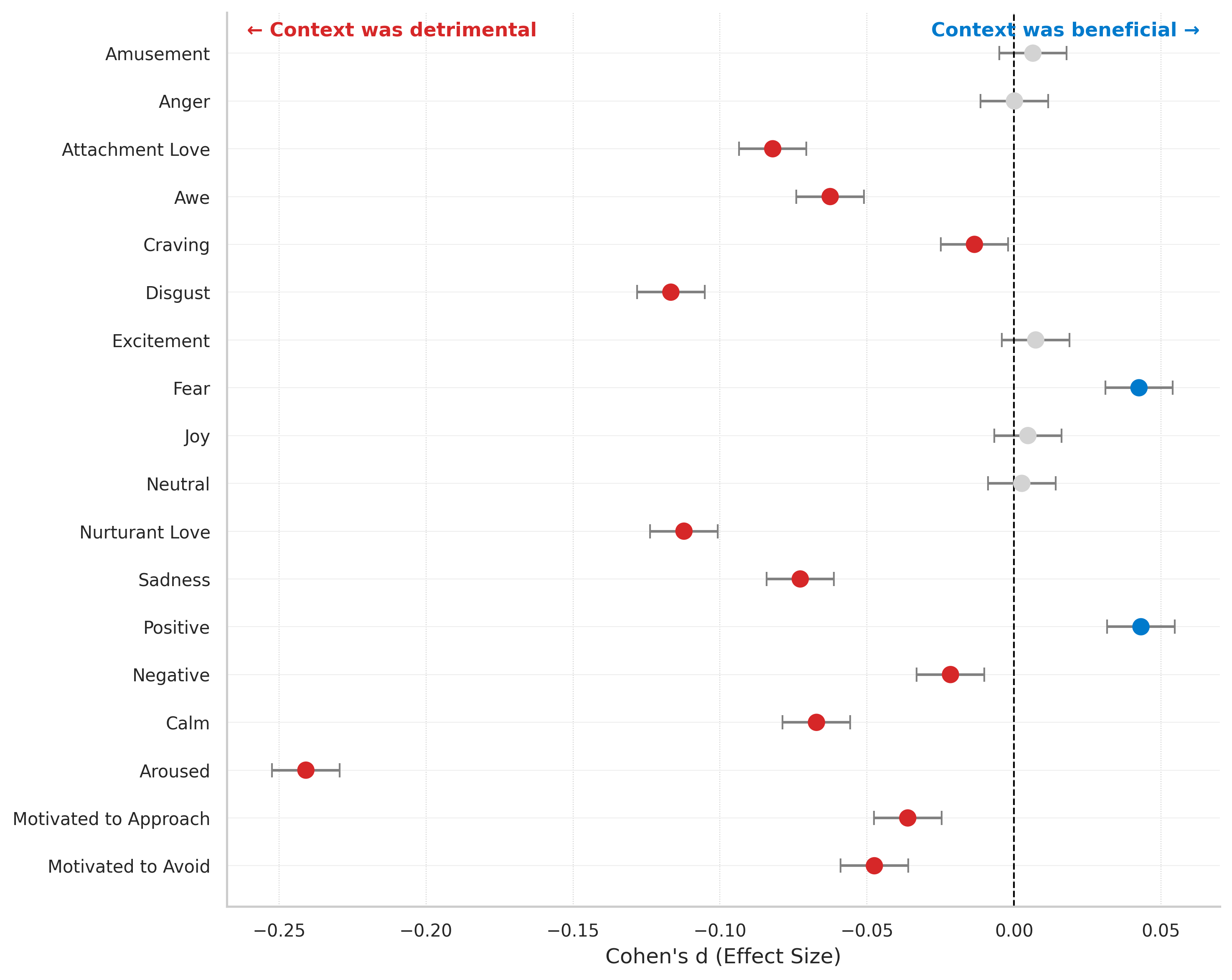}
        }
        \hfill 
        \subfloat[Gemma with ``Demographic Background'']{
            \includegraphics[width=0.5\linewidth]{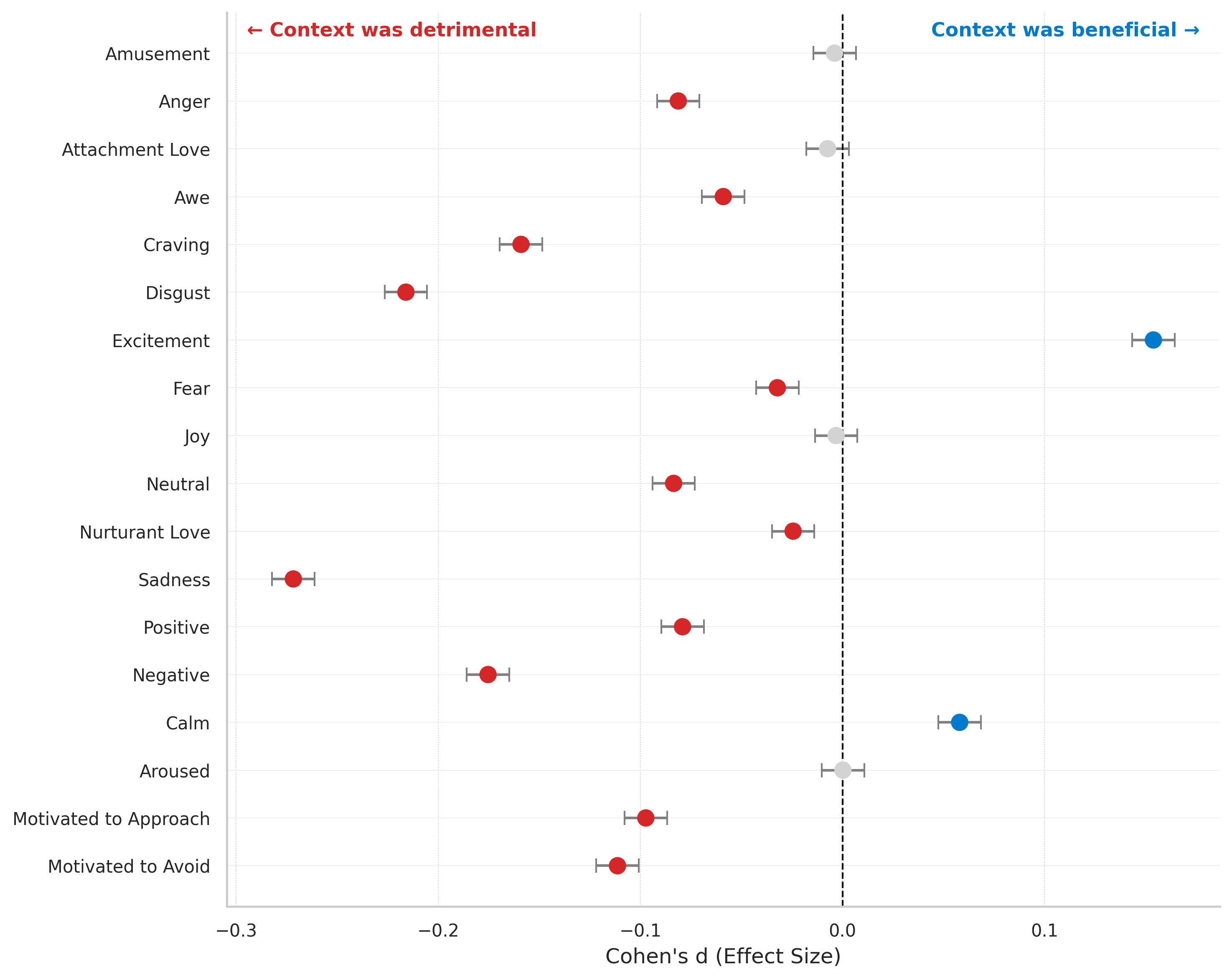}
        }   
    \end{adjustbox}
    \begin{adjustbox}{width=\textwidth,center}
        \subfloat[Gemini with ``Emotional Background'']{
            \includegraphics[width=0.5\linewidth]{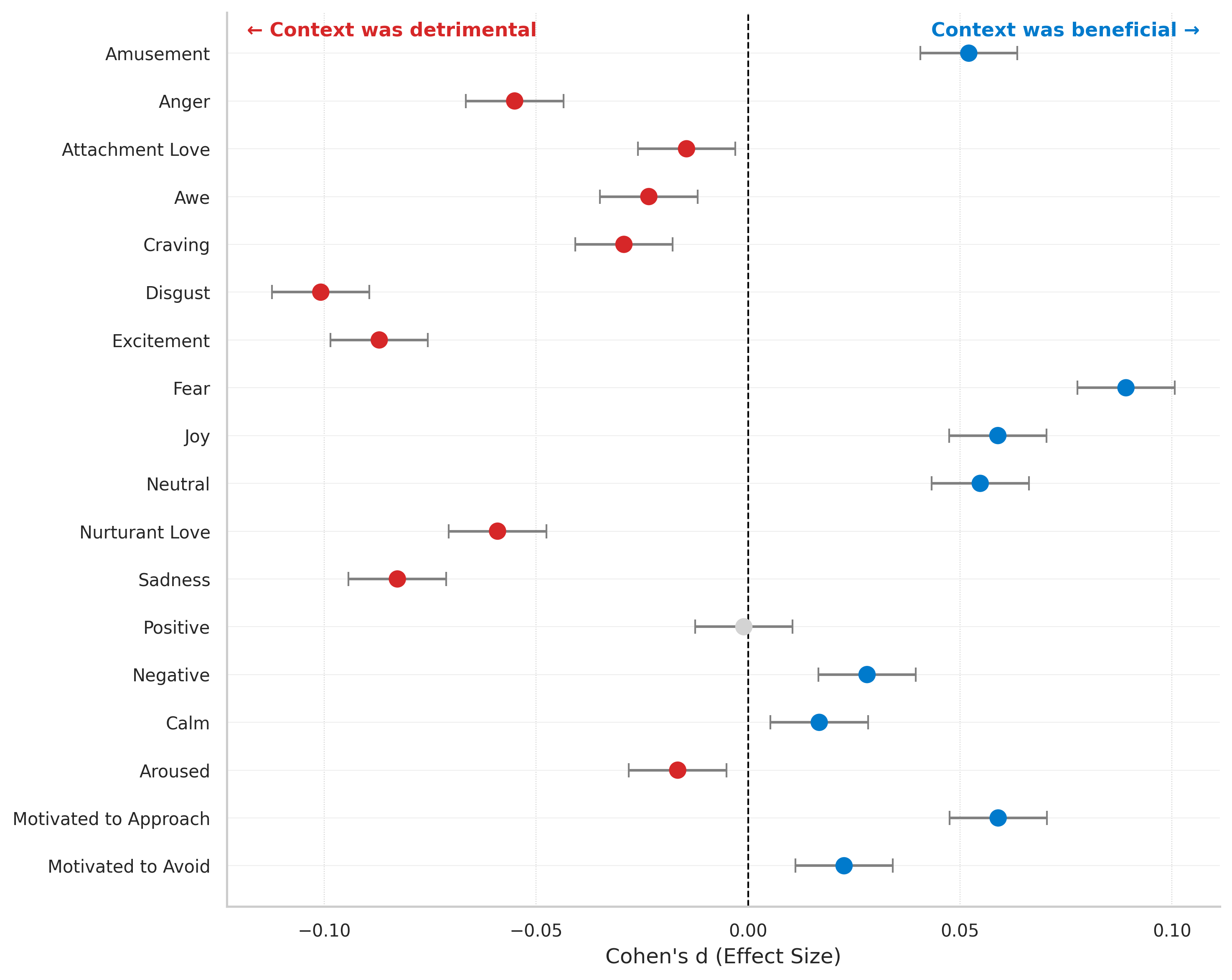}
        }
        \hfill
        \subfloat[Gemma with ``Emotional Background'']{
            \includegraphics[width=0.5\linewidth]{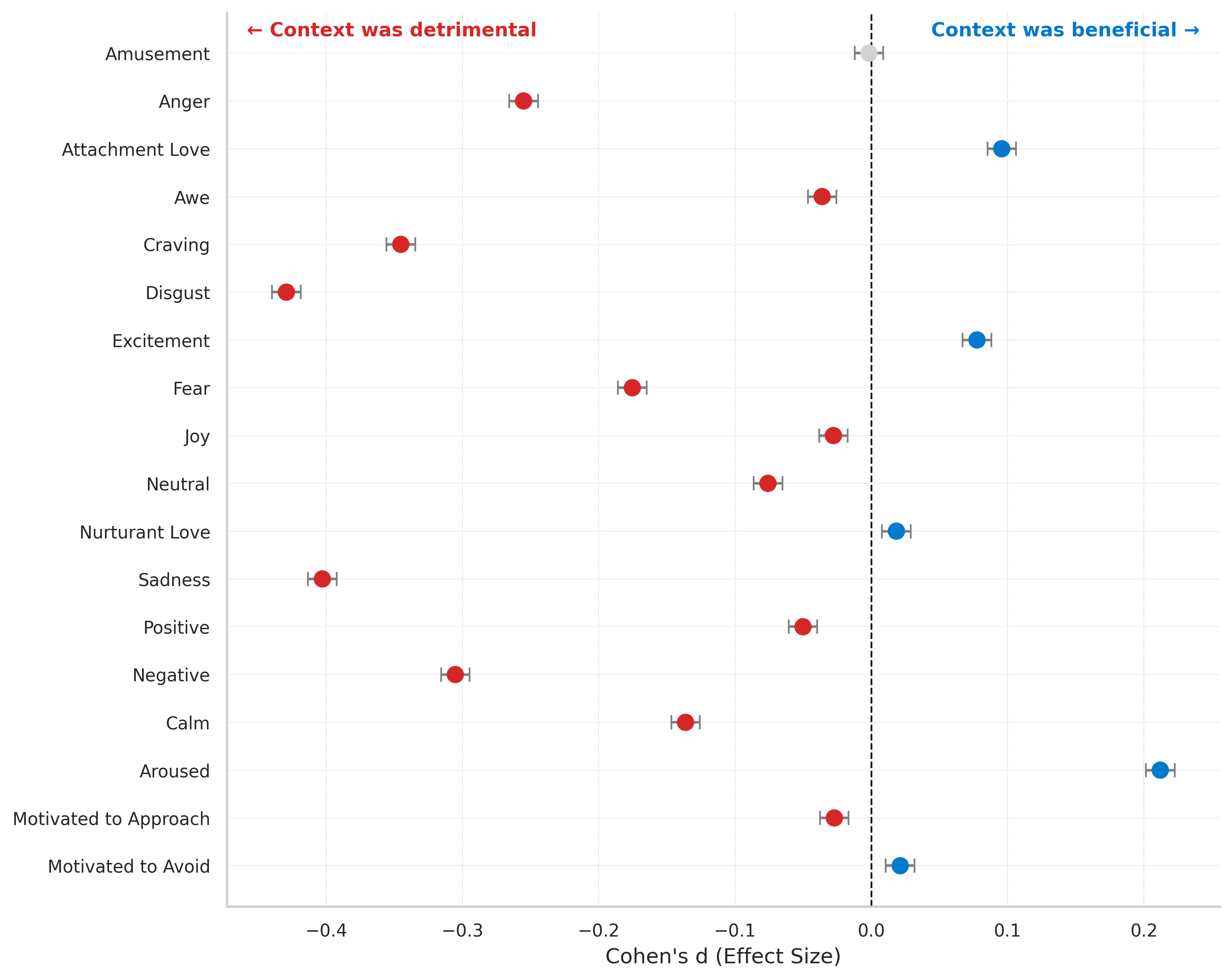}
        } 
    \end{adjustbox}
    \caption{
        Effect of participant background information on prediction accuracy for Gemini-2.5-Flash (a, c, e) and Gemma-3 27B (b, d, f). Rows denote background type. Plots report Cohen’s $d$ for the changes in MAE (positive = improved accuracy); error bars show 95\% CIs, with effects significant when the CI excludes zero.
    }
    \label{fig:context_effects_side_by_side}
\end{figure*}

To investigate whether providing models with specific human rater metadata improves their affective predictions, we conducted experiments using three types of background information: full context, demographic only, and emotional state only. We evaluated this on the LAI-GAI dataset using the Gemini-2.5-Flash and Gemma-3 27B models. As illustrated in Fig.~\ref{fig:context_effects_side_by_side}, the effects of rater-conditioned prompting were generally small and inconsistent across both models and all context variations. We quantified the impact using Cohen's $d$ for changes in Mean Absolute Error (MAE), where positive values indicate an improvement in prediction accuracy. While a few specific affective dimensions showed minor, statistically significant improvements (indicated by 95\% confidence intervals excluding zero), the overall trend suggests that explicitly injecting rater context yields only marginal and highly variable gains in zero-shot affective alignment.

\section{Human Inter-Rater Ceiling}
\label{sec:human_ceiling}

Because the normative ratings are themselves aggregates of noisy human judgments, no external predictor can be expected to correlate perfectly with them. We therefore estimated a human \emph{noise ceiling} from split-half inter-rater reliability: raters are randomly split into two halves, the half-means are correlated across images, and the resulting coefficient bounds the alignment any model can achieve. For \textbf{LAI-GAI} we use the official split-half reliabilities published with the dataset release~\cite{behnke2025using} (macro-average $r=.93$); our own split-half computation reproduces these values to within $\le.006$. For \textbf{IAPS}, whose public norms provide per-image means, standard deviations, and rater counts ($n=60$) but not rater-level responses, we estimate the reliability by variance decomposition, $R=(V_{\text{obs}}-\overline{SD^2/n})/V_{\text{obs}}$, and map it to the split-half scale for comparability. We omit \textbf{NAPS}, as our copy of its norms contains only means.

Table~\ref{tab:human_ceiling} reports the per-scale ceiling alongside the best model (Gemini-3-Flash) and the percentage of the ceiling it attains. Agreement is high for valence and discrete emotions but lowest for arousal, where the two rater halves themselves agree at only $r=.69$ (LAI-GAI) and $.85$ (IAPS). The model recovers $98\%$ or more of the ceiling for disgust, sadness, joy, and valence, but only $56$--$68\%$ for arousal, indicating that the arousal gap reflects a genuine limitation rather than label noise alone.

\begin{table*}[t]
    \centering
    \footnotesize
    \caption{Human split-half inter-rater ceiling per scale, the best model's correlation (Gemini-3-Flash), and the percentage of the ceiling attained. LAI-GAI ceilings are the official values~\cite{behnke2025using}; IAPS ceilings are estimated by variance decomposition. NAPS is omitted (our norms provide only means).}
    \label{tab:human_ceiling}
    \begin{tabular}{lccc}
        \toprule
        \textbf{Scale} & \textbf{Human ceiling} & \textbf{Gemini-3-Flash $r$} & \textbf{\% of ceiling} \\
        \midrule
        \multicolumn{4}{c}{\textbf{LAI-GAI}} \\
        \midrule
        Amusement        & .94 & .65 & 69 \\
        Anger            & .97 & .88 & 91 \\
        Attach.\ Love    & .93 & .86 & 93 \\
        Awe              & .89 & .75 & 85 \\
        Craving          & .94 & .89 & 95 \\
        Disgust          & .98 & .96 & 98 \\
        Excitement       & .94 & .63 & 67 \\
        Fear             & .97 & .93 & 96 \\
        Joy              & .97 & .95 & 98 \\
        Neutral          & .86 & .76 & 88 \\
        Nurt.\ Love      & .92 & .83 & 90 \\
        Sadness          & .98 & .96 & 98 \\
        Positive         & .98 & .97 & 99 \\
        Negative         & .99 & .97 & 98 \\
        Calm             & .96 & .86 & 90 \\
        Aroused          & .69 & .39 & 56 \\
        Motiv.\ Appr.\   & .95 & .92 & 97 \\
        Motiv.\ Avoid    & .97 & .96 & 99 \\
        \textbf{Macro avg.} & \textbf{.93} & \textbf{.84} & \textbf{89} \\
        \midrule
        \multicolumn{4}{c}{\textbf{IAPS}} \\
        \midrule
        Anger            & .96 & .84 & 87 \\
        Disgust          & .97 & .86 & 89 \\
        Fear             & .95 & .88 & 92 \\
        Happiness        & .96 & .87 & 90 \\
        Sadness          & .97 & .87 & 90 \\
        Surprise         & .92 & .59 & 65 \\
        Emot.\ Val.\     & .97 & .87 & 90 \\
        Arousal          & .85 & .58 & 68 \\
        \textbf{Macro avg.} & \textbf{.94} & \textbf{.80} & \textbf{84} \\
        \bottomrule
    \end{tabular}
\end{table*}

\section{Statistical Significance Testing}
\label{sec:stat_tests}

To quantify the reliability of the reported differences, we computed bootstrap $95\%$ confidence intervals ($B=10{,}000$ resamples over images) for every weighted-F1 (Task~1) and per-scale Pearson $r$ (Task~2) value. Model pairs were compared with paired tests on their shared image set: a label-swap permutation test for classification weighted F1, and a paired bootstrap (shared resampling draws) for the Task~2 mean correlation. Table~\ref{tab:cls_ci} reports the per-model weighted-F1 intervals; Tables~\ref{tab:cls_pairwise} and~\ref{tab:reg_pairwise} report the key pairwise comparisons. Full per-scale $r$ intervals are released with the evaluation code.

\begin{table*}[t]
    \centering
    \footnotesize
    \begin{threeparttable}
    \caption{Weighted-F1 ($\%$) with bootstrap $95\%$ confidence intervals per dataset.}
    \label{tab:cls_ci}
    \begin{tabular}{lccc}
        \toprule
        \textbf{Model} & \textbf{LAI-GAI} & \textbf{NAPS} & \textbf{IAPS} \\
        \midrule
        GPT-4.1\tnote{*}          & 65.2 (60.6--69.7) & 79.4 (75.9--82.8) & 65.6 (62.0--69.1) \\
        GPT-5.4                   & 63.0 (58.1--67.8) & 83.1 (79.8--86.4) & 67.3 (63.7--70.8) \\
        Gemini-2.5-Flash\tnote{*} & 70.5 (66.2--74.8) & 76.2 (72.4--79.9) & 66.9 (63.4--70.5) \\
        Gemini-3-Pro              & 61.8 (57.0--66.6) & 82.9 (79.3--86.2) & 66.2 (62.6--69.8) \\
        Gemini-3-Flash            & 61.0 (56.2--65.9) & 84.2 (80.8--87.5) & 68.5 (64.9--72.1) \\
        Qwen-VL-2.5-72B           & 68.3 (64.0--72.6) & 80.4 (76.9--83.9) & 65.3 (61.7--68.9) \\
        Qwen-VL-3-Think-30B       & 69.5 (65.2--73.6) & 77.4 (73.6--81.0) & 63.5 (59.9--67.2) \\
        Intern-VL-3.5-38B         & 67.4 (63.0--71.7) & 79.7 (76.3--83.0) & 60.0 (56.3--63.7) \\
        Gemma-3-27B               & 74.8 (70.8--78.8) & 74.0 (70.2--77.8) & 57.9 (54.4--61.6) \\
        Kimi-VL-A3B-16B           & 70.7 (66.4--74.7) & 75.8 (72.2--79.4) & 61.9 (58.3--65.5) \\
        Qwen-VL-3-Think-8B        & 62.4 (57.9--66.8) & 78.0 (74.5--81.5) & 60.8 (57.2--64.4) \\
        Intern-VL-3.5-8B          & 63.0 (58.0--67.7) & 75.9 (72.1--79.7) & 60.7 (57.0--64.4) \\
        Qwen-VL-2.5-7B            & 59.8 (55.2--64.3) & 77.3 (73.3--81.0) & 64.7 (61.0--68.4) \\
        GLM-V-4.1-9B              & 59.4 (54.9--63.7) & 71.7 (68.0--75.5) & 59.9 (56.3--63.5) \\
        LLaVA-Mistral-7B          & 57.3 (52.2--62.3) & 67.9 (63.9--71.8) & 56.2 (52.3--59.9) \\
        LLaVA-Vicuna-7B           & 35.5 (30.8--39.9) & 50.0 (45.5--54.2) & 37.3 (33.4--41.0) \\
        \bottomrule
    \end{tabular}
    \begin{tablenotes}
        \item[*] Computed on non-blocked images (see Section~\ref{sec:content_safety}).
    \end{tablenotes}
    \end{threeparttable}
\end{table*}

\begin{table*}[t]
    \centering
    \footnotesize
    \begin{threeparttable}
    \caption{Classification weighted-F1 pairwise comparisons (paired permutation test on shared images).}
    \label{tab:cls_pairwise}
    \begin{tabular}{llccc}
        \toprule
        \textbf{Data} & \textbf{Comparison} & \textbf{$\Delta$F1} & \textbf{95\% CI} & \textbf{$p$} \\
        \midrule
        NAPS    & Qwen-72B vs GPT-4.1            & $+1.0$  & $[-1.9, 3.9]$ & $.49$ \\
        NAPS    & Gemini-3-Flash vs Qwen-72B    & $+3.8$  & $[1.0, 6.6]$  & $.009$ \\
        NAPS    & GPT-5.4 vs GPT-4.1            & $+3.7$  & $[1.2, 6.4]$  & $.003$ \\
        IAPS    & Gemini-3-Flash vs Gemini-2.5\tnote{$\dagger$} & $+4.9$ & $[2.2, 7.6]$ & $<.001$ \\
        IAPS    & GPT-5.4 vs GPT-4.1            & $+3.2$  & $[0.9, 5.6]$  & $.007$ \\
        LAI-GAI & Gemma-27B vs Gemini-2.5        & $+3.9$  & $[-0.2, 8.1]$ & $.065$ \\
        LAI-GAI & Gemini-2.5 vs Gemini-3-Flash   & $+10.3$ & $[6.1, 14.6]$ & $<.001$ \\
        \bottomrule
    \end{tabular}
    \begin{tablenotes}
        \item[$\dagger$] Computed on the 634 images both models processed.
    \end{tablenotes}
    \end{threeparttable}
\end{table*}

\begin{table*}[t]
    \centering
    \footnotesize
    \caption{Task~2 mean-correlation comparisons of Gemini-3-Flash against each Task~2 model (paired bootstrap on shared images).}
    \label{tab:reg_pairwise}
    \begin{tabular}{llccc}
        \toprule
        \textbf{Data} & \textbf{vs.\ Model} & \textbf{$\Delta\bar{r}$} & \textbf{95\% CI} & \textbf{$p$} \\
        \midrule
        IAPS    & Gemini-2.5-Flash & $+.058$ & $[.042, .075]$ & $<.001$ \\
        IAPS    & GPT-4.1          & $+.013$ & $[.002, .023]$ & $.029$ \\
        IAPS    & Qwen-VL-2.5-72B  & $+.117$ & $[.100, .136]$ & $<.001$ \\
        IAPS    & Gemma-3-27B      & $+.089$ & $[.075, .103]$ & $<.001$ \\
        IAPS    & Intern-VL-3.5-8B & $+.126$ & $[.106, .146]$ & $<.001$ \\
        IAPS    & Qwen-VL-2.5-7B   & $+.166$ & $[.145, .186]$ & $<.001$ \\
        \midrule
        NAPS    & Gemini-2.5-Flash & $+.068$ & $[.051, .085]$ & $<.001$ \\
        NAPS    & GPT-4.1          & $+.052$ & $[.036, .067]$ & $<.001$ \\
        NAPS    & Qwen-VL-2.5-72B  & $+.130$ & $[.110, .149]$ & $<.001$ \\
        NAPS    & Gemma-3-27B      & $+.114$ & $[.094, .134]$ & $<.001$ \\
        NAPS    & Intern-VL-3.5-8B & $+.203$ & $[.177, .228]$ & $<.001$ \\
        NAPS    & Qwen-VL-2.5-7B   & $+.270$ & $[.239, .301]$ & $<.001$ \\
        \midrule
        LAI-GAI & Gemini-2.5-Flash & $-.001$ & $[-.007, .006]$ & $.96$ \\
        LAI-GAI & GPT-4.1          & $+.030$ & $[.024, .037]$ & $<.001$ \\
        LAI-GAI & Qwen-VL-2.5-72B  & $+.015$ & $[.006, .024]$ & $<.001$ \\
        LAI-GAI & Gemma-3-27B      & $+.016$ & $[.007, .025]$ & $<.001$ \\
        LAI-GAI & Intern-VL-3.5-8B & $+.034$ & $[.021, .047]$ & $<.001$ \\
        LAI-GAI & Qwen-VL-2.5-7B   & $+.031$ & $[.020, .042]$ & $<.001$ \\
        \bottomrule
    \end{tabular}
\end{table*}

\end{document}